\input harvmac
\input epsf

% we generate labeldefs.tmp by "writedefs" below -- sheer
\input labeldefs.tmp  

% write out defs so we can forward refer to things (see harvsamp.tex)
  \writedefs

\overfullrule=0pt
%\draftmode

\def\Title#1#2{\rightline{#1}\ifx\answ\bigans
\nopagenumbers\pageno0\vskip1in
\else\pageno1\vskip.8in\fi \centerline{\titlefont #2}\vskip .5in}
\newcount\figno
\figno=0
\def\fig#1#2#3{
\par\begingroup\parindent=0pt\leftskip=1cm\rightskip=1cm\parindent=0pt
\baselineskip=11pt
\global\advance\figno by 1
\midinsert
\epsfxsize=#3
\centerline{\epsfbox{#2}} 
\vskip 12pt 
{\bf Figure\ \the\figno: } #1\par
\endinsert\endgroup\par
}
\def\figlabel#1{\xdef#1{\the\figno}}
\def\encadremath#1{\vbox{\hrule\hbox{\vrule\kern8pt\vbox{\kern8pt
\hbox{$\displaystyle #1$}\kern8pt}
\kern8pt\vrule}\hrule}}

%\MaldacenaRE
\lref\MaldacenaRE{
  J.~M.~Maldacena,
  ``The large N limit of superconformal field theories and supergravity,''
  Adv.\ Theor.\ Math.\ Phys.\  {\bf 2}, 231 (1998)
  [Int.\ J.\ Theor.\ Phys.\  {\bf 38}, 1113 (1999)]
  [arXiv:hep-th/9711200].
  %%CITATION = IJTPB,38,1113;%%
}

%\GubserBC
\lref\GubserBC{
  S.~S.~Gubser, I.~R.~Klebanov and A.~M.~Polyakov,
  ``Gauge theory correlators from non-critical string theory,''
  Phys.\ Lett.\  B {\bf 428}, 105 (1998)
  [arXiv:hep-th/9802109].
  %%CITATION = PHLTA,B428,105;%%
}

%\WittenQJ
\lref\WittenQJ{
  E.~Witten,
  ``Anti-de Sitter space and holography,''
  Adv.\ Theor.\ Math.\ Phys.\  {\bf 2}, 253 (1998)
  [arXiv:hep-th/9802150].
  %%CITATION = 00203,2,253;%%
}

%\HeemskerkPN
\lref\HeemskerkPN{
  I.~Heemskerk, J.~Penedones, J.~Polchinski and J.~Sully,
  ``Holography from Conformal Field Theory,''
  JHEP {\bf 0910}, 079 (2009)
  [arXiv:0907.0151 [hep-th]].
  %%CITATION = JHEPA,0910,079;%%
}
%\BanksDD
\lref\BanksDD{
  T.~Banks, M.~R.~Douglas, G.~T.~Horowitz and E.~J.~Martinec,
  ``AdS dynamics from conformal field theory,''
  arXiv:hep-th/9808016.
  %%CITATION = HEP-TH/9808016;%%
}

%\HamiltonJU
\lref\HamiltonJU{
  A.~Hamilton, D.~N.~Kabat, G.~Lifschytz and D.~A.~Lowe,
  ``Local bulk operators in AdS/CFT: A boundary view of horizons and
  locality,''
  Phys.\ Rev.\  D {\bf 73}, 086003 (2006)
  [arXiv:hep-th/0506118].
  %%CITATION = PHRVA,D73,086003;%%
}

%\HamiltonAZ
\lref\HamiltonAZ{
  A.~Hamilton, D.~N.~Kabat, G.~Lifschytz and D.~A.~Lowe,
  ``Holographic representation of local bulk operators,''
  Phys.\ Rev.\  D {\bf 74}, 066009 (2006)
  [arXiv:hep-th/0606141].
  %%CITATION = PHRVA,D74,066009;%%
}
%\KochCY
\lref\KochCY{
  R.~d.~M.~Koch, A.~Jevicki, K.~Jin and J.~P.~Rodrigues,
  ``$AdS_4/CFT_3$ Construction from Collective Fields,''
Phys.\ Rev.\ D {\bf 83}, 025006 (2011).
[arXiv:1008.0633 [hep-th]].
%%CITATION = arXiv:1008.0633%%
}

%\DasVW
\lref\DasVW{
  S.~R.~Das and A.~Jevicki,
  ``Large N collective fields and holography,''
Phys.\ Rev.\ D {\bf 68}, 044011 (2003).
[hep-th/0304093].
%%CITATION = hep-th/0304093%%
}

%\HamiltonFH
\lref\HamiltonFH{
  A.~Hamilton, D.~N.~Kabat, G.~Lifschytz and D.~A.~Lowe,
  ``Local bulk operators in AdS/CFT: A holographic description of the black
  hole interior,''
  Phys.\ Rev.\  D {\bf 75}, 106001 (2007)
  [Erratum-ibid.\  D {\bf 75}, 129902 (2007)]
  [arXiv:hep-th/0612053].
  %%CITATION = PHRVA,D75,106001;%%
}

%\HamiltonWJ
\lref\HamiltonWJ{
  A.~Hamilton, D.~N.~Kabat, G.~Lifschytz and D.~A.~Lowe,
  ``Local bulk operators in AdS/CFT and the fate of the BTZ singularity,''
  arXiv:0710.4334 [hep-th].
  %%CITATION = ARXIV:0710.4334;%%
}

%\BalasubramanianRI
\lref\BalasubramanianRI{
  V.~Balasubramanian, S.~B.~Giddings and A.~E.~Lawrence,
  ``What do CFTs tell us about anti-de Sitter spacetimes?,''
  JHEP {\bf 9903}, 001 (1999)
  [arXiv:hep-th/9902052].
  %%CITATION = JHEPA,9903,001;%%
}

%\BenaJV
\lref\BenaJV{
  I.~Bena,
  ``On the construction of local fields in the bulk of AdS(5) and other
  spaces,''
  Phys.\ Rev.\  D {\bf 62}, 066007 (2000)
  [arXiv:hep-th/9905186].
  %%CITATION = PHRVA,D62,066007;%%
}

%\KlebanovJA
\lref\KlebanovJA{
  I.~R.~Klebanov and A.~M.~Polyakov,
  ``AdS dual of the critical O(N) vector model,''
  Phys.\ Lett.\  B {\bf 550}, 213 (2002)
  [arXiv:hep-th/0210114].
  %%CITATION = PHLTA,B550,213;%%
}

\lref\wittentalk{
E.~Witten, Spacetime Reconstruction, Talk at JHS 60 Conference, Caltech,
3-4 November 2001, http://quark.caltech.edu/jhs60/witten/1.html}

%\WittenQJ
\lref\WittenQJ{
  E.~Witten,
  ``Anti-de Sitter space and holography,''
  Adv.\ Theor.\ Math.\ Phys.\  {\bf 2}, 253 (1998)
  [arXiv:hep-th/9802150].
  %%CITATION = 00203,2,253;%%
}

%\RajuBY
\lref\RajuBY{
  S.~Raju,
  ``BCFW for Witten Diagrams,''
  arXiv:1011.0780 [hep-th].
  %%CITATION = ARXIV:1011.0780;%%
}

%\GaryMI
\lref\GaryMI{
  M.~Gary and S.~B.~Giddings,
  ``The flat space S-matrix from the AdS/CFT correspondence?,''
  Phys.\ Rev.\  D {\bf 80}, 046008 (2009)
  [arXiv:0904.3544 [hep-th]].
  %%CITATION = PHRVA,D80,046008;%%
}

%\RehrenJN
\lref\RehrenJN{
  K.~-H.~Rehren,
  ``Algebraic holography,''
Annales Henri Poincare {\bf 1}, 607 (2000).
[hep-th/9905179].
%%CITATION = hep-th/9905179%%
}

%\DuetschHC
\lref\DuetschHC{
  M.~Duetsch and K.~-H.~Rehren,
  ``Generalized free fields and the AdS - CFT correspondence,''
Annales Henri Poincare {\bf 4}, 613 (2003).
[math-ph/0209035].
%%CITATION = math-ph/0209035%%
}

%\DolanUT
\lref\DolanUT{
  F.~A.~Dolan and H.~Osborn,
  ``Conformal four point functions and the operator product expansion,''
  Nucl.\ Phys.\  B {\bf 599}, 459 (2001)
  [arXiv:hep-th/0011040].
  %%CITATION = NUPHA,B599,459;%%
}

%\DolanHV
\lref\DolanHV{
  F.~A.~Dolan and H.~Osborn,
  ``Conformal partial waves and the operator product expansion,''
  Nucl.\ Phys.\  B {\bf 678}, 491 (2004)
  [arXiv:hep-th/0309180].
  %%CITATION = NUPHA,B678,491;%%
}

%\HawkingDH
\lref\HawkingDH{
  S.~W.~Hawking and D.~N.~Page,
  ``Thermodynamics Of Black Holes In Anti-De Sitter Space,''
  Commun.\ Math.\ Phys.\  {\bf 87}, 577 (1983).
  %%CITATION = CMPHA,87,577;%%
}

%\BloeteQM
\lref\BloeteQM{
  H.~W.~J.~Bloete, J.~L.~Cardy and M.~P.~Nightingale,
  ``Conformal Invariance, the Central Charge, and Universal Finite Size
  Amplitudes at Criticality,''
  Phys.\ Rev.\ Lett.\  {\bf 56}, 742 (1986).
  %%CITATION = PRLTA,56,742;%%
}

%\AharonyFS
\lref\AharonyFS{
  O.~Aharony, J.~R.~David, R.~Gopakumar, Z.~Komargodski and S.~S.~Razamat,
  ``Comments on worldsheet theories dual to free large N gauge theories,''
  Phys.\ Rev.\  D {\bf 75}, 106006 (2007)
  [arXiv:hep-th/0703141].
  %%CITATION = PHRVA,D75,106006;%%
}

%\HofmanUG
\lref\HofmanUG{
  D.~M.~Hofman,
  ``Higher Derivative Gravity, Causality and Positivity of Energy in a UV
  complete QFT,''
  Nucl.\ Phys.\  B {\bf 823}, 174 (2009)
  [arXiv:0907.1625 [hep-th]].
  %%CITATION = NUPHA,B823,174;%%
}

%\AharonyRQ
\lref\AharonyRQ{
  O.~Aharony and Z.~Komargodski,
  ``The space-time operator product expansion in string theory duals of field
  theories,''
  JHEP {\bf 0801}, 064 (2008)
  [arXiv:0711.1174 [hep-th]].
  %%CITATION = JHEPA,0801,064;%%
}

%\KovtunKW
\lref\KovtunKW{
  P.~Kovtun and A.~Ritz,
  ``Black holes and universality classes of critical points,''
  Phys.\ Rev.\ Lett.\  {\bf 100}, 171606 (2008)
  [arXiv:0801.2785 [hep-th]].
  %%CITATION = PRLTA,100,171606;%%
}

%\deBoerWK
\lref\deBoerWK{
  J.~de Boer, K.~Papadodimas and E.~Verlinde,
  ``Holographic Neutron Stars,''
  arXiv: 0907.2695 [hep-th].
  %%CITATION = ARXIV:0907.2695;%%
}

%\ArsiwallaBT
\lref\ArsiwallaBT{
  X.~Arsiwalla, J.~de Boer, K.~Papadodimas and E.~Verlinde,
  ``Degenerate Stars and Gravitational Collapse in AdS/CFT,''
  arXiv:1010.5784 [hep-th].
  %%CITATION = ARXIV:1010.5784;%%
}

%\CastroCE
\lref\CastroCE{
  A.~Castro, A.~Lepage-Jutier and A.~Maloney,
  ``Higher Spin Theories in AdS3 and a Gravitational Exclusion Principle,''
  arXiv:1012.0598 [hep-th].
  %%CITATION = ARXIV:1012.0598;%%
}

%\MaldacenaDE
\lref\MaldacenaDE{
  J.~M.~Maldacena, A.~Strominger and E.~Witten,
  ``Black hole entropy in M-theory,''
  JHEP {\bf 9712}, 002 (1997)
  [arXiv:hep-th/9711053].
  %%CITATION = JHEPA,9712,002;%%
}

%\BakasRY
\lref\BakasRY{
  I.~Bakas and E.~Kiritsis,
  ``Bosonic Realization of a Universal W Algebra and Z(infinity)
  Parafermions,''
  Nucl.\ Phys.\  B {\bf 343}, 185 (1990)
  [Erratum-ibid.\  B {\bf 350}, 512 (1991)].
  %%CITATION = NUPHA,B343,185;%%
}

%\BakasXU
\lref\BakasXU{
  I.~Bakas and E.~Kiritsis,
  ``Grassmannian Coset Models and Unitary Representations of W(infinity),''
  Mod.\ Phys.\ Lett.\  A {\bf 5}, 2039 (1990).
  %%CITATION = MPLAE,A5,2039;%%
}

%\BakasFS
\lref\BakasFS{
  I.~Bakas and E.~Kiritsis,
  ``Beyond The Large N Limit: Nonlinear W(Infinity) As Symmetry Of The Sl(2,R)
  / U(1) Coset Model,''
  Int.\ J.\ Mod.\ Phys.\  A {\bf 7S1A}, 55 (1992)
  [Int.\ J.\ Mod.\ Phys.\  A {\bf 7}, 55 (1992)]
  [arXiv:hep-th/9109029].
  %%CITATION = IMPAE,A7,55;%%
}

%\BakasRB
\lref\BakasRB{
  I.~Bakas and E.~Kiritsis,
  ``Target space description of W(infinity) symmetry in coset models,''
  Phys.\ Lett.\  B {\bf 301}, 49 (1993)
  [arXiv:hep-th/9211083].
  %%CITATION = PHLTA,B301,49;%%
}

%\BerkoozIZ
\lref\BerkoozIZ{
  M.~Berkooz and H.~L.~Verlinde,
  ``Matrix theory, AdS/CFT and Higgs-Coulomb equivalence,''
  JHEP {\bf 9911}, 037 (1999)
  [arXiv:hep-th/9907100].
  %%CITATION = JHEPA,9911,037;%%
}

%\deBoerUN
\lref\deBoerUN{
  J.~de Boer, S.~El-Showk, I.~Messamah and D.~Van den Bleeken,
  ``A bound on the entropy of supergravity?,''
  JHEP {\bf 1002}, 062 (2010)
  [arXiv:0906.0011 [hep-th]].
  %%CITATION = JHEPA,1002,062;%%
}

%\AharonyUG
\lref\AharonyUG{
  O.~Aharony, O.~Bergman, D.~L.~Jafferis and J.~Maldacena,
  ``N=6 superconformal Chern-Simons-matter theories, M2-branes and their
  gravity duals,''
  JHEP {\bf 0810}, 091 (2008)
  [arXiv:0806.1218 [hep-th]].
  %%CITATION = JHEPA,0810,091;%%
}

%\PolyakovAF
\lref\PolyakovAF{
  A.~M.~Polyakov,
  ``Gauge fields and space-time,''
  Int.\ J.\ Mod.\ Phys.\  A {\bf 17S1}, 119 (2002)
  [arXiv:hep-th/0110196].
  %%CITATION = IMPAE,A17S1,119;%%
}

%\GiddingsJQ
\lref\GiddingsJQ{
  S.~B.~Giddings,
  ``Flat-space scattering and bulk locality in the AdS/CFT  correspondence,''
  Phys.\ Rev.\  D {\bf 61}, 106008 (2000)   
  [arXiv:hep-th/9907129].
  %%CITATION = PHRVA,D61,106008;%%
}

%\PolchinskiRY
\lref\PolchinskiRY{
  J.~Polchinski,
  ``S-matrices from AdS spacetime,''
  arXiv:hep-th/9901076.
  %%CITATION = HEP-TH/9901076;%%
}

%\'tHooftJZ
\lref\tHooftJZ{
  G.~'t Hooft,
  ``A Planar Diagram Theory for Strong Interactions,''
  Nucl.\ Phys.\  B {\bf 72}, 461 (1974).
  %%CITATION = NUPHA,B72,461;%%
}

%\LuninYV
\lref\LuninYV{
  O.~Lunin and S.~D.~Mathur,
  ``Correlation functions for M(N)/S(N) orbifolds,''
  Commun.\ Math.\ Phys.\  {\bf 219}, 399 (2001)
  [arXiv:hep-th/0006196].
  %%CITATION = CMPHA,219,399;%%
}

%\DouglasRC
\lref\DouglasRC{
  M.~R.~Douglas, L.~Mazzucato and S.~S.~Razamat,
  ``Holographic dual of free field theory,''
  arXiv:1011.4926 [hep-th].
  %%CITATION = ARXIV:1011.4926;%%
}

%\SeibergWF
\lref\SeibergWF{
  N.~Seiberg,
  ``Emergent spacetime,''
  arXiv:hep-th/0601234.
  %%CITATION = HEP-TH/0601234;%%
}

%\GaberdielWB
\lref\GaberdielWB{
  M.~R.~Gaberdiel and T.~Hartman,
  ``Symmetries of Holographic Minimal Models,''
  arXiv:1101.2910 [hep-th].
  %%CITATION = ARXIV:1101.2910;%%
}

%\BerensteinAA
\lref\BerensteinAA{
  D.~Berenstein,
  ``Large N BPS states and emergent quantum gravity,''
  JHEP {\bf 0601}, 125 (2006)
  [arXiv:hep-th/0507203].
  %%CITATION = JHEPA,0601,125;%%
}

%\GaberdielVE
\lref\GaberdielVE{
  M.~R.~Gaberdiel,
  ``Constraints on extremal self-dual CFTs,''
  JHEP {\bf 0711}, 087 (2007)
  [arXiv:0707.4073 [hep-th]].
  %%CITATION = JHEPA,0711,087;%%
}

%\GaberdielXB
\lref\GaberdielXB{
  M.~R.~Gaberdiel, S.~Gukov, C.~A.~Keller, G.~W.~Moore and H.~Ooguri,
  ``Extremal N=(2,2) 2D Conformal Field Theories and Constraints of
  Modularity,''
  arXiv:0805.4216 [hep-th].
  %%CITATION = ARXIV:0805.4216;%%
}

%\HellermanBU
\lref\HellermanBU{
  S.~Hellerman,
  ``A Universal Inequality for CFT and Quantum Gravity,''
  arXiv:0902.2790 [hep-th].
  %%CITATION = ARXIV:0902.2790;%%
}

%\VerlindeHP
\lref\VerlindeHP{
  E.~P.~Verlinde,
  ``On the Origin of Gravity and the Laws of Newton,''
  arXiv:1001.0785 [hep-th].
  %%CITATION = ARXIV:1001.0785;%%
}

%\HellermanQD
\lref\HellermanQD{
  S.~Hellerman and C.~Schmidt-Colinet,
  ``Bounds for State Degeneracies in 2D Conformal Field Theory,''
  arXiv:1007.0756 [hep-th].
  %%CITATION = ARXIV:1007.0756;%%
}

%\deBoerPN
\lref\deBoerPN{
  J.~de Boer, M.~Kulaxizi and A.~Parnachev,
  ``AdS7/CFT6, Gauss-Bonnet Gravity, and Viscosity Bound,''
  JHEP {\bf 1003}, 087 (2010)
  [arXiv:0910.5347 [hep-th]].
  %%CITATION = JHEPA,1003,08y7;%%
}

%\deBoerGX
\lref\deBoerGX{
  J.~de Boer, M.~Kulaxizi and A.~Parnachev,
  ``Holographic Lovelock Gravities and Black Holes,''
  JHEP {\bf 1006}, 008 (2010)
  [arXiv:0912.1877 [hep-th]].
  %%CITATION = JHEPA,1006,008;%%
}

%\KulaxiziJT
\lref\KulaxiziJT{
  M.~Kulaxizi and A.~Parnachev,
  ``Energy Flux Positivity and Unitarity in CFTs,''
  arXiv:1007.0553 [hep-th].
  %%CITATION = ARXIV:1007.0553;%%
}

%\LuninPW
\lref\LuninPW{
  O.~Lunin and S.~D.~Mathur,
  ``Three-point functions for M(N)/S(N) orbifolds with N = 4 supersymmetry,''
  Commun.\ Math.\ Phys.\  {\bf 227}, 385 (2002)
  [arXiv:hep-th/0103169].
  %%CITATION = CMPHA,227,385;%%
}

%\SezginRT
\lref\SezginRT{
  E.~Sezgin and P.~Sundell,
  ``Massless higher spins and holography,''
  Nucl.\ Phys.\  B {\bf 644}, 303 (2002)
  [Erratum-ibid.\  B {\bf 660}, 403 (2003)]
  [arXiv:hep-th/0205131].
  %%CITATION = NUPHA,B644,303;%%
}

%\PetkouZZ
\lref\PetkouZZ{
  A.~C.~Petkou,
  ``Evaluating the AdS dual of the critical O(N) vector model,''
  JHEP {\bf 0303}, 049 (2003)
  [arXiv:hep-th/0302063].
  %%CITATION = JHEPA,0303,049;%%
}

%\LeighGK
\lref\LeighGK{
  R.~G.~Leigh and A.~C.~Petkou,
  ``Holography of the N = 1 higher-spin theory on AdS(4),''
  JHEP {\bf 0306}, 011 (2003)
  [arXiv:hep-th/0304217].
  %%CITATION = JHEPA,0306,011;%%
}

%\PakmanZZ
\lref\PakmanZZ{
  A.~Pakman, L.~Rastelli and S.~S.~Razamat,
  ``Diagrams for Symmetric Product Orbifolds,''
  JHEP {\bf 0910}, 034 (2009)
  [arXiv:0905.3448 [hep-th]].
  %%CITATION = JHEPA,0910,034;%%
}

%\LiuTH
\lref\LiuTH{
  H.~Liu,
  ``Scattering in anti-de Sitter space and operator product expansion,''
  Phys.\ Rev.\  D {\bf 60}, 106005 (1999)
  [arXiv:hep-th/9811152].
  %%CITATION = PHRVA,D60,106005;%%
}

%\RattazziPE
\lref\RattazziPE{
  R.~Rattazzi, V.~S.~Rychkov, E.~Tonni and A.~Vichi,
  ``Bounding scalar operator dimensions in 4D CFT,''
  JHEP {\bf 0812}, 031 (2008)
  [arXiv:0807.0004 [hep-th]].
  %%CITATION = JHEPA,0812,031;%%
}

%\RychkovIJ
\lref\RychkovIJ{
  V.~S.~Rychkov and A.~Vichi,
  ``Universal Constraints on Conformal Operator Dimensions,''
  Phys.\ Rev.\  D {\bf 80}, 045006 (2009)
  [arXiv:0905.2211 [hep-th]].
  %%CITATION = PHRVA,D80,045006;%%
}

%\CaraccioloBX
\lref\CaraccioloBX{
  F.~Caracciolo and V.~S.~Rychkov,
  ``Rigorous Limits on the Interaction Strength in Quantum Field Theory,''
  Phys.\ Rev.\  D {\bf 81}, 085037 (2010)
  [arXiv:0912.2726 [hep-th]].
  %%CITATION = PHRVA,D81,085037;%%
}

%\PolandWG
\lref\PolandWG{
  D.~Poland and D.~Simmons-Duffin,
  ``Bounds on 4D Conformal and Superconformal Field Theories,''
  arXiv:1009.2087 [hep-th].
  %%CITATION = ARXIV:1009.2087;%%
}

%\RattazziGJ
\lref\RattazziGJ{
  R.~Rattazzi, S.~Rychkov and A.~Vichi,
  ``Central Charge Bounds in 4D Conformal Field Theory,''
  arXiv:1009.2725 [hep-th].
  %%CITATION = ARXIV:1009.2725;%%
}

%\RattazziYC
\lref\RattazziYC{
  R.~Rattazzi, S.~Rychkov and A.~Vichi,
  ``Bounds in 4D Conformal Field Theories with Global Symmetry,''
  arXiv:1009.5985 [hep-th].
  %%CITATION = ARXIV:1009.5985;%%
}

%\HoffmannTR
\lref\HoffmannTR{
  L.~Hoffmann, A.~C.~Petkou and W.~Ruhl,
  ``A note on the analyticity of AdS scalar exchange graphs in the crossed
  channel,''
  Phys.\ Lett.\  B {\bf 478}, 320 (2000)
  [arXiv:hep-th/0002025].
  %%CITATION = PHLTA,B478,320;%%
}

%\HoffmannMX
\lref\HoffmannMX{
  L.~Hoffmann, A.~C.~Petkou and W.~Ruhl,
  ``Aspects of the conformal operator product expansion in AdS/CFT
  correspondence,''
  Adv.\ Theor.\ Math.\ Phys.\  {\bf 4}, 571 (2002)
  [arXiv:hep-th/0002154].
  %%CITATION = 00203,4,571;%%
}

%\DiazNM
\lref\DiazNM{
  D.~E.~Diaz and H.~Dorn,
  ``On the AdS higher spin / O(N) vector model correspondence: Degeneracy  of
  the holographic image,''
  JHEP {\bf 0607}, 022 (2006)
  [arXiv:hep-th/0603084].
  %%CITATION = JHEPA,0607,022;%%
}

%\PetkouFB
\lref\PetkouFB{
  A.~C.~Petkou and N.~D.~Vlachos,
  ``Finite-size effects and operator product expansions in a CFT for $d > 2$,''
  Phys.\ Lett.\  B {\bf 446}, 306 (1999)
  [arXiv:hep-th/9803149].
  %%CITATION = PHLTA,B446,306;%%
}

%\PetkouFC
\lref\PetkouFC{
  A.~C.~Petkou and N.~D.~Vlachos,
  ``Finite-size and finite-temperature effects in the conformally invariant
  O(N) vector model for $2 < d < 4$,''
  arXiv:hep-th/9809096.
  %%CITATION = HEP-TH/9809096;%%
}

%\AharonySX
\lref\AharonySX{
  O.~Aharony, J.~Marsano, S.~Minwalla, K.~Papadodimas and M.~Van Raamsdonk,
  ``The Hagedorn / deconfinement phase transition in weakly coupled large N
  gauge theories,''
  Adv.\ Theor.\ Math.\ Phys.\  {\bf 8}, 603 (2004)
  [arXiv:hep-th/0310285].
  %%CITATION = 00203,8,603;%%
}

%\KiritsisHY
\lref\KiritsisHY{
  E.~Kiritsis,
  ``Product CFTs, gravitational cloning, massive gravitons and the space of
  gravitational duals,''
  JHEP {\bf 0611}, 049 (2006)
  [arXiv:hep-th/0608088].
  %%CITATION = JHEPA,0611,049;%%
}

%\AharonyHZ
\lref\AharonyHZ{
  O.~Aharony, A.~B.~Clark and A.~Karch,
  ``The CFT/AdS correspondence, massive gravitons and a connectivity index
  conjecture,''
  Phys.\ Rev.\  D {\bf 74}, 086006 (2006)
  [arXiv:hep-th/0608089].
  %%CITATION = PHRVA,D74,086006;%%
}

%\KiritsisXJ
\lref\KiritsisXJ{
  E.~Kiritsis and V.~Niarchos,
  ``(Multi)Matrix Models and Interacting Clones of Liouville Gravity,''
  JHEP {\bf 0808}, 044 (2008)
  [arXiv:0805.4234 [hep-th]].
  %%CITATION = JHEPA,0808,044;%%
}

%\NiarchosQB
\lref\NiarchosQB{
  V.~Niarchos,
  ``Multi-String Theories, Massive Gravity and the AdS/CFT Correspondence,''
  Fortsch.\ Phys.\  {\bf 57}, 646 (2009)
  [arXiv:0901.2108 [hep-th]].
  %%CITATION = FPYKA,57,646;%%
}

%\KiritsisAT
\lref\KiritsisAT{
  E.~Kiritsis and V.~Niarchos,
  ``Interacting String Multi-verses and Holographic Instabilities of Massive
  Gravity,''
  Nucl.\ Phys.\  B {\bf 812}, 488 (2009)
  [arXiv:0808.3410 [hep-th]].
  %%CITATION = NUPHA,B812,488;%%
}

%\KiritsisXC
\lref\KiritsisXC{
  E.~Kiritsis and V.~Niarchos,
  ``Large-N limits of 2d CFTs, Quivers and AdS3 duals,''
  arXiv:1011.5900 [hep-th].
  %%CITATION = ARXIV:1011.5900;%%
}

%\SusskindDQ
\lref\SusskindDQ{
  L.~Susskind and E.~Witten,
  ``The holographic bound in anti-de Sitter space,''
  arXiv:hep-th/9805114.
  %%CITATION = HEP-TH/9805114;%%
}
%\oSBORNcr
\lref\OsbornCR{
  H.~Osborn and A.~C.~Petkou,
  ``Implications of Conformal Invariance in Field Theories for General
  Dimensions,''
  Annals Phys.\  {\bf 231}, 311 (1994)
  [arXiv:hep-th/9307010].
  %%CITATION = APNYA,231,311;%%
}

\lref\Jost{
R.~Jost, 
``The General Theory of Quantized Fields'', 
AMS, Providence, RI, 1965.
}

%\WittenKT
\lref\WittenKT{
  E.~Witten,
  ``Three-Dimensional Gravity Revisited,''
  arXiv:0706.3359 [hep-th].
  %%CITATION = ARXIV:0706.3359;%%
}

%\GiombiWH
\lref\GiombiWH{
  S.~Giombi and X.~Yin,
  ``Higher Spin Gauge Theory and Holography: The Three-Point Functions,''
  JHEP {\bf 1009}, 115 (2010)
  [arXiv:0912.3462 [hep-th]].
  %%CITATION = JHEPA,1009,115;%%
}

%\SundborgUE
\lref\SundborgUE{
  B.~Sundborg,
  ``The Hagedorn Transition, Deconfinement and N=4 SYM Theory,''
  Nucl.\ Phys.\  B {\bf 573}, 349 (2000)
  [arXiv:hep-th/9908001].
  %%CITATION = NUPHA,B573,349;%%
}

%\WittenZW
\lref\WittenZW{
  E.~Witten,
  ``Anti-de Sitter space, thermal phase transition, and confinement in  gauge
  theories,''
  Adv.\ Theor.\ Math.\ Phys.\  {\bf 2}, 505 (1998)
  [arXiv:hep-th/9803131].
  %%CITATION = 00203,2,505;%%
}

%\PakmanZZ
\lref\PakmanZZ{
  A.~Pakman, L.~Rastelli and S.~S.~Razamat,
  ``Diagrams for Symmetric Product Orbifolds,''
  JHEP {\bf 0910}, 034 (2009)
  [arXiv:0905.3448 [hep-th]].
  %%CITATION = JHEPA,0910,034;%%
}

%\AharonyTI
\lref\AharonyTI{
  O.~Aharony, S.~S.~Gubser, J.~M.~Maldacena, H.~Ooguri and Y.~Oz,
  ``Large N field theories, string theory and gravity,''
  Phys.\ Rept.\  {\bf 323}, 183 (2000)
  [arXiv:hep-th/9905111].
  %%CITATION = PRPLC,323,183;%%
}

%\MaldacenaDS
\lref\MaldacenaDS{
  J.~M.~Maldacena and L.~Susskind,
  ``D-branes and Fat Black Holes,''
  Nucl.\ Phys.\  B {\bf 475}, 679 (1996)
  [arXiv:hep-th/9604042].
  %%CITATION = NUPHA,B475,679;%%
}

%\deBoerFK
\lref\deBoerFK{
  J.~de Boer, F.~Denef, S.~El-Showk, I.~Messamah and D.~Van den Bleeken,
  ``Black hole bound states in AdS3 x S2,''
  JHEP {\bf 0811}, 050 (2008)
  [arXiv:0802.2257 [hep-th]].
  %%CITATION = JHEPA,0811,050;%%
}

%\KlebanovJA
\lref\KlebanovJA{
  I.~R.~Klebanov and A.~M.~Polyakov,
  ``AdS dual of the critical O(N) vector model,''
  Phys.\ Lett.\  B {\bf 550}, 213 (2002)
  [arXiv:hep-th/0210114].
  %%CITATION = PHLTA,B550,213;%%
}

%\VanRaamsdonkAR
\lref\VanRaamsdonkAR{
  M.~Van Raamsdonk,
  ``Comments on quantum gravity and entanglement,''
  arXiv:0907.2939 [hep-th].
  %%CITATION = ARXIV:0907.2939;%%
}

%\VanRaamsdonkPW
\lref\VanRaamsdonkPW{
  M.~Van Raamsdonk,
  ``Building up spacetime with quantum entanglement,''
  Gen.\ Rel.\ Grav.\  {\bf 42}, 2323 (2010)
  [arXiv:1005.3035 [hep-th]].
  %%CITATION = GRGVA,42,2323;%%
}

%\FestucciaSA
\lref\FestucciaSA{
  G.~Festuccia and H.~Liu,
  ``The arrow of time, black holes, and quantum mixing of large N Yang-Mills
  theories,''
  JHEP {\bf 0712}, 027 (2007)
  [arXiv:hep-th/0611098].
  %%CITATION = JHEPA,0712,027;%%
}

%\LiuTY
\lref\LiuTY{
  H.~Liu and A.~A.~Tseytlin,
  ``On four-point functions in the CFT/AdS correspondence,''
  Phys.\ Rev.\  D {\bf 59}, 086002 (1999)
  [arXiv:hep-th/9807097].
  %%CITATION = PHRVA,D59,086002;%%
}

%\FreedmanBJ
\lref\FreedmanBJ{ 
  D.~Z.~Freedman, S.~D.~Mathur, A.~Matusis and L.~Rastelli,
  ``Comments on 4-point functions in the CFT/AdS correspondence,''
  Phys.\ Lett.\  B {\bf 452}, 61 (1999)
  [arXiv:hep-th/9808006].
  %%CITATION = PHLTA,B452,61;%%
}

%\D'HokerPJ
\lref\DHokerPJ{
  E.~D'Hoker, D.~Z.~Freedman, S.~D.~Mathur, A.~Matusis and L.~Rastelli,
  ``Graviton exchange and complete 4-point functions in the AdS/CFT
  correspondence,''
  Nucl.\ Phys.\  B {\bf 562}, 353 (1999)
  [arXiv:hep-th/9903196].
  %%CITATION = NUPHA,B562,353;%%
}

%\D'HokerNI
\lref\DHokerNI{
  E.~D'Hoker, D.~Z.~Freedman and L.~Rastelli,
  ``AdS/CFT 4-point functions: How to succeed at z-integrals without  really
  trying,''
  Nucl.\ Phys.\  B {\bf 562}, 395 (1999)
  [arXiv:hep-th/9905049].
  %%CITATION = NUPHA,B562,395;%%
}

%\D'HokerJP
\lref\DHokerJP{
  E.~D'Hoker, S.~D.~Mathur, A.~Matusis and L.~Rastelli,
  ``The operator product expansion of N = 4 SYM and the 4-point functions  of
  supergravity,''
  Nucl.\ Phys.\  B {\bf 589}, 38 (2000)
  [arXiv:hep-th/9911222].
  %%CITATION = NUPHA,B589,38;%%
}

%\MinasianQN
\lref\MinasianQN{
  R.~Minasian, G.~W.~Moore and D.~Tsimpis,
  ``Calabi-Yau black holes and (0,4) sigma models,''
  Commun.\ Math.\ Phys.\  {\bf 209}, 325 (2000)
  [arXiv:hep-th/9904217].
  %%CITATION = CMPHA,209,325;%%
}

%\ZamolodchikovTI
\lref\ZamolodchikovTI{
  A.~B.~Zamolodchikov,
  ``Renormalization Group and Perturbation Theory Near Fixed Points in
  Two-Dimensional Field Theory,''
  Sov.\ J.\ Nucl.\ Phys.\  {\bf 46}, 1090 (1987)
  [Yad.\ Fiz.\  {\bf 46}, 1819 (1987)].
  %%CITATION = YAFIA,46,1819;%%
}

%\ZamolodchikovGT
\lref\ZamolodchikovGT{
  A.~B.~Zamolodchikov,
  ``Irreversibility of the Flux of the Renormalization Group in a 2D Field
  Theory,''
  JETP Lett.\  {\bf 43}, 730 (1986)
  [Pisma Zh.\ Eksp.\ Teor.\ Fiz.\  {\bf 43}, 565 (1986)].
  %%CITATION = ZFPRA,43,565;%%
}

%\FitzpatrickZM
\lref\FitzpatrickZM{
  A.~L.~Fitzpatrick, E.~Katz, D.~Poland and D.~Simmons-Duffin,
  ``Effective Conformal Theory and the Flat-Space Limit of AdS,''
  arXiv:1007.2412 [hep-th].
  %%CITATION = ARXIV:1007.2412;%%
}

%\GaryAE
\lref\GaryAE{
  M.~Gary, S.~B.~Giddings and J.~Penedones,
  ``Local bulk S-matrix elements and CFT singularities,''
  Phys.\ Rev.\  D {\bf 80}, 085005 (2009)
  [arXiv:0903.4437 [hep-th]].
  %%CITATION = PHRVA,D80,085005;%%
}

%\HeemskerkTY
\lref\HeemskerkTY{
  I.~Heemskerk and J.~Sully,
  ``More Holography from Conformal Field Theory,''
  JHEP {\bf 1009}, 099 (2010)
  [arXiv:1006.0976 [hep-th]].
  %%CITATION = JHEPA,1009,099;%%
}

%\PenedonesUE
\lref\PenedonesUE{
  J.~Penedones,
  ``Writing CFT correlation functions as AdS scattering amplitudes,''
  arXiv:1011.1485 [hep-th].
  %%CITATION = ARXIV:1011.1485;%%
}

%\GopakumarNS
\lref\GopakumarNS{
  R.~Gopakumar,
  ``From free fields to AdS,''
  Phys.\ Rev.\  D {\bf 70}, 025009 (2004)
  [arXiv:hep-th/0308184].
  %%CITATION = PHRVA,D70,025009;%%
}

%\GopakumarQB
\lref\GopakumarQB{
  R.~Gopakumar,
  ``From free fields to AdS. II,''
  Phys.\ Rev.\  D {\bf 70}, 025010 (2004)
  [arXiv:hep-th/0402063].
  %%CITATION = PHRVA,D70,025010;%%
}

%\GopakumarYS
\lref\GopakumarYS{
  R.~Gopakumar,
  ``Free field theory as a string theory?,''
  Comptes Rendus Physique {\bf 5}, 1111 (2004)
  [arXiv:hep-th/0409233].
  %%CITATION = CRPOB,5,1111;%%
}

%\GopakumarFX
\lref\GopakumarFX{
  R.~Gopakumar,
  ``From free fields to AdS. III,''
  Phys.\ Rev.\  D {\bf 72}, 066008 (2005)
  [arXiv:hep-th/0504229].
  %%CITATION = PHRVA,D72,066008;%%
}

%\VasilievDN
\lref\VasilievDN{
  M.~A.~Vasiliev,
  ``Higher-spin gauge theories in four, three and two dimensions,''
  Int.\ J.\ Mod.\ Phys.\  D {\bf 5}, 763 (1996)
  [arXiv:hep-th/9611024].
  %%CITATION = IMPAE,D5,763;%%
}

%\VasilievBA
\lref\VasilievBA{
  M.~A.~Vasiliev,
  ``Higher spin gauge theories: Star-product and AdS space,''
  arXiv:hep-th/9910096.
  %%CITATION = HEP-TH/9910096;%%
}

%\VasilievEV
\lref\VasilievEV{
  M.~A.~Vasiliev,
  ``Nonlinear equations for symmetric massless higher spin fields in
  (A)dS(d),''
  Phys.\ Lett.\  B {\bf 567}, 139 (2003)
  [arXiv:hep-th/0304049].
  %%CITATION = PHLTA,B567,139;%%
}

%\GaberdielPZ
\lref\GaberdielPZ{
  M.~R.~Gaberdiel and R.~Gopakumar,
  ``An AdS3 Dual for Minimal Model CFTs,''
  arXiv:1011.2986 [hep-th].
  %%CITATION = ARXIV:1011.2986;%%
}

%\PolyakovGS
\lref\PolyakovGS{
  A.~M.~Polyakov,
  ``Nonhamiltonian approach to conformal quantum field theory,''
  Zh.\ Eksp.\ Teor.\ Fiz.\  {\bf 66}, 23 (1974).
  %%CITATION = ZETFA,66,23;%%
}

%\IntriligatorJJ
\lref\IntriligatorJJ{
  K.~A.~Intriligator and B.~Wecht,
  ``The exact superconformal R-symmetry maximizes a,''
  Nucl.\ Phys.\  B {\bf 667}, 183 (2003)
  [arXiv:hep-th/0304128].
  %%CITATION = NUPHA,B667,183;%%
}

%\HofmanAR
\lref\HofmanAR{
  D.~M.~Hofman and J.~Maldacena,
  ``Conformal collider physics: Energy and charge correlations,''
  JHEP {\bf 0805}, 012 (2008)
  [arXiv:0803.1467 [hep-th]].
  %%CITATION = JHEPA,0805,012;%%
}

%\DrukkerNC
\lref\DrukkerNC{
  N.~Drukker, M.~Marino and P.~Putrov,
  ``From weak to strong coupling in ABJM theory,''
  arXiv:1007.3837 [hep-th].
  %%CITATION = ARXIV:1007.3837;%%
}

\Title{\vbox{\baselineskip12pt
%\hbox{arXiv:xxxx.xxxx}
%\hbox{CERN-PH-TH/2011-}\hbox{IPhT-../...}
}}
{\vbox{\centerline {Emergent spacetime and holographic CFTs}}}
\vskip-10pt
\centerline{Sheer El-Showk$^{a,b}$  and  Kyriakos Papadodimas$^c$}

\bigskip

\centerline{$^a$ Institut de Physique Theeorique,}
\centerline{CEA Saclay, CNRS URA 2306,}
\centerline{F-91191 Gif-sur-Yvette, France}
\bigskip
\centerline{$^b$ Institute
  for Theoretical Physics  University of Amsterdam,}
\centerline{Science Park  904, Postbus 94485, 1090 GL Amsterdam, The Netherlands}
\centerline{\it sheer.el-showk@cea.fr}
\bigskip
\centerline{$^c$ Theory Group, Physics Department, CERN}
\centerline{CH-1211 Geneva 23, Switzerland}
\centerline{\it kyriakos.papadodimas@cern.ch}
\bigskip

\centerline{\bf Abstract}

\noindent We discuss universal properties of conformal field
theories with holographic duals. A central feature of these theories
is the existence of a low-lying sector of operators whose correlators
factorize.  We demonstrate that factorization can only hold in the large
central charge limit.  Using conformal invariance and factorization we
argue that these operators are naturally represented as fields in AdS
as this makes the underlying linearity of the system manifest.  In
this class of CFTs the solution of the conformal bootstrap conditions
can be naturally organized in structures which coincide with Witten
diagrams in the bulk. The large value of the central charge suggests
that the theory must include a large number of new operators not
captured by the factorized sector.  Consequently we may think of the
AdS hologram as an effective representation of a small sector of the
CFT, which is embedded inside a much larger Hilbert space
corresponding to the black hole microstates.

\bigskip
\noindent
\Date{January , 2011}
\listtoc  \writetoc

\newsec{Introduction}

One of the most important ideas considered in the last decades is the
possibility that space and time may be emergent concepts. In string
theory we know examples of dualities in which spacetime, or at least
some of its dimensions can be reconstructed from certain underlying
quantum systems \foot{See \SeibergWF\ for a recent review.}. Besides
refining the technical aspects of such dualities, it is important to
identify the general principles governing the universality class of
quantum systems from which a macroscopic semi-classical spacetime can
emerge.

In this paper we want to revisit this question in a simplified context. We will
focus on the emergent nature of gravitational theories in anti-de Sitter space.
After the discovery of the AdS/CFT correspondence \MaldacenaRE, \GubserBC,
\WittenQJ\ it has been understood that such theories are ``dual'' to conformal
field theories living in lower dimensions. At the moment it is only the field
theory side of these dualities which can be non-perturbatively defined. From the
field theory point of view the bulk anti-de Sitter spacetime, with its
gravitational interactions, is an emergent concept. Is this phenomenon isolated
to very special conformal field theories or is it generic? And if so, what are
the general principles behind the emergence of a ``dual AdS spacetime''?  We
propose to explore these issues by posing the following question:
\smallskip
{\it Which conformal field
theories have a holographic description and why?}
\smallskip
\noindent To address this question we will review properties of
well-known CFTs with holographic duals and attempt to recast them in a
general language that abstracts away the specifics of the given
theory.  The philosophy of our approach will be the following: we want
to explore why and how, for a certain class of CFTs, an observer
unaware of the AdS/CFT correspondence, would naturally end-up
``rediscovering'' it. That is, using only the consistency of the CFT,
we will try to understand why it would be natural to formulate it as
an effective gravitational theory in AdS.

The general approach of our paper is closely related to various other
works which appeared in the
past \BanksDD, \BalasubramanianRI, \BerkoozIZ, \wittentalk, \BerensteinAA. Similar
questions have also been addressed recently
in \HeemskerkPN, \HeemskerkTY, \FitzpatrickZM, \PenedonesUE\ on which
we heavily base some of our discussions. The emergence of gravity has
also been recently addressed in \VerlindeHP\ from another point of
view and also in \VanRaamsdonkAR, \VanRaamsdonkPW. Many of the
statements in this paper are well known to the experts and we do not
have many new technical results to report. Nonetheless we hope that
our presentation may be useful.

\bigskip
\bigskip

To summarize the main points, in this paper we argue that a CFT is
naturally described holographically in AdS space if it has the
following basic properties:
\medskip
i) It has large central charge i.e. many degrees of freedom.

ii) It has a small number of operators of low conformal dimension.

iii) The correlators of the low lying operators factorize\foot{As we will explain later, condition iii) requires i), but not the other way around.}.
\medskip
\noindent Intuitively these conditions mean that we should be looking for CFTs
with a large number of degrees of freedom i.e. many fundamental
fields, but few weakly coupled light\foot{In the sense of having small
conformal dimension.} operators. We will try to argue that in theories
with these properties the effective interactions of the low-lying
operators are naturally described in terms of a dual gravitational
theory in anti-de Sitter space. These conditions, as well as the
meaning of the holographic dual theory, will be made more precise in
the rest of the paper.

Conditions i), ii), iii) are automatically satisfied for large $N$
gauge theories in the 't Hooft limit: for these theories the central
charge scales like $N^2$, the low-lying operators are the gauge
invariant combinations of traces (whose number is small
i.e. $N$-independent) and factorization of single trace operators is
guaranteed by the large $N$ 't Hooft factorization. We want to argue
that the aforementioned conditions are the central universal features
of CFTs with holographic duals. We expect that any CFT with these
properties, even if it is not a standard gauge theory, will have a
natural holographic description. From this point of view it would seem
that the main role of gauge invariance for holography of large $N$
gauge theories is that it is an efficient way to engineer quantum
field theories in which conditions i), ii), iii) are met.

In the rest of the paper we start from the bottom-up. We start by
considering CFTs which contain operators whose correlators
factorize. Such operators have certain features of ordinary free
fields, though they do not obey linear equations of motion. We will
refer to them as ``generalized free fields''. By uplifting these
operators to one dimension higher they can be extended to ordinary
free fields, thus making the underlying linearity of the system
manifest. In this sense a CFT observer would have a reason to
introduce one extra dimension in exchange for having linear equations
of motion.

When introducing small interactions around such generalized free
fields we have to make sure that they satisfy the ``conformal
bootstrap'' conditions. We argue it is natural to reorganize the
perturbed CFT correlators in terms of algebraic structures which
coincide with what we would normally call ``Witten diagrams'' in the
bulk. For this we study in some detail the relation between Conformal
Partial Waves (CPWs) on the boundary and Witten diagrams in the
bulk. We will try to argue that a physicist, unaware of AdS/CFT, who
would try to solve the conformal bootstrap conditions around a
``generalized free CFT'' would reorganize the solution in terms of
objects which look like the Witten diagrams and might thus be led to
the formulation of a gravitational theory in AdS.

Finally we will see that such generalized free CFTs are not fully
consistent by themselves and can only be understood as being small
sectors of much bigger underlying conformal field theories. The reason
is that in order to have factorization of correlators it is necessary
for the central charge of the CFT to be large. This implies that at
large conformal dimension the theory must have a large number of
operators that do not appear in the naive spectrum of generalized free
fields.  From the bulk point of view these operators correspond to the
black hole microstates.

The approach of our paper seems to be more consistent with the idea
that the bulk is the effective description of a small semi-independent
sector of low-lying operators in the conformal field theory i.e. of
the ``confined phase'' in the language of gauge theories. According to
the logic followed in our presentation it seems that there would be no
reason for operators with large conformal dimension, i.e. those
corresponding to black hole microstates, to have representatives in
the bulk in terms of local fields (though they might be represented as
states living, in a sense, on black hole horizons).

The plan of the paper is as follows.  In Section \mainconditions\ we
give an overview of the main ideas of the paper, including some
motivational examples, eschewing any technical details.
Section \cftconsistency\ provides some background on standard CFT
consistency conditions as well as introducing constraints following
from consistency of the thermal theory.  In Section \gff\ we introduce
the notion of a Generalized Free Field and study the structure of CFTs
composed of such operators; in particular it is argued that such
theories cannot define consistent local CFTs unless augmented by
additional operators.  By recasting them in a higher dimensional
language where their linear structure is manifest it is argued in
Section \emergence\ that such generalized free CFTs are, in a sense,
inherently holographic.  This is further developed in
Section \interactions\ where perturbations around such theories are
studied.  The latter are most naturally organized in a structure that
mimics Witten diagrams in a putative AdS bulk.  In
Section \finitetemp\ we return to the spectrum and argue that
factorization implies a divergent entropy density.  The relationship
to Cardy's formula in two dimensions and the extended nature of the
``Cardy regime'' in AdS duals is also discussed.
Section \discussions\ closes with some discussion and proposals for
future directions.  A series of appendices contain technical
background and results.  Appendix A provides background on the
Conformal Partial Wave decomposition of CFT correlation functions
while Appendix B includes a computation of the coefficients of such an
expansion applied to Witten diagrams.  Appendix C reviews the
different notions of central charges in $d > 2$.

\newsec{Main Picture: conditions for holographic CFTs}
\seclab\mainconditions

In this section we summarize the main picture without going into
technical details.  A CFT is by itself well-defined even at strong
coupling (for example by simulating it on a lattice) and in principle
there is no need to invoke a holographic dual description. The reason
that for some CFTs we do it nevertheless is that introducing the
``dual spacetime'' makes the description of the theory (or perhaps a
sector of the theory) simpler. In general a given physical theory may
admit more than one equivalent mathematical description. We do not
have a sharp quantitative criterion to decide which of them is the
``simplest'' one, however we will follow the naive intuition that if a
physical theory admits a description in terms of a small number of
weakly interacting local fields living on a spacetime then this
description is a natural one to use.

To proceed we first have to explain what kind of dual gravitational
theories we have in mind. At the moment an independent definition of
what is meant by ``quantum gravity in AdS'' in not available. We only
know what semiclassical gravity or perturbative string theory means,
which implies that the Planck length must be taken to be much smaller
than the effective curvature radius of the AdS space. So we will focus
on analyzing CFTs which can potentially describe weakly coupled dual
holographic theories (in the sense of small $g_s$, not necessarily
small curvature). These holographic theories may look like
two-derivative gravity, like weakly coupled string theory at high
(string) curvatures, M-theory, higher spin gravity or something more
exotic. Since we want to be general, we will take the following as the
minimal requirement for a holographic bulk theory:

{\it We will assume that as we take the weak coupling limit\foot{i.e. the classical limit, not necessarily a small curvature limit.} in the
bulk, the number of light modes stays finite.  }

By this we mean that the number of modes below any given (and fixed)
energy scale remains finite as we send the coupling, measured by the
ratio of the Planck to the AdS length scale, to zero. This condition
is satisfied in all known examples of AdS/CFT with a weak coupling
limit (again, in the sense of small $g_s$) and if violated it seems
unlikely that any useful notion of spacetime, classical or stringy,
would emerge.

Let us translate these statements into expectations for the dual
CFT. First of all, since we want to consider theories with a weak
coupling limit we should be looking for {\it families} of CFTs
characterized by some parameter\foot{The parameter $g$ does not have
to be a continuous variable; in particular it may not be an ordinary
coupling constant. For example, it may be a discrete label in a
sequence of theories.} $g$, with the property that as $g\rightarrow 0$
we reach the free limit in the bulk. In standard large $N$ gauge
theories this parameter would be $g\sim {1\over N}$ . According to the
condition that we mentioned above, we will assume that in this limit
the CFT has a finite number of operators of low conformal
dimension\foot {To be more precise, we assume that for any given
$\Delta$ the number of operators with dimension lower that $\Delta$
remains finite as $g\rightarrow 0$. It is important to take the limit
in this way, that is first fix $\Delta$ and then send $g\rightarrow
0$. }. Moreover we will assume that in the limit $g\rightarrow 0$ the
correlators of these low-lying operators factorize (in a sense that will be made more precise later). We believe that
this is the most crucial element of a CFT with a holographic dual.

Note that while this may sound innocuous it is a rather non-trivial
constraint: the basic axioms of a CFT imply that such a set of
factorized operators can only arise as a small subsector of a much
bigger conformal field theory. The basic reason for this is that the
assumption of factorization of correlators conflicts with a universal
interaction in conformal field theories, the one mediated by the
exchange of the stress-energy tensor between operators. The coupling
of the stress energy tensor to other conformal primaries is fixed by
the Ward identities and, as we will see, the only way to have it
decoupled (which is necessary to have factorization) is that, as
$g\rightarrow 0$ the central charge\foot{Defined by the 2-point
function of the stress-energy tensor.} $c$ of the CFT goes to
infinity. For simplicity, and to be consistent with the case of large
$N$ gauge theories, in the rest of the paper we will make the
identification $g= {1\over \sqrt{c}}$.  The fact that the central
charge goes to infinity means that the number of degrees of freedom in
the theory is large.  Since we assumed that we have a finite number of
operators at low conformal dimension, this large number of degrees of
freedom has to emerge at large conformal dimension.

\fig{Typical spectrum of CFTs with holographic duals. The shaded 
part at large conformal dimension represents a large ($c$-dependent)
number of heavy operators, which correspond to black hole
microstates. The red dots represent a small number ($c$-independent)
light operators, whose correlators factorize. These operators are
represented holographically as free (or weakly interacting) fields in
AdS. This is only a cartoon and the spectrum at intermediate values of
conformal dimension may have complicated form.}
{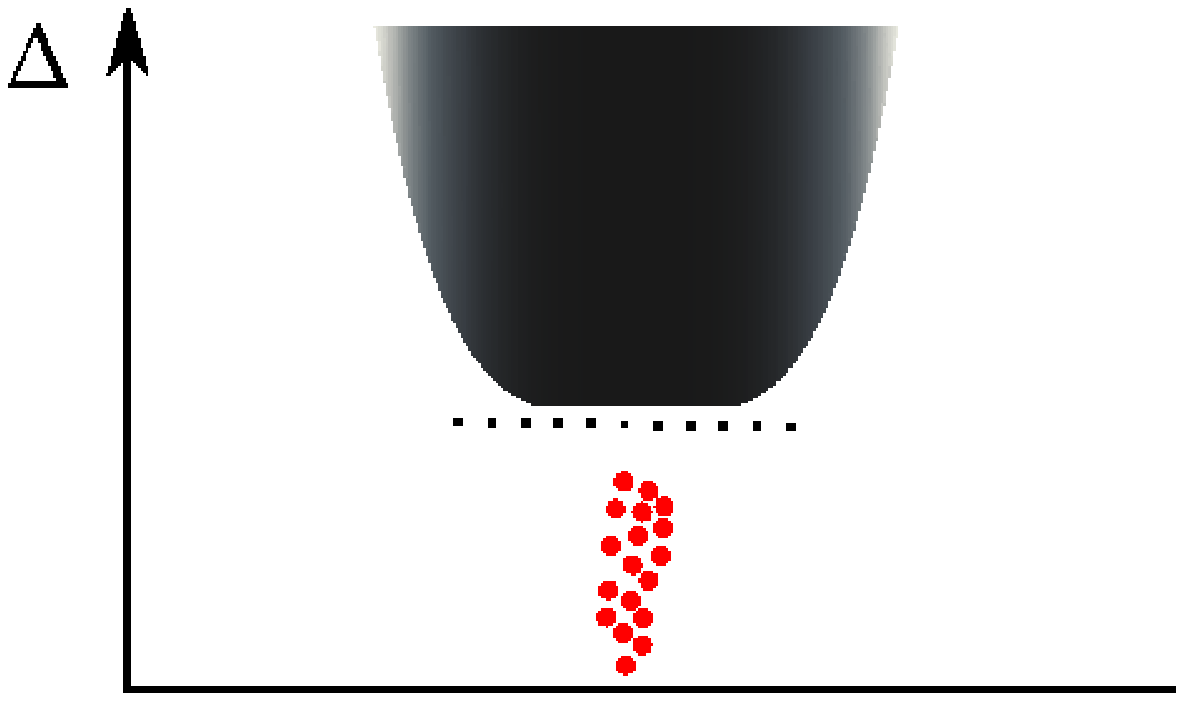}{4.truein}
\figlabel{\bootstrapone}

Now we can phrase our criterion more precisely. We are looking for
CFTs with large central charge $c$ with the property that in the
$c\rightarrow \infty$ limit their spectrum has the following
structure: at conformal dimensions of order ${\cal O}(1)$, as far as
the $c$-scaling is concerned, the number of operators remains finite,
while above a certain conformal dimension $\Delta_*$ (which scales to
infinity as $c\rightarrow \infty$) we have a $c$-dependent
proliferation of states\foot{This is to be understood as a qualitative
statement, since the change of degeneracies from ${\cal O}(1)$ to
${\cal O}(c)$ does not have to be sharp, there may be intermediate
regimes.}. The theory has a small low-lying sector separated from the
huge number of operators with large conformal dimension of order
$\Delta_*$ and higher as depicted in figure 1.

In the bulk the quantity $g$ plays the role of $\hbar$, hence in the
limit $c\rightarrow \infty$ we have $\hbar\rightarrow 0$. The
low-lying operators of the CFT are dual to the (perturbative)
supergravity/closed string modes whose numbers remains finite as we
send $\hbar \rightarrow 0$. On the other hand the large number of
states of conformal dimension above $\Delta_*$ corresponds to black
hole microstates whose degeneracy blows up in the classical limit
$\hbar\rightarrow 0$ as can be seen from the Bekenstein-Hawking
entropy formula.

Moreover we assume that in the limit $c\rightarrow \infty$ the
low-lying operators factorize i.e. they become generalized free
fields\foot{While we use the term ``field'' for these operators it
should be clear that they are not ``fundamental'' fields of the
Lagrangian over which one should path integrate.}.  As we will try to
argue in the rest of the paper, the effective dynamics of the
low-lying generalized free fields can be most naturally represented in
an AdS space with semi-local\foot{By this we mean that the AdS theory
is not necessarily a local effective field theory with a small number
of light fields.  It could, for example, be a highly curved (but weakly
coupled) string theory.} interactions. From this point of view the
hologram seems to be an effective representation of a small part of
the Hilbert space of the CFT (a similar point of view is espoused
in \FitzpatrickZM). Of course according to the strong version of the
AdS/CFT correspondence, it may become possible in the future to define
the bulk side independently, including all of the ${\cal O}(c)$
degrees of freedom.

\subsec{Some comments}

In this subsection we mention some clarifications.

\bigskip
1. From the known examples of AdS/CFT it is clear that the holographic
representation of a CFT can have various degrees of complexity. The
simplest case is the one where the bulk theory is classical gravity,
possibly with a few other light fields, on a background of the form
AdS $\times$ M, where M is some internal manifold whose size is large
compared to the Planck and other UV scales (such as the string
scale). These theories are the ones where there is a parametric
separation between the mass of fields with spin up to two and those
with higher spin, and the relevant conditions in the CFT were
discussed in \HeemskerkPN. A next level of complexity corresponds to
theories with a significantly larger number of light fields in the
bulk, in which the aforementioned separation between spins does not
exist. Examples of this kind are higher-spin gravity or classical
string theory on highly curved backgrounds. In the latter class we
have, for example, the ${\cal N}=4$ SYM at large $N$ and finite/small
't Hooft coupling $\lambda$. Finally there may exist even more exotic
bulk theories, such as those dual to CFTs at finite $N$ - i.e. fully
quantum gravity, or dual to CFTs where there is no qualitative
separation in the degeneracies of the spectrum between small and large
conformal dimensions (see ``counter-examples'' subsection below).

When we pose the question ``which CFTs are holographic?'' we have to specify
what kind of bulk theories we are referring to. In this paper we decided to focus
on the universality class of bulk theories which are semi-classical but may be
highly curved/stringy (so, for example, we would include the ${\cal N}=4$ SYM at
small 't Hooft coupling, but not at finite $N$). The reason for doing so is
because this class of theories seems general enough to include various
perturbative string theory backgrounds as well as some of the more general
models mentioned in section \examples. At the same time we did not want to be
too general, since, had we gone into the quantum gravity regime (finite central
charge), we would have to deal with the problem of defining (independently of
the CFT) what we mean by quantum gravity in an AdS space of Planckian
size\foot{By this we do not mean to claim that there are no such bulk theories,
as this would be inconsistent with the strong version of AdS/CFT, but rather that we do
not know any such theory yet and hence it is not possible for us to say anything
concrete about what to expect from such theories.}.

\medskip

2. The ``jump'' in the degeneracies of the spectrum that we refer to
should be distinguished from the gap discussed
in \HeemskerkPN. In that paper the gap was between operators
corresponding to supergravity fields and those dual to ``closed string
states'' i.e. it was a gap within the low-lying generalized free field
sector, from our point of view. The change in the qualitative
properties of the spectrum\foot{We would like to thank
E. Kiritsis for helpful comments.} that we are talking about is between
operators with conformal dimension of order ${\cal O}(1)$ (dual to
perturbative gravitational and stringy states) and operators of
dimension ${\cal O}(c)$ which are dual to black hole-like states
i.e. between the generalized free field sector and the other operators
in the CFT\foot{Of course the two statements are not in contradiction
since the goal of \HeemskerkPN\ was to identify CFTs with local
gravity duals and not more general stringy ones. Also the property of
the spectrum that we are talking about is presumably assumed
in \HeemskerkPN, though not explicitly discussed.}.

\medskip

3. The generalized free fields have been discussed before in the
literature, however we should emphasize that our point of view is
quite different. We claim that such theories cannot be considered as
fundamental CFTs, but only as limits of certain small sectors of much
larger CFTs\foot{Here we are referring to works done in the 60s and 70s in the context of axiomatic QFT (see for example \Jost\ and also more recently to the work on ``algebraic holography'', for instance \RehrenJN, \DuetschHC,
where generalized free fields are also used, but seen as fundamental theories.}.

\medskip

4. The coupling $g=1/\sqrt{c}$ is a universal coupling in conformal
perturbation theory mediated by the exchange of the stress-energy
tensor, as it follows strictly from conformal invariance and the Ward
identities.  One might then argue that the additional ``black hole
microstates'' with $c$-dependent masses and degeneracies are, in some
sense, the associated non-perturbative states.  As generalized free
theories do not have a local $d$-dimensional Lagrangian description
this analogy can only be made precise in the bulk dual (where it is
manifest).

\medskip

5. In large $N$ gauge theories the conditions i), ii), iii) are
automatically satisfied because of two important physical effects:
confinement/deconfinement (which guarantees the right qualitative
properties of the spectrum) and the 't Hooft large $N$ factorization.
But as we mentioned in the introduction it is perhaps true that these
two properties, spectrum and factorization, are of more fundamental
importance for holography than gauge invariance itself.

\medskip

6. Our approach has an important limitation: we will discuss the
emergence of the bulk from the CFT but the structure that we will
describe will be an effective theory in AdS, with potentially a very
large (though $c$-independent) number of fields. The missing element
is finding the ``organizing principle'' behind this effective AdS
theory. For example if we were to follow our logic for the ${\cal
N}=4$ SYM at weak 't Hooft coupling we would end up with an AdS theory
with a large number of fields (corresponding to the stringy modes,
whose number grows exponentially with conformal dimension), but
without knowing that there is an underlying worldsheet string theory
governing these fields. In this sense the approach initiated by
R. Gopakumar \GopakumarNS, \GopakumarQB, \GopakumarYS, \GopakumarFX\ and
also \AharonyFS, \AharonyRQ\ is more ambitious, since it attempts a
direct reconstruction of the bulk worldsheet theory from the boundary
correlators.

\medskip

7. On the other hand it is possible that there exist CFTs with
effective holographic duals which do not correspond to string
theories, in the sense of perturbative worldsheet theories (examples
are M-theory backgrounds or the duals of $O(N)$
models \KlebanovJA). For such CFTs it is not clear what kind of
structure we would expect in the bulk so we restrict ourselves to
constructing an effective AdS theory.

\medskip

8. It would be useful to find a precise way to quantify, from the CFT point of
view, the extent to which the dual AdS theory is local. One criterion seems to
be the number of single-particle operators at low conformal dimension. It is
natural to expect that the fewer operators there are at low conformal dimension,
the sharper the notion of locality and of classical geometry in AdS will be.
This can be seen by considering three different classes of examples with
increasing complexity: theories with classical gravity duals, higher spin
gravity theories and highly curved (weakly coupled) string theories. The notion
of geometry is most clear in the first case, it becomes more complicated in the
second, and in the third the only notion of spacetime is the one provided by the
worldsheet conformal field theory (i.e. spacetime is ``stringy'').
Correspondingly the number\foot{and growth with conformal dimension.} of single
trace operators is small in the first case, where as emphasized in \HeemskerkPN\
there is a parametric separation in the dimensions of operators with spin 2 and
higher, larger in the second and exponentially growing with conformal dimension
in the third. There may of course exist other possible forms of the spectrum in
holographic CFTs that have not been discovered yet.

\subsec{Examples}
\subseclab\examples

To motivate and provide context for the rest of our discussion let us
consider the general qualitative features of CFTs known to admit AdS
duals. Most of the known examples fall broadly into four main
categories: 1) large $N$ gauge theories, 2) large $N$ symmetric
orbifolds, 3) large $N$ vector $O(N)$ models and 4) particular
$\sigma$-models on targets spaces of parametrically high-dimension
(i.e. the MSW CFT \MaldacenaDE).

\hfil \break
\noindent {\bf Large $N$ Gauge Theories}

Large $N$ gauge theories in the 't Hooft limit are the canonical
example of quantum field theories with holographic duals. This was
already suggested by the observation of 't Hooft that the $1/N$
expansion in gauge theories can be interpreted as a genus expansion of
a dual string theory \tHooftJZ. After the AdS/CFT
correspondence \MaldacenaRE\ the ${\cal N}=4$ SYM at large $N$ became
the best studied example of a theory with a gravitational dual, along
with a very large number of other holographic gauge theories analyzed
in the last decade.
 
Let us see how large $N$ gauge theories satisfy the criteria i), ii),
iii) that we presented. For simplicity we will only consider the case
where the gauge theories are conformal and we will also restrict
ourselves to gauge theories with a weak coupling limit in their moduli
space\foot{We will also assume that the number of flavor fields does
not scale with $N$, otherwise the scaling of various quantities
mentioned below must be modified. We thank E. Kiritsis for comments
about these issues.}.

First we study the spectrum at low conformal
dimension\foot{i.e. conformal dimension of order 1.}.  If we consider
the $N\rightarrow \infty$ limit then according to arguments of 't
Hooft the theory simplifies considerably with correlators of
gauge-invariant operators factorizing.  Such factorization implies
that the spectrum of the theory should enjoy a particularly simple
form and at weak coupling we can compute the spectrum
explicitly \PolyakovAF, \SundborgUE, \AharonySX. The requirement of
gauge invariance restricts the spectrum of local operators to be
composed of traces, thus suppressing the dependence on $N$.
Single-trace operators can be constructed from the fundamental fields
$\Phi_i$ of the gauge theory as $\Tr[\Phi_{i_1}\dots \Phi_{i_n}]$ with
$n \ll N$, whose degeneracy depends only on $n$. Due to large $N$
factorization the spectrum has the form of a freely generated Fock
space of multi-trace operators, where the basic excitations are the
single-trace operators.

To summarize, we argued that at low conformal dimension the theory has
a small number of operators (i.e. their degeneracy does not scale with
$N$), whose correlators factorize due to large $N$
combinatorics. Hence the conditions ii) and iii) are satisfied.

When, however, the length and number of traces become large (i.e. when
either starts to scale with $N$), trace relations correct the
spectrum, introducing an $N$-dependence.  This is an essential feature
since otherwise the finite temperature partition function would
encounter a Hagedorn-type divergence at temperatures of order 1,
caused by the exponential growth of single-trace operators as a
function of their conformal dimension. Moreover, even if, for some
reason, there was no exponential growth in the degeneracy of
single-trace operators, and if the degeneracy were to continue to
scale naively as a multiparticle gas of single-trace operators, then
the high temperature behaviour of this theory would be
inconsistent with conformal invariance (see
Section \spartition). Hence factorization should break down when the
conformal dimension of operators becomes large (that is, $N$
dependent). Despite the very gauge-theoretic nature of this structure
we will see that something similar occurs quite generally in any large
$c$ CFT satisfying assumptions ii) and iii).

Let us now consider what happens at large conformal
dimension. Although the central charge is not a unique notion in
higher dimensional CFTs, the various notions of central charge (see
e.g. Section \cardyone\ and Appendix C) are all proportional to $N^2$
for gauge theories.  Intuitively these $N^2$ degrees of freedom
correspond to the $N^2$ components of the fundamental fields of the
theory. This is a reliable estimate if the theory admits a weak
coupling limit\foot{Notice however that this is not true beyond the 't
Hooft limit, for theories which are strongly coupled. For example in
the ABJM theory \AharonyUG\ at $k=1$, while the number of fundamental
fields of the theory scales like $N^2$ the prediction from gravity is
that the entropy density scales like $N^{3/2}$.  In this case the
strong coupling dynamics invalidates the naive free-field intuition
about the number of degrees of freedom \DrukkerNC.}. We thus expect that the
entropy density at very high temperatures will grow like $s\sim N^2
T^{d-1}$ which is consistent with condition (i). The spectrum of
operators of the gauge theory is related, via the state-operator map,
to the spectrum of states of the CFT on ${\bf S}^{d-1}\times {\bf
R}$. The high temperature entropy density mentioned above implies that
the entropy of operators of conformal dimension $\Delta$ grows like
$S \sim N^{2/d} \Delta^{(d-1)/d}$. This indicates a very large number
of operators of large conformal dimension as depicted in figure 1.

These two regimes (of small and large conformal dimension) correspond
to different phases of the theory. When we consider large $N$ gauge
theories in the canonical ensemble we encounter a phase transition
associated with deconfinement; the free Fock space states (and their
interacting generalizations) dominate in the low temperature confining
phase (dual to free gravitons in AdS) and the high temperature
deconfined quark-gluon plasma phase is associated with heavy black
hole microstates with an $N$-dependent entropy. At strong coupling in
the bulk this is the Hawking-Page phase transition \HawkingDH, which
was first related to deconfinement in the gauge theory
in \WittenQJ. In general large $N$ gauge theories a range of
intermediate behaviours is possible \AharonySX\ encapsulating various
other phase transitions (e.g. an intermediate stringy regime with
Hagedorn behaviour) but for our purposes the picture presented here is
sufficient.

\hfil \break
\noindent {\bf Symmetric Orbifolds}

A second class of theories with holographic duals are large $N$
symmetric orbifold CFTs in two dimensions. Perhaps the best studied
examples are 2-d $\sigma$-models with symmetric orbifold target spaces
(${\cal M}^N/S_N$ with ${\cal M} = K3$ or $T^4$ as the best known
examples), which are realized on bound states of D1 and D5 branes in
IIB string theory \AharonyTI. More generally starting with {\it any}
two dimensional conformal field theory CFT$_a$ we can construct the
symmetric orbifold ${\rm CFT}_N \equiv ({\rm CFT}_a)^N/S_N$. At large
$N$ the theory ${\rm CFT}_N$ satisfies all the criteria that we listed
and thus should have some sort of holographic dual (most likely
stringy).

Let us see how the conditions i), ii), iii) are satisfied in this
case. While the central charge of the theory ${\rm CFT}_N$ is $N$
times the central charge of ${\rm CFT}_a$ the low-lying spectrum does
not enjoy the naive $N$ dependence one might expect from a theory with
order $N$ degrees of freedom.  This is because the orbifold projection
restricts the spectrum only to symmetrized states so e.g. if CFT$_a$
is a $\sigma$-model on $\cal M$ the supersymmetric zero mode
wavefunctions are not $N$ copies of the cohomology of $\cal M$ but a
single copy. Thus at low conformal dimension the degeneracy of
operators is small, in the sense that it does not grow with
$N$\foot{For supersymmetric CFTs the statements about the spectrum
have to be understood in the NS sector which is dual to AdS$_3$ in
global coordinates.}.

The correlators also enjoy a large $N$ expansion implying
factorization at $N
\rightarrow \infty$ \LuninYV, \LuninPW, \PakmanZZ\ and consequently we
expect a simple structure for the multi-particle spectrum.  These
low-lying states are separated from the heavy ones by a phase
transition. The phase transition in this system has a rather different
origin from that of gauge theories.  Here it is associated with the
appearance of long strings \MaldacenaDS.  Orbifold CFTs have twisted
sectors coming from long strings with winding number $w$.  These
strings can be described by inserting twist operators with
$\Delta \sim w$.  Because these strings are long they have a lower
mass gap and hence larger degeneracy at fixed temperature than a
shorter string.  The degeneracy of the short strings is
$N$-independent since we must symmetrize over the strings whereas a
single string with $w \sim N$ (the max winding number) feels an
``effective'' temperature $T \gg 1$ even if the real temperature is
order one because the mass gap on this string goes as $1/N$.  Hence
Cardy's formula can be used, giving an entropy density $s \sim N T$.
Thus at a temperature of order one\foot{More precisely it occurs at
temperature $T= 1/(2\pi)$ which corresponds to a torus which is
invariant under the modular transformation $\tau \rightarrow
-{1\over \tau}$. Here we assume that the CFT is defined on a spatial
circle of length $2\pi$.} the increased entropy of the long string
overcomes their energy cost and there is a phase transition from a
short string to a long string phase.  This implies that the degeneracy
of operators with large conformal dimension does indeed scale with
$c\sim N$. Note that while the qualitative structure is similar to a
gauge theory the actual mechanism for the phase transition is quite
different.

\hfil\break
\noindent{\bf $O(N)$ \& Related Vector Models }

Another example of a CFT which satisfies our criteria i), ii), iii) is the
$O(N)$ vector model.  The central charge of the CFT grows with $N$,
while at low conformal dimension the number of local operators which
are $O(N)$ invariant is $N$-independent i.e. small. Large $N$
factorization of ``single-particle'' correlators also holds.

It is believed \KlebanovJA\ that this theory (perhaps appropriately gauged) is
dual to a higher spin gravitational theory in anti-de Sitter space \VasilievDN,
\VasilievBA, \VasilievEV. See \SezginRT, \PetkouZZ, \LeighGK, \DasVW, \KochCY, \DouglasRC\ for
related work. Further evidence for this conjecture has been provided recently
with the computation of 3-point functions on both sides of the duality
\GiombiWH\ demonstrating that they agree.

The example of the $O(N)$ model is particularly interesting because
the bulk theory is not classical gravity, in the sense that it
contains an infinite number of fields and there is no parametric
separation between the graviton and the higher spin fields, while at
the same time it is not a string theory\foot{At least not a
conventional one, for example the growth of single-trace operators in
the CFT does not exhibit a Hagedorn (exponential) growth with
conformal dimension, unlike what happens in adjoint-valued gauge
theories.}. This suggests that the landscape of holographic CFTs may
contain various exotic possibilities.

Recently \GaberdielPZ\ considered coset WZW models and showed that
these CFTs provide a two-dimensional analog of vector $O(N)$ models.
In particular they are dual to higher spin theories on AdS$_3$.
Moreover, such CFTs come in families parametrized by a coupling
constant $0 \leq \lambda \leq 1$ analogous to the 't Hooft coupling of
gauge theories.  Unlike gauge theories, however, this coupling is
bounded from above so the associated bulk geometry is never a standard
gravity dual.  The authors of \GaberdielPZ\ nonetheless present
evidence for a complete equivalence of the spectrum in the
$N\rightarrow \infty$ limit. See also \GaberdielWB.  This example
suggests that the notion of a bulk geometry as providing an effective
description of a CFT at large $c$ is not restricted to ``string''
bulks with a weak-coupling gravity regime.  Rather, as we will argue,
a large $c$ CFT is naturally described by an AdS theory in one higher
dimension with a possibly large, but $c$-independent, number of
(approximately free) fields.

\hfil \break
\noindent {\bf The MSW CFT}

Perhaps the most poorly understood example of a theory with a known
gravitational dual is the ${\cal N}=(0,4)$ CFT described
in \MaldacenaDE\ (see also \MinasianQN) as the low-energy dimensional
reduction (to 1+1d) of the theory on a smooth M5-brane wrapping ${\cal
M}\times S^1 \times {\cal R}_t^1$ with ${\cal M}$ a holomorphic
divisor (four-cycle) in a CY $X$ with a Poincare dual $[p] \in
H^2(X)$, $S^1$ the M-theory circle of radius $R$, and ${\cal R}_t^1$
time.

While little is known about this theory it is clear from the
M5-decoupling limit that it it is dual to M-theory on
AdS$_3\times$S$^2\times$CY.  The central charge of this theory is
$c \sim \int_X p\wedge p \wedge p$, the dimension of the moduli space
of the divisor.  If the M5-brane wraps a cycle ${\cal M}$ whose size
is small relative to the (M-theory Planck) volume of the CY, $V_X$,
its dynamics can be captured by a field theory which, moreover, can be
reduced to a 1+1d $\sigma$-model in the limit $R^6 \gg V_X$.  The
latter is always satisfied in the near-horizon AdS$_3$ decoupling
limit so the dual is an effective CFT$_2$ with at least two
parameters, $c$ and $V_X$.  $V_X$ is in a supergravity hypermultiplet
and seems to play the role of a coupling constant in the
$\sigma$-model.  This is also consistent with the validity of 5-d
supergravity only in the regime $c \gg V_X$.

As emphasized in \MaldacenaDE\ the M5 under consideration can be
thought of as a {\it single} very large, smooth brane which does not
intersect itself so we may restrict attention to the {\it abelian} M5
brane theory living on $\cal M$.  Thus, unlike the previous examples,
there does not seem to be any large symmetry group truncating the
spectrum at low conformal dimension.  Rather, at a generic point, the
theory seems to be nothing more than a $\sigma$-model on a
parametrically high-dimensional target space (a complex torus bundle
over ${\cal CP}^{c\over 6}$ \MinasianQN).  Thus at weak coupling one
might imagine a degeneracy of order $c$ in the low-lying states of the
NS sector (from the $c$ approximately free oscillator in the
$\sigma$-model) but this is not consistent with the known spectrum of
gravitons in the dual (global) AdS$_3$ theory
\deBoerUN. This problem even plagues the supersymmetric spectrum as half-BPS
states correspond to arbitrary excitations on the non-supersymmetric
side of the theory.

It may be the case that some unknown symmetry\foot{As mentioned in \MinasianQN\
this theory suffers monodromies around singular points in the divisor moduli
space and these may constrain the spectrum.} truncates the spectrum as occurred
in other examples but, lacking this, we will consider alternative explanations
here\foot{We would like to thank J. de Boer for discussions on several of the
possibilities mentioned below.}.  The regime of validity of supergravity corresponds to
the CY being small on the scale set by the divisor implying a strongly coupled
theory.  We can thus expect large corrections to the dimensions of generic
operators in the theory.  It is not surprising then that the large degeneracy of
states expected by associating $c$ with the number of degrees of
freedom\foot{Recall that in 2-d $c$ is directly related to the local entropy
density via Cardy's argument.} is not realized at low conformal dimension as the
strong-coupling of the theory generically corrects the conformal dimensions of
operators.  While BPS states are expected to be protected from such corrections
the amount of supersymmetry present is not enough to prevent long multiplets
from forming from shorter ones and indeed this is observed in orbifold conformal
field theories dual to AdS$_3\times$S$^3$. Unfortunately, as we discuss in
section \needgauge, there are reasons why this sort of truncation of the
spectrum is not entirely satisfactory.

An alternative explanation may be provided by the existence of winding
modes along the torus factors in the target space.  Turning on
three-form flux on the original $M5$ wrapping the divisor freezes most
of its deformation moduli resulting in a much lower dimensional target
space.  Such configurations, however, necessarily carry $M2$ charge so
this argument would not apply to the sectors of the theory without
such charges.  

Another possibility is simply that the $\sigma$-model description of the CFT
captures only its grossest features (i.e. symmetries and central charges) but is
otherwise too naive.  While this model was used successfully in \MaldacenaDE\ to
reproduce the (subleading) black hole entropy this actually used very few
detailed properties of the theory.  Thus it is possible that the naive
$\sigma$-model description of this theory is incorrect.  For instance, the
divisor moduli space (the CFT target space) is known to have singular loci where
the five-brane self-intersects, generating new light degrees of freedom which
may furnish the missing gauge symmetry.  One would still have to explain  what
error is being made in the $\sigma$-model description away from the singular
locus.

While we do not understand how the low-energy spectrum of the MSW theory is
truncated to match the spectrum of AdS$_3$ we can observe that even in a trivial
free-field realization this theory seems to satisfy Cardy's formula already at a
$h \sim c$ \deBoerFK\ (whereas the standard Cardy regime is $h \gg c$ with $h$
the conformal dimension).  On the other hand without a known mechanism to excise
the $c$-dependence of the entropy for $h \ll c$ this will not imply any actual
phase transition.

\hfil \break
\noindent {\bf Other examples}

It is believed that the theory on $N$ coincident M5 branes in eleven
dimensional M-theory flows in the IR to an isolated six dimensional
superconformal field theory. The central charge of this CFT scales
like $N^3$. In the large $N$ limit the CFT is holographically dual to
M-theory on AdS$_4\times$S$^7$.  Unfortunately the field theory side
of this duality is not well understood but from the bulk it is clear
that the boundary CFT should satisfy the properties i), ii), iii) that
we have mentioned.

Another class of examples are direct products\foot{Here we assume that
the number of factors in the product is small compared to the central
charge of each of the factors.} of CFTs, each of which has a
holographic description, and their small deformations. Examples of
this kind were studied
in \KiritsisHY, \AharonyHZ, \KiritsisXJ, \KiritsisAT, \NiarchosQB\ and
it was argued that in a sense they correspond to multiple emergent AdS
throats. These results are consistent with our discussion, as should
become evident in later sections. We focus on the case where there is
only one copy of AdS space in order to simplify the analysis.

Recently a new interesting class of large $N$ WZW and coset 2d CFTs
and their holographic description was considered in \KiritsisXC, also
based on earlier work \BakasRY, \BakasXU, \BakasFS, \BakasRB\ where
aspects of the large $N$ limit of these CFTs was considered. It would
be interesting to explore how they fit into our general discussion.

\subsec{Counter-Examples}
\subseclab\cexamples

Let us consider theories that do not exhibit the properties i), ii),
iii) high-lighted above, that we claim are necessary conditions for
the existence of a gravitational dual.

\hfil\break
\noindent{\bf CFTs with small central charge}

In this paper we are interested in theories with weakly coupled
holographic duals; that is with duals where the Planck length is
parametrically smaller than the radius of the AdS space. This can only
happen when the central charge of the CFT is parametrically large. The
reason that we restrict ourselves to this class is that since we have
no independent definition of quantum gravity in an AdS space of Planck
size we see no way of addressing the question of which CFTs have
holographic duals of this kind.  Rather we adopt the perspective that
the bulk provides an ``effective'' description of large $c$ CFTs
irrespective of whether it itself is (non-perturbatively)
well-defined.

Of course according to the strong version of the AdS/CFT
correspondence even the ${\cal N}=4$ SYM with small gauge group, for
example $SU(2)$, should have some kind of highly quantum AdS string
theory dual, which should be, presumably, independently definable. We
do not have any specific reason to doubt this, so CFTs with small
central charge are not really counter-examples, but they simply do not
fall into the class of weakly coupled holographic duals that we
decided to consider in this paper.

\hfil\break
\noindent{\bf Direct products of small CFTs}

Another counterexample is to take a CFT with small central charge and
construct the direct product\foot{i.e. without imposing any orbifold
symmetrization.} of a large number $N$ of copies of it. Such a CFT has
large central charge but the low-lying spectrum differs quite sharply
from what we expect in an AdS dual. The number of operators of low
conformal dimension grows with $N$, which would indicate that the
number of light fields in the dual AdS space would blow up as we take   
the weak coupling limit, i.e. $\hbar\rightarrow 0$. It is doubtful
that such a dual theory would be very useful. Moreover there is no
sense in which such a theory factorizes as $N\rightarrow\infty$.  Nor
is it likely that a small perturbation of this theory will have the
correct properties\foot{Notice that these problems do not arise if we
consider the product of a small number of large CFTs as discussed at
the end of the previous subsection.}.

\hfil\break
\noindent{\bf Cyclic orbifolds in two dimensions\foot{This counterexample was
suggested to us by J. de Boer.}}

Starting with any two dimensional conformal field theory CFT$_a$ we
can construct the cyclic orbifold $({\rm CFT}_a)^N/Z_N$. Like the
symmetric orbifold $({\rm CFT}_a)^N /S_N$, the cyclic orbifold has
central charge which is $N$ times that of CFT$_a$ so it satisfies
condition i). However its spectrum at low conformal dimension does not
satisfy the condition ii): the low-lying spectrum {\it is not}
$N$-independent, contrary to what happens in symmetric
orbifolds. Moreover these theories do not satisfy the factorization
condition iii) i.e. they do not have a good $1/N$ expansion. Hence it
seems unlikely that such CFTs have reasonable holographic duals.

\subsec{Do we need gauge-symmetry?}
\subseclab\needgauge 

The collection of examples and counter-examples above leads to a broad
picture of which structures are necessary in order for a bulk dual to
emerge but it remains to be determined exactly which features are
essentially related to holography and which are specific to certain
models or classes of models.

In particular what is clearly necessary, as mentioned in the
introduction, is:

i) A family of CFTs with $c \rightarrow \infty$ with a bulk dual
emerging as a perturbative description about the $c=\infty$ point.

ii) These CFTs contain operators $\{{\cal O}_\Delta\}$ whose conformal
dimension is $c$-independent, $\Delta \sim {\cal O}(c^0)$, and whose
degeneracy is also independent of $c$ (while we'll see that conformal
invariance imposes a $c$-dependence on the spectrum of operators with
$\Delta \sim {\cal O}(c)$).

iii) Correlators of the low-lying operators $\{{\cal O}_\Delta\}$
above should factorize in the $c\rightarrow \infty$ limit.

While we may impose these as {\it necessary} conditions for an
abstract CFT to have a bulk dual (and we will try to argue in the
following that these requirements are also sufficient) we
have not addressed the question of what theories exhibit these
properties or, put another way, what underlying structure results in
CFTs satisfying i), ii), iii).  Clearly the examples above demonstrate
that certain theories with a large gauge symmetry fall into the class
covered by i), ii), iii) but one of the hopes of this work is to
generalize beyond this canonical example.

Unfortunately, aside from abstract conformal field theories\foot{The
data for which can be specified in terms of OPE coefficients and
conformal dimensions of operators as will be described in the next
section.}, it is rather difficult to define theories away from weakly
coupled points.  Let us then consider the following.  Suppose that our
putative dual CFT has a marginal deformation that takes it to a weakly
coupled point. At this point must the CFT be a gauge theory\foot{By
gauge theory we mean also more general examples like symmetric
orbifolds that have a large gauge symmetry.}?

An argument in favor of this would be the following.  The spectrum of
a free theory with central charge $c$ will in general have a
$c$-dependent degeneracy\foot{This follows simply from the dependence
of both $c$ ($C_T$ in \OsbornCR) and the free energy on the number of
free scalar, vector and tensor fields \OsbornCR.} at all conformal
dimensions (including the lowest levels).  To avoid this we need to
manually truncate the spectrum by decoupling a large number of states.
This type of decoupling is characteristic of a gauge symmetry which
implies that parts of the spectrum are unphysical.  We might try to
imagine different mechanisms that achieve this truncation but we know
of no other candidate in a free theory.

In a strongly coupled theory, on the other hand, it is quite possible
that generic operators are massive with a mass scale set by the
coupling $\lambda$.  Thus for theories without a weakly coupled point
the problem cannot be made as sharp but one might still imagine that
at $\Delta \sim \lambda$ the $c$-dependence of the spectrum should
re-emerge (though such naive arguments may not hold in a strongly
coupled theory).  Since $\lambda$ is unrelated to $c$ this does not
reflect the general structure we expect in the bulk in which the
spectrum remains independent of $c$ for all $\Delta \ll c$ in the
$c\rightarrow
\infty$ limit.

It would be interesting to explore this question further by
considering examples such as the MSW CFT where gauge-invariance is
certainly not evident but in the following we will sidestep this issue
by simply assuming the correct form of the spectrum without concerning
ourselves with the mechanism that implements this.

\newsec{Consistency Requirements for a CFT}
\seclab\cftconsistency

Before we proceed we review some basic consistency requirements that
have to be satisfied by any unitary conformal field theory.

\subsec{Conformal Bootstrap}

Let us say we are given a candidate set of correlation functions for
all local operators of a QFT. What are the conditions that they have
to satisfy if they are to come from a well-defined conformal field
theory? The basic ingredients of a conformal field theory are the
conformal primary operators ${\cal O}_i$, which are labeled by their
conformal dimension $\Delta_i$ and spin. The normalization of the
operators is arbitrary and we can choose a basis so that their 2-point
functions, in the case of scalar operators, have the form
$$
\langle {\cal O}_i (x)\, {\cal O}_j(y) \rangle = {\delta_{ij} \over |x-y|^{2\Delta_i}}
$$
A basic property of a CFT is that there is an operator product
expansion (OPE): the product of two local operators can be expressed
as a sum over other local operators which has a finite radius of
convergence. In general
\eqn\genope{
{\cal O}_i(x) \,{\cal O}_j(y) =\sum_k c_{ij}^k(x-y) {\cal B}_k(y) }
where the sum runs over all local operators ${\cal B}_k$, not
necessarily primary, and $c_{ij}^k(x-y)$ are functions dependent on
the dimensions and spins of the operators involved, as well as on the
dynamics of the theory. The equality \genope\ holds in the sense that
we can replace the product on the LHS by the sum on the RHS inside
correlation functions, as long as there are no other operators at
smaller distances from $y$ than $|x-y|$.

The expression \genope\ can be greatly simplified by imposing the
requirement of conformal invariance. Then it can be shown that the
coefficients $c_{ij}^k(x-y)$ for the descendant operators can be
uniquely determined by kinematics from those between conformal
primaries. So all the dynamical information is contained in the OPE
coefficients between primaries\foot{For simplicity we write down only
the contribution from scalar primaries. For intermediate primaries
with nonzero spin the form of the OPE is more complicated but still
fixed in terms of a single constant.}
\eqn\basicopef{
{\cal O}_i(x) \, {\cal O}_j(0) = \sum_k C_{ij}^k {\cal O}_k{1\over
|x|^{\Delta_i+\Delta_j - \Delta_k}} +{\rm descendants} } where now
$C_{ij}^k$ are position independent constants. The full OPE (including
the descendant contributions) can be reconstructed as
\eqn\difopintr{
{\cal O}_i(x) {\cal O}_j(y) = \sum_k C_{ij}^k \, \hat{\cal
F}_k(x-y,\partial_y) \,{\cal O}_k(y) } where the differential operator
$ \hat{\cal F}_k(x-y,\partial_y)$ only depends on kinematics i.e. on
the dimensions and spins of the primaries ${\cal O}_i,{\cal O}_j,
{\cal O}_k$ and not on the dynamics of the CFT\foot{This differential
operator can be determined by multiplying both sides of
\difopintr\ with ${\cal O}_k(z)$ and demanding that the resulting 3-point
function on the LHS is reproduced by the differential operator acting
on the 2-point function on the RHS, see for example \DolanUT.}.

The OPE can be used to compute higher $n$-point functions of local
operators. Starting with an $n$-point function, we can first take two
of the operators which are close to each other and replace them with
their complete OPE. In this way we can rewrite the $n$-point function
as an (infinite) sum over $(n-1)$-point functions.  Similarly the
$(n-1)$-point functions can be reduced by an OPE to sums over
$(n-2)$-point functions and so on, until we get to the basic 2- and
3-point functions.  Hence the entire CFT can be reconstructed from
knowing the spectrum of conformal primaries ${\cal O}_i$, i.e. their
conformal dimensions $\Delta_i$ and spins, and the 3-point functions
$C_{ij}^k$.

A natural question is whether every choice of the data $(\Delta_i,
C_{ij}^k)$ corresponds to a consistent conformal field theory. The
answer is negative and this can be understood as follows. When we use
successive OPEs to reduce the $n$-point functions to 2- and 3-point
functions there is an ambiguity in the choice of the order that we
perform the OPE. To end up with a consistent theory the answer for the
$n$-point function should be independent of the order in which the
OPEs are performed. This leads to certain conditions that have to be
satisfied by the data $(\Delta_i,C_{ij}^k)$.

This can be seen in the simplest way by considering a four point
function
$$
\langle {\cal O}_1(x_1){\cal O}_2(x_2){\cal O}_3(x_3){\cal O}_4(x_4) \rangle
$$
which can evaluated by doing the OPE between the two operators at
$x_1$ and $x_2$ and at the same time between $x_3$ and $x_4$, or by
doing the OPE between $x_1$ and $x_4$ and at the same time between
$x_2$ and $x_3$ \foot{There is also the channel $(13)\rightarrow (24)$
but it is sufficient to check the conformal bootstrap between the
other two channels mentioned above.}. In the first case, i.e. in the
channel $(12)\rightarrow (34)$, the 4-point function becomes
\eqn\directope{
\langle {\cal O}_1(x_1){\cal O}_2(x_2){\cal O}_3(x_3){\cal O}_4(x_4) \rangle
= \sum_k C_{12}^k C_{34}^k \hat{{\cal
F}}_k(x_1-x_2,\partial_{x_2})\hat{{\cal F}}_k(x_3-x_4,\partial_{x_4})
{1\over |x_2-x_4|^{2\Delta_k}} } As we explained, the only terms in
this expression which are dependent on the dynamics are the OPE
coefficients $C_{12}^k,C_{34}^k$. So we introduce the following
functions
\eqn\cblockdef{
{\bf G}_k^{12,34}(x_1,x_2,x_3,x_4) \equiv\hat{{\cal
F}}_k(x_1-x_2,\partial_{x_2})\hat{{\cal F}}_k(x_3-x_4,\partial_{x_4})
{1\over |x_2-x_4|^{2\Delta_k}} } 
The functions ${\bf G}_k^{12,34}$ are
the so-called conformal blocks, or conformal partial waves
(CPWs). They correspond to the contribution of an operator ${\cal
O}_k$ and all of its descendants to the double OPE between $(12)$ and
$(34)$. The conformal blocks are determined by kinematics of the
conformal group. Explicit expressions for the CPWs in the case of
$d=4$ can be found in \DolanUT, \DolanHV\ and are summarized in
appendix A. Once the kinematic factors have been absorbed into the
CPWs, all of the dynamics of the theory lies in the information about
conformal dimensions and the 3-point functions $C_{ij}^k$.

We evaluated the 4-point function by performing the double OPE in the
``direct channel'' $(12)\rightarrow(34)$. Alternatively we could have
performed the OPE in the ``crossed channel'' $(14)\rightarrow (23)$
and we would have derived a similar expression as \directope\ with the
roles of the points $2$ and $4$ interchanged.  Demanding that the
resulting 4-point function is the same gives the following condition
\eqn\bootstrapb{
\sum_k C_{12}^k C_{34}^k\,\, {\bf G}_k^{12,34}(x_1,x_2,x_3,x_4) = \sum_k C_{14}^k C_{23}^k \,\, {\bf G}_k^{14,23} (x_1,x_4,x_2,x_3)}
which is called the conformal bootstrap condition (also
called``crossing symmetry'', if the four external operators are the
same) and is depicted schematically in figure 1. Notice that we have
such a condition for each choice of the four external operators.
\fig{Conformal bootstrap condition for the 4-point function.}
{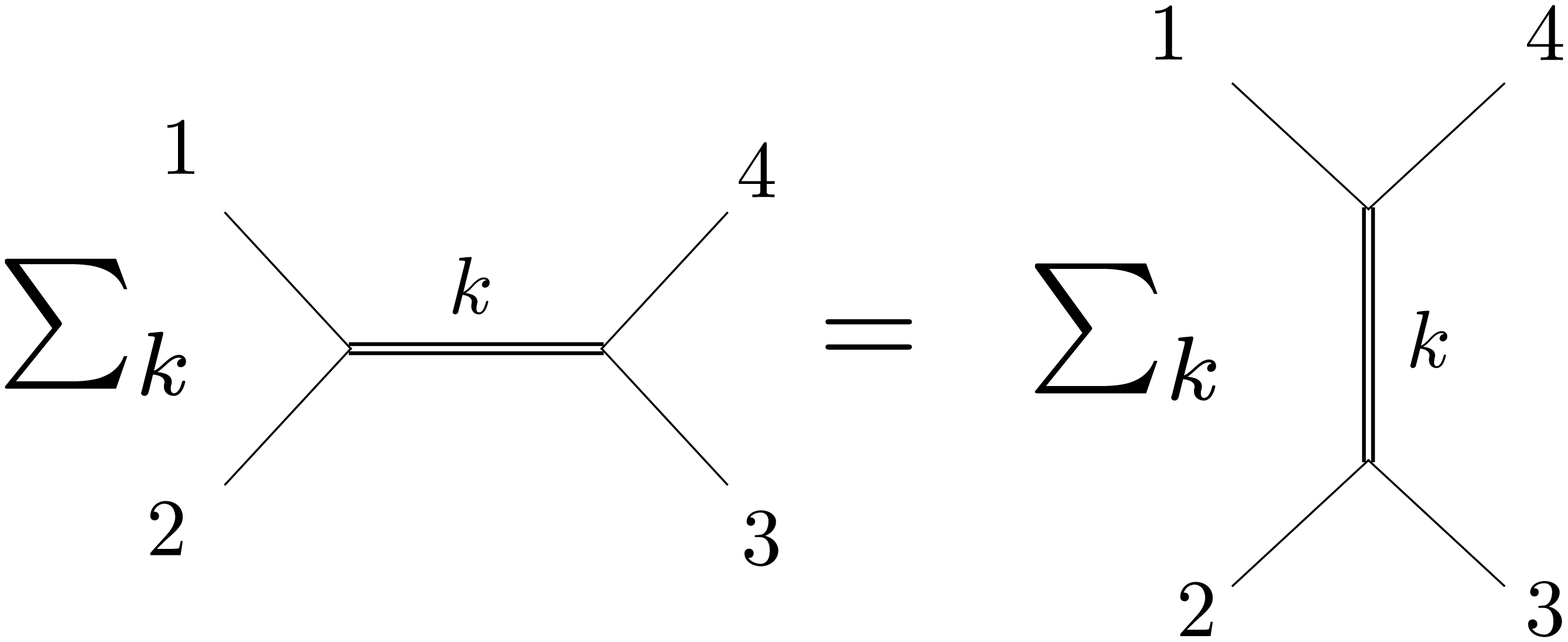}{4.truein}
\figlabel{\bootstrapone}
Any consistent CFT must satisfy the bootstrap condition\foot{Let us
notice that if we have $N$ conformal primaries, then the number of
data $(\Delta_i, C_{ij}^k)$ grows like $N^3+N$. On the other hand the
number of conditions of the form \bootstrapb\ grows like $N^4$. This
suggests that it might be possible to fully determine the dynamics of
all CFTs simply by imposing the bootstrap conditions. In practice it
is difficult to solve these conditions, especially since $N$ is
infinite. The program of solving CFTs via the ``conformal bootstrap''
approach has had some success in special two dimensional CFTs (for
example in the case of the minimal models).}. It can be shown that if
the bootstrap condition is satisfied for the 4-point function then the
reduction of all higher $n$-point functions via successive OPEs will
be consistent, independent of the order that we choose to perform the
OPE.

Notice that the unitarity of the theory requires that the coefficients
$C_{ijk}$ are real. See \RattazziPE, \RychkovIJ, \CaraccioloBX, \PolandWG, 
\RattazziGJ, \RattazziYC\ for recent work were the unitarity of the CFT 
in $d>2$ was used to derive certain constraints for the data
$\Delta_i,C_{ijk}$, in special cases.

\subsec{Modular Invariance and consistency at finite temperature}
\subseclab\sgenmod 

The conditions \bootstrapb\ are necessary conditions for the
consistency of the CFT.  However they are not sufficient. This can be
seen even in the simpler case of two dimensional CFTs. While the
conformal bootstrap equations guarantee the consistency of correlators
on the plane, the same is not true if we place the theory on a
nontrivial manifold, for example the torus or the cylinder\foot{The
requirement of having a consistent theory on certain nontrivial
manifolds (i.e. ${\bf R}^{d-1}\times {\bf S}^1$ or ${\bf
S}^{d-1}\times {\bf S}^1$) corresponds, from a physical point of view,
to demanding consistency of the CFT at finite temperature. }. Then
either by modular invariance or by arguments similar to those
described below we find new constraints on the theory which go
beyond the conformal bootstrap conditions.

In two dimensions modular invariance is derived by considering the CFT
on a torus. Conformal invariance together with invariance under large
diffeomorphisms of the torus implies that the path integral of the
theory must be invariant under the transformation $\tau \rightarrow
-1/\tau$ of the Teichmuller parameter. This implies an invariance of
the thermal partition function under the transformation $T \rightarrow
1\,/\,(L^2 T)$ of the temperature, where $L$ is the length of the
spatial circle on which the CFT is defined. This relates the spectrum
of large conformal dimension operators (which dominate at large $T$)
to that of low lying spectrum (which dominate for small $T$).

Modular invariance is a very strong constraint on the spectrum of a CFT and is
moreover independent of the conformal bootstrap conditions on the plane.  To see
this let us consider a two dimensional CFT consisting only of the Virasoro
module of the identity operator (which includes the stress tensor $T(z)$). All
$n$-point functions of this theory, for example correlators of the form
$\langle\, T(z_1)\,...\,T(z_n)\,\rangle$ can be directly computed because of
holomorphy\foot{An $n$-point function of this form is meromorphic with respect
to any of its arguments, so it can be determined by its singularities. It has
poles whenever one of the operators approaches the other insertions. These poles
are fixed by the $TT$ OPE which is completely determined by conformal invariance
and only contains $T$ and its descendants. Hence the poles of this correlator
are related to $(n-1)$-point functions of $T$. Recursively all $n$-point
functions of this type can be reconstructed from the $TT$ 2-point function and
the $TT$ OPE. The answer depends only on the value of the central charge $c$.}
and they only depend on the central charge $c$. It is straightforward to show
that they satisfy the bootstrap equations if we assume that fields only fuse
within the $T$-module. So the theory seems to be consistent on the plane, in the
sense that crossing symmetry is satisfied. On the other hand its partition
function is not modular invariant\foot{Defining $q=e^{-\beta}$ the partition
function of this theory would be $ Z(q) = q^{-c/24} \prod_{n=2}^\infty {1\over
1-q^n} $ which is not modular invariant, so it cannot be a consistent theory.},
hence such a theory cannot exist.

In higher dimensions there is no direct analogue of modular
invariance. If we consider the CFT on a $d$-dimensional torus ${\bf
T}^d$ then the partition function has to be invariant under $SL(d,{\bf
Z})$ large diffeomorphisms but it is not clear how to translate this
invariance to a condition for the spectrum or correlators of the
CFT\foot{We would like to thank J. de Boer for discussions related to
this.}. Nonetheless demanding the consistency of the CFT at finite
temperature does indeed introduce some new constraints which go beyond
the bootstrap conditions.  We refer to such constraints as
``generalized modular invariance'' though they are clearly much more
difficult to analyze than modular invariance of a two dimensional CFT
on a torus.

To begin, let us consider a conformal field theory on ${\bf
R}^{d-1}\times {\bf S}^1$ where the size of ${\bf S}^1$ is $\beta
= {1 \over T}$. Notice that on this space, unlike what happens on
${\bf R}^d$, operators can have nonzero expectation values: we have
introduced the scale $\beta$, so an operator ${\cal O}_k$ of dimension
$\Delta_k$ can have an expectation value of the form
\eqn\thermalvev{
\langle {\cal O}_k \rangle_\beta = {A_k \over \beta^{\Delta_k}}
} 
where the constant $A_k$ is independent of $\beta$.

These thermal 1-point functions can be computed from the 3-point
functions $C_{ij}^k$ as follows: consider the CFT living on a sphere
${\bf S}^{d-1}$ of radius 1, at inverse temperature $\beta$. By
definition we have
$$
\langle {\cal O}_k\rangle_\beta' = {\sum_{|\psi\rangle} \langle \psi
|\, {\cal O}_k\,|\psi\rangle e^{-\beta E_{\psi}}\over Z(\beta)}
$$
where $Z(\beta)\equiv \sum_{|\psi\rangle} e^{-\beta E_\psi}$ is the
thermal partition function of the system. The sum runs over all states
of the Hilbert space, $E_{\psi}$ is the energy of the state
$|\psi\rangle$ and the prime refers to the fact that the 1-point
function is evaluated on ${\bf S}^{d-1}\times {\bf S}^1$
unlike \thermalvev\ which was on ${\bf R}^{d-1}\times {\bf
S}^1$. Using the state operator map $|\psi\rangle \leftrightarrow
{\cal O}$ this can be thought of as a sum over all local operators
${\cal O}$ of the CFT (not necessarily conformal primaries) with the
identifications $E_\psi\leftrightarrow
\Delta_{\cal O}$ and $\langle \psi|\, {\cal O}_k\,|\psi\rangle\leftrightarrow C_{{\cal O}
{\cal O}{\cal O}_k}$ (assuming we work in a basis where the ${\cal
O}$'s are hermitian). The high temperature limit of the 1-point
function on ${\bf S}^{d-1}\times {\bf S}^1$ should be the same as that
on ${\bf R}^{d-1}\times {\bf S}^1$, in other words
$\lim_{\beta\rightarrow 0} {\langle {\cal O}_k\rangle'\over \langle
{\cal O}_k\rangle}=1$, so we finally find
\eqn\onefromthree{
A_k = \lim_{\beta\rightarrow 0} \left(\beta^{\Delta_k} {\sum_{\cal O}
 C_{{\cal O} {\cal O}{\cal O}_k} e^{-\beta \Delta_{\cal O}}\over
\sum_{\cal O} e^{-\beta \Delta_{\cal O}}}\right)} Thus in principle if we know all the data $(\Delta_k,C_{ij}^k)$ of
the zero temperature CFT, we can compute the thermal 1-point
functions $A_k$.

Let us now consider correlation functions of operators at finite
temperature starting with a thermal 2-point function $\langle {\cal
O}_i(x) {\cal O}_j(y)\rangle_\beta$. If the points $x$ and $y$ are
close to each other then we can consider the OPE between the two
operators. Notice that the OPE is a short distance expansion (or in
other words an operator statement) so it should hold even when the
theory is placed at finite temperature. The only difference is that
operators on the RHS can have nontrivial 1-point functions. In other
words, if we know the exact OPE of two operators
$$
{\cal O}_i (x) {\cal O}_j(y) = \sum_k C_{ij}^k(x-y) {\cal O}_k(y)
$$
then we can express the thermal 2-point function of the operators in
terms of 1-point functions as
\eqn\thermaltwoope{
\langle {\cal O}_i (x) {\cal O}_j(y) \rangle_\beta = 
\sum_k C_{ij}^k(x-y) \langle {\cal O}_k\rangle_\beta = 
\sum_k C_{ij}^k(x-y) {A_k \over \beta^{\Delta_k}}
} where the constants $A_k$ were introduced in \thermalvev. Notice
that this has to be understood as an expansion for $x\rightarrow y$,
which may not converge when $|x-y|>\beta$, but which can (presumably)
be analytically continued everywhere\foot{See \PetkouFB, \PetkouFC\
for some applications of the finite temperature OPE in $d>2$
dimensional CFTs.}. Of course the analytic continuation can be done
only after summing over $k$ and not term-by-term, so in practice it
may be difficult to perform.

The 2-point function constructed in this way has to be periodic around
the thermal circle (i.e. to satisfy the KMS condition)
\eqn\periodcond{
\langle {\cal O}_i (\tau,{\bf x}) {\cal O}_j(y) \rangle_\beta  = \langle {\cal O}_i (\tau+\beta,{\bf x}) {\cal O}_j(y) \rangle_\beta 
} where ${\bf x}$ is the coordinate on ${\bf R}^{d-1}$. This already
imposes some new conditions among the 3-point functions $C_{ij}^k$ and
the conformal dimensions $\Delta_i$ via
equations \onefromthree, \thermaltwoope\ and \periodcond.

Moreover when we consider the limit where $|x-y|\rightarrow \infty$,
i.e. when the two points are widely separated in the spatial
directions, we expect that the 2-point function will factorize to a
product of 1-point functions
\eqn\factcond{
\lim_{|x-y|\rightarrow \infty} \langle {\cal O}_i (x) {\cal O}_j(y) \rangle_\beta  = \langle {\cal O}_i \rangle_\beta  \langle {\cal O}_j \rangle_\beta = {A_i A_j \over \beta^{\Delta_i + \Delta_j}}
}This condition gives more non-trivial constrains for the data
$(\Delta_i, C_{ij}^k)$ \foot{The constants $A_k$ can be computed
from \onefromthree\ and then the LHS by the (analytic continuation
of) \thermaltwoope) }.

\fig{The thermal 2-point function $\langle{\cal O}_i(x) {\cal
O}_j(y)\rangle_\beta$ can be evaluated by the OPE in the limit $x\rightarrow y$,
in terms of thermal 1-point functions. In the opposite limit where
$|x-y|\rightarrow \infty$ it should factorize to the product of
1-point functions $\langle{\cal O}_i\rangle_\beta \langle {\cal
O}_j\rangle_\beta$.}  {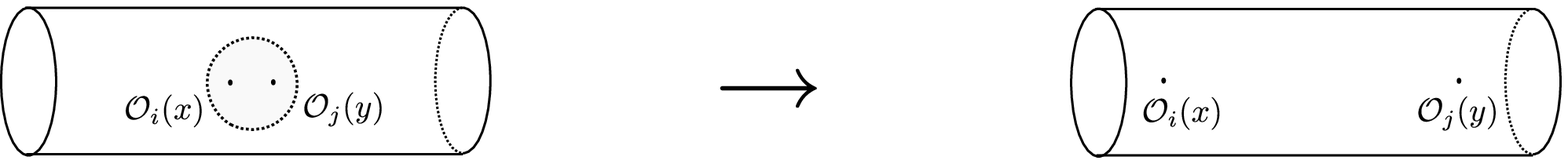}{6.truein}
\figlabel{\bootstraptwo}
Finally we have the following condition: let us consider a small
variation of the temperature. The temperature is related to the size
of the thermal circle, or equivalently to the value of the metric
component $g_{00}$. Changing the temperature can be understood as
changing $g_{00}$. We know that by its definition, the operator
$T_{00}$ is ``dual'' to the component $g_{00}$ in the sense that
$g_{00}$ is the source for $T_{00}$. Changing the metric
infinitesimally is equivalent to inserting the stress tensor in the
correlators.  Hence we arrive at the following relation
\eqn\cardy{
{\partial \langle {\cal O}_k\rangle_{\beta} \over \partial \beta} =-
{1\over \beta}\int d^d x \, \, \langle T_{00}(x)\, {\cal
O}_k(0)\rangle_{\beta}^c} where the integral\foot{The integral has to
be regularized and (temperature-independent) divergences have to be
removed.} is over ${\bf R}^{d-1}\times {\bf S}^1$ and the superscript
$c$ in the 2-point function stands for ``connected''\foot{The
connected 2-point function of two operators is defined as $ \langle
{\cal O}_i(x) \,{\cal O}_j(y)\rangle_{\beta}^c \equiv \langle {\cal
O}_i(x) \,{\cal O}_j(y)\rangle_{\beta} -\langle {\cal
O}_i(x)\rangle_\beta \langle {\cal O}_j(y)\rangle_\beta$.}. The LHS of
this equation can be immediately evaluated from \thermalvev\ while the
RHS from the (analytic continuation of) the OPE \thermaltwoope. The
set of equations \cardy\ (one equation for each operator ${\cal O}_k$)
give additional nontrivial conditions that have to be satisfied by any
conformal field theory in any dimension.

To summarize, in this section we have formulated some additional
general constraints \periodcond, \factcond\ and \cardy\ for consistent
CFTs which go beyond the conformal bootstrap conditions, though we
have not analyzed these constraints sufficiently to understand the
extent to which they are all independent. As we will see in the next
subsection they impose some non-trivial conditions on the spectrum of
operators, reminiscent of those coming from modular invariance in two
dimensional CFTs.

Before we close this section let us mention that the constraints from
modular invariance of the boundary CFT, in the context of the AdS$_3$/CFT$_2$
correspondence, have been recently discussed in \WittenKT, \GaberdielVE, 
\GaberdielXB, \HellermanBU, \HellermanQD, \CastroCE.

\subsec{``Cardy formula'' and higher dimensional CFTs}
\subseclab\cardyone

Notice that in a two-dimensional CFT if we apply the condition \cardy\
by taking the operator ${\cal O}_k$ to be $T_{00}$ itself, then we can
derive the Cardy formula without direct reference to modular
invariance: on general grounds the thermal expectation value of the
stress tensor on ${\bf R}\times {\bf S}^1$ (i.e. the thermal energy
density) has the form
$$
\langle T_{00}\rangle_\beta ={\pi \over 6} \,\widetilde{c}\,T^{2}
$$
where $T$ is the temperature. The constant $\widetilde{c}$ is a priori
unrelated to the central charge $c$, defined by the 2-point function
of the stress-energy tensor on the plane.  Standard thermodynamic
arguments imply that the associated entropy density is
$$
s = {\pi \over 3}\,\widetilde{c} \, T
$$

Now, the thermal 2-point function $\langle T_{00}(x)
T_{00}(y)\rangle_\beta$ in a 2d CFT can be exactly computed by the
cylinder-to-plane exponential map and is proportional to the central
charge $c$. By imposing condition \cardy\ and performing the integral
one can then show \BloeteQM\ that in two dimensions
$$
\widetilde{c} = c
$$
This leads to the Cardy formula, which determines the entropy of the
CFT at high temperatures in terms of the central charge $c$.

Can this argument be used in higher dimensional CFTs to relate the
thermal entropy density to the ``central charge'' of the CFT? In a
higher dimensional CFT the entropy density has the form
$s\sim \widetilde{c} \,T^{d-1}$. Let us call $c$ the constant which
appears in the 2-point function of the stress tensor in flat space.
While equation \cardy\ is still true for higher dimensional CFTs, the
argument described above cannot be applied and there is no analogue of
the Cardy formula. In particular, the constants $c$ and
$\widetilde{c}$ can in general be different (and {\it do} differ in
known examples\foot{For example in the large $N$ $SU(N)$ ${\cal N}=4$
SYM the ratio $\widetilde{c}/ c$ is believed to continuously
interpolate between the value $4/3$ at $\lambda\ll 1$ and the value
$1$ at $\lambda\gg 1$. Also in free four-dimensional CFTs the ratio
$\widetilde{c}/ c$ can take various values, see \KovtunKW\ for a
recent discussion.}).  The problem with applying the previous argument
is that there is no way to evaluate the thermal 2-point function
$\langle T_{00}(x) T_{00}(y)\rangle_\beta$ in a simple way, unlike
what happens in 2d CFTs. In higher dimensions there is no analogue of
the exponential map between ${\bf R}^{d-1}\times {\bf S}^1$ and ${\bf
R}^d$ and the thermal 2-point function of the stress tensor is not
fixed by conformal invariance. However, as we discuss in
section \subcardy, we may still be able to use the condition \cardy\
to derive some qualitative statements about the high-temperature
entropy of the CFT, in the spirit of the Cardy formula.

\newsec{Generalized Free CFTs and holography}
\seclab\gff

We now return to our goal of identifying the class of $d$-dimensional
CFTs for which a holographic gravitational theory provides an
effective description, at least in some regime.

As we explained in the previous sections, the minimum structure
necessary in order for a CFT to have a (weakly coupled) bulk dual is
that it contains in its Hilbert space a sector which consists of {\it
Generalized Free Fields } (GFF) i.e. operators whose correlators
factorize.  By analyzing a CFT containing such fields we are naturally
led to the notion of ``multiparticle states''.  These GFF and the
multiparticle states are defined in an expansion about large central
charge, $c \rightarrow \infty$, and satisfy crossing symmetry to
zeroth order in a $1/c$ expansion.  Even at this order, however, we
will find an inconsistency emerging from attempting to reconcile the
spectrum of this theory with other constraints from conformal
invariance: such CFTs fail to satisfy the ``generalized modular
invariance'' conditions mentioned in \sgenmod.

For finite but large $c$ we argue that our analysis of the spectrum
receives corrections for operators of large dimension (states with
$\Delta \sim c$) resolving the inconsistency.  This hints at an
essential property of such theories that is manifest in their dual
bulk description: they must be completed by the addition of ``black
hole states'' satisfying $\Delta
\sim c$.  We will return to this notion in later sections.

\subsec{Generalized free fields: factorization for $\Delta>{d-2 \over 2}$}

Let us now explain what is special about ``generalized free
fields''. Intuitively we think of a field as being ``free'' when it
obeys linear equations of motion. A basic consequence of linearity is
that it allows the superposition of solutions, or superposition of
excitations of the field. Similarly, we think of a ``weakly
interacting'' field as one which is a small deformation of a free one,
i.e.  when it obeys equation of motion which have small nonlinearities
around linear equations of motion. The nonlinearities are controlled
by a coupling constant $g$, and as $g\rightarrow 0$ the field becomes
linear.

In a CFT in $d$ spacetime dimensions, the condition that a scalar
operator ${\cal O}$ is free is equivalent to the fact that its
conformal dimension is $\Delta={d-2 \over 2}$. This can be shown from
the conformal algebra. To see whether $\nabla^2 {\cal O} =0$ we
consider the norm of the state $P_\mu P^\mu |{\cal O}\rangle$.  Using
the conformal algebra we find that
\eqn\confeom{
\|P_\mu P^\mu |{\cal O}\rangle\|^2 = \langle {\cal O}| K_\nu K^\nu 
P_\mu P^\mu |{\cal O}\rangle = 8\, d\, \Delta\left(\Delta-{d-2 \over
2}\right)
\||{\cal O}\rangle\|^2
} so the condition $\nabla^2{\cal O}=0$ is equivalent to
$\Delta={d-2 \over 2}$.

However there is another notion of a free field, characterized by the
fact that its correlation functions factorize to products of 2-point
functions.  For a standard free field with $\Delta={d-2 \over 2}$ the
factorization is an immediate consequence of the equation of motion
$\nabla^2 {\cal O}=0$. However, more generally we may have fields with
$\Delta>{d-2 \over 2}$ whose correlators factorize. These are called
``generalized free fields''. The reason that such fields should be
called free is that there is a sense in which we can superimpose
excitations created by these fields, as a result of their
factorization property. Generalized free fields have been discussed
before, in various contexts, in the literature \Jost.  However, as we
will explain later, our perspective is different since we argue that
such fields can only exist in a limiting sense (so some statements in
the older literature may be less relevant from our point of view).

More precisely let us define a generalized free field\foot{By the term
``field'' we mean a local operator in the CFT, not necessarily a
fundamental field of the Lagrangian. If the CFT is a gauge theory we
are obviously talking about gauge invariant operators.} $\CO$ of
dimension $\Delta>{d-2\over 2}$ as one whose correlators take the form
$$
\langle \CO(x_1) ... \CO(x_{n})\rangle = 
\langle \CO(x_1) \CO(x_2) \rangle ... \langle \CO(x_{n-1}) \CO(x_n) \rangle +
{\rm permutations}
$$
In particular $n$-point functions for $n$ odd vanish\foot{Conformal
invariance implies that one point functions vanish in a conformally
invariant vacuum.}. One peculiar property of these fields is that
while they are free, in the sense that the obey some kind of
superposition principle\foot{This is due to factorization. For
example, the energy of states created by acting repeatedly with the
field ${\cal O}$ is additive. In gauge theory language this is
expressed by the fact that the conformal dimension of a multi-trace
operator is equal to the sum of the conformal dimensions of its
constituent single-trace operators, at infinite $N$.}, they do not
obey linear equations of motion.  In particular, because $\CO$ has the
wrong dimension to be a free field we cannot describe it in terms of a
local free Lagrangian in the $d$-dimensional flat space where the CFT
lives.

For example if we choose the normalization of the operators
appropriately then the 2-point function is
\eqn\basictwo{
\langle {\cal O}(x) {\cal O}(y) \rangle = {1\over |x-y|^{2\Delta}}}
which does not obey a linear differential equation, unless $\Delta =
{d-2 \over 2}$.  As we will see in section \emergence\ the lack of
linear equations for such fields on the boundary is an important point
related to their holographic description.

\subsec{Conformal bootstrap for generalized free fields and multiparticle spectrum}

Let us now make the assumption that our CFT contains in its spectrum a
GFF i.e.  a scalar operator ${\cal O}$ of dimension $\Delta>{d-2\over
2}$ whose correlators factorize.  What conclusions can we draw from
this?

It is easy to see that the bootstrap conditions cannot be satisfied if
we assume that ${\cal O}$ is the only operator in the CFT. In
particular we will see that we must introduce composite operators made
out of products of ${\cal O}$. These are the equivalent of multi-trace
operators in gauge theories.  Thus the existence of a generalized free
field, combined with crossing symmetry automatically implies a Fock
space structure for the Hilbert space.

For example let us consider the 4-point function
\eqn\fourpointf{
\langle{\cal O}(x_1) {\cal O}(x_2){\cal O}(x_3){\cal O}(x_4)\rangle = {1 \over |x_{12}|^{2\Delta}\, |x_{34}|^{2\Delta}} + {\rm permutations}
} By considering a conformal partial wave expansion of the correlator
in the channel $(12) \rightarrow (34)$ we can infer that the conformal
field theory must contain a tower of conformal primary operators with
conformal dimension $2\Delta + 2n + l$ and spin $l$ \DolanUT. These
operators can be written as $\CO_{n,l}^{(2)} = \,\,: {\cal
O}\hat{\partial}_{[\mu_1}...\hat{\partial}_{\mu_l]} (\hat{\nabla}^2
)^n {\cal O}:$ where the brackets denote the symmetric traceless
combination and
$\hat{\partial}\equiv \overrightarrow{\partial}-\overleftarrow{\partial}$
in order to project out descendant contributions. Intuitively these
operators can be understood as ``two-particle'' states made out of the
field ${\cal O}$.

In other words the OPE of ${\cal O}$ with itself has the form
\eqn\opesingle{
{\cal O}(x) {\cal O}(0) = {1\over |x|^{2\Delta}}\, +\, \sum_{n,l} C_{n,l} \,
\left( |x|^{2n + l} \CO_{n,l}^{(2)} + \,\,{\rm descendants} \right)
} with OPE coefficients $C_{n,l}$ computed in \DolanUT\ and we have
suppressed Lorentz indices for simplicity.

Notice that the existence of the operators ${\cal O}^{(2)}_{n,l}$ is a
very nontrivial statement from the CFT point of view: in a general CFT
if we have an operator ${\cal O}$ with dimension $\Delta$ there is no
reason to expect the existence of an operator $:{\cal O} {\cal O}:$ of
dimension $2\Delta$. It is precisely the fact that our field ${\cal
O}$ is, in some sense, ``free'' that is responsible for the existence
of such operators.  By analogy with large $N$ gauge theories or via a
presumed gravitational dual, we will refer to these new operators as
multi-particle operators.  Likewise the original $\CO$ we will refer
to as a single-particle operator.

Correlation functions (or OPEs) of the 2-particle operators ${\cal
O}_{n,l}^{(2)}$ can be computed by taking limits of correlators of
${\cal O}$. For example, to compute the 3-point function of 2-particle
operators
\eqn\threepd{
\langle  {\cal O}_{n_1,l_1}^{(2)}(x_1) \,\,{\cal O}_{n_2,l_2}^{(2)}(x_2)\,\,{\cal
O}_{n_3,l_3}^{(2)}(x_3)\rangle } we start by considering the 6-point
function of the single-particle operator ${\cal O}$, which by the
assumption of factorization is
\eqn\sixpt{
\langle \CO(x_1)...\CO(x_6) \rangle = {1\over |x_{12}|^{2\Delta}
|x_{34}|^{2\Delta} |x_{56}|^{2\Delta} } + {\rm permutations} } By
acting with the appropriate combination of derivatives, considering
the $x_{12}, x_{34}, x_{56} \rightarrow 0$ limit and using the
OPE \opesingle\ for each of the pairs of points $(12), (34),(56)$ we
can isolate the desired 3-point function \threepd. Similarly we can
also determine the OPE between 2-particle operators. Working
iteratively this way (making further subtractions before taking the
limits) we can determine the OPEs of various two-particle states.
Thus the OPEs of the two-particle states are completely
determined by the OPE of the single-particle states.

Similarly by considering the conformal partial wave expansion of
correlators of 2-particle operators with ${\cal O}$ we can infer the
existence of operators with the quantum numbers corresponding to
``3-particle'' operators $:{\cal O}{\cal O}{\cal O}:$ etc. Following
this procedure we can inductively show that the conformal field theory
must contain a sector of operators which has the structure of a freely
generated Fock space, where the basic excitation has the quantum
numbers of the operator ${\cal O}$ and of its conformal
descendants. We want to emphasize that this follows from the single
assumption that correlators of the field ${\cal O}$ factorize.

More generally we consider a CFT which has more than one generalized
free fields, which we call ${\cal O}_i$. By definition these are
operators whose correlators factorize and they define the set of
single-particle states in the theory.  By considering the conformal
partial wave decomposition of various correlators of such operators we
can show that the theory will have to contain the ``multi-particle''
operators which can be constructed out of the main building blocks
${\cal O}_i$. For example if we have two generalized free fields
${\cal O}_1,{\cal O}_2$ of conformal dimension $\Delta_1,\Delta_2$,
then by considering the 4-point function
$$
\langle{\cal O}_1(x_1) {\cal O}_2(x_2){\cal O}_1(x_3){\cal O}_2(x_4)\rangle
= {1\over |x_{13}|^{2\Delta_1}|x_{24}|^{2\Delta_2}}
$$
and decomposing it in CPWs in the $(12)\rightarrow (34)$ channel we
can see that there must be a scalar operator with dimension
$\Delta_1+\Delta_2$ which we denote by $:{\cal O}_1{\cal O}_2:$ i.e. a
``two-particle'' operator, and so on.

\subsec{Partition functions of generalized free fields}
\subseclab\spartition

While the freely generated Fock space structure described above looks
reasonable at first sight, it suffers from some pathologies related to
the conditions of ``generalized modular invariance'' that we reviewed
in section \sgenmod. For simplicity let us assume that there is only
one generalized free field ${\cal O}$ in the theory. The ``single
particle states'' are operators of the form
$$
{\cal O},\,\,\partial_i {\cal O},\,\,\partial_i \partial_j {\cal O} ...
$$
of dimension $\Delta,\,\Delta+1,\,\Delta+2...$. The 2-particle states
are
$$
:{\cal O}{\cal O}:,\,\, :{\cal O}\partial_i {\cal O}:,\,\,...
$$
with dimensions $2\Delta,2\Delta+1,...$.

The single particle states form a representation of the conformal
group $SO(2,d)$. Let us call this representation $V_{\Delta}$. The
2-particle states also form a representation, which can be understood
as $Sym(V_{\Delta}
\otimes V_{\Delta})$, where $Sym$ stands for 
symmetrization reflecting the fact that the excitations behave like
identical bosons. More generally the $N$-particle states are the
representation $Sym(\otimes^N V_{\Delta})$.

Let us now consider the generalized free CFT on ${\bf S}^{d-1}\times
{\rm time}$ and compute its thermal partition function\foot{In this
section we ignore contribution from the Casimir energy on the sphere.}
$$
Z(q) = \sum_{operators} q^{\Delta},\qquad q= e^{-\beta}
$$
where $\beta={1\over T}$ is the inverse temperature and we take the
radius of the ${\bf S}^{d-1}$ to be equal to one (we will restore it
in the final formula). In principle we have to sum over one-, two-,
three- etc. particle states. However since the Hilbert space is a
freely generated Fock space, we can use a standard statistical
mechanics argument to determine the full partition function, starting
from the ``single-particle partition function''.  We first define the
single-particle partition function as
$$
Z_1(q) = \sum_{\rm single\, particle} q^\Delta
$$
Then the multi-particle partition function is given by the
formula\foot{The way to understand this formula the following: each of
the single particle states behaves as a simple harmonic oscillator
which can be excited arbitrary number of times so its partition
function is ${1\over 1-q^{\Delta}}$. Since all oscillators are
independent the full partition function is $Z(q) =
\prod_{\Delta} {1\over 1-q^{\Delta}}$. Taking the logarithm, using the
expansion $\log(1-q) =-\sum_{n=1}^\infty {q^n\over n}$ and
interchanging the order of summation over $\Delta$ and $n$ we arrive
at \multi.}
\eqn\multi{
\log Z(q) = \sum_{n=1}^\infty {1\over n} Z_1(q^n)}
Let us apply this formula to the generalized free CFT, where for
simplicity we will assume that there is only one generalized free
field ${\cal O}$.

In the case of a generalized free field ${\cal O}$ with $\Delta >
{d-2 \over 2}$ all single particle states can be constructed by acting
on ${\cal O}$ with an arbitrary number of derivatives $\partial_i,i =
1,...,d$, where $d$ is the spacetime dimension of the CFT. The
derivatives are all independent.  Adding a derivative increases the
conformal dimension by $1$. Since the derivatives are independent we
can think of them as $d$ independent harmonic oscillators. Then the
single particle partition function is simply
\eqn\spp{
Z_1(q) = q^{\Delta} {1\over (1-q)^d} } 
and the full partition function of the CFT is given by formula \multi.

Let us now try to study the high temperature limit which corresponds
to $\beta\rightarrow 0$ or $q\rightarrow 1$. We have
$$
\log Z (q) =  \sum_{n=1}^\infty {1\over n}  {q^{n \Delta}\over (1-q^n)^d}
$$
When $q\rightarrow 1$ the partition function goes like
$$
\log Z (q) \approx {1\over (1-q)^d} \sum_{n=1}^\infty  
{1\over n^{d+1}}
$$
or using $q = e^{-{1\over T}}$
$$
\log Z(q) \approx \zeta(d+1) \,\,T^d 
$$
where $\zeta$ is the Zeta function. Now we consider the free energy
$F$ defined by $\log Z= - F/T$. Reinstating the dependence on the
radius $R$ of the ${\bf S}^{d-1}$ sphere we find that at high
temperatures
\eqn\freeenergygff{
F \approx -\zeta(d+1) \,R^d\,T^{d+1} } This result seems problematic
in two (related) ways: first it is not extensive since it does not
scale with the volume $R^{d-1}$ and second the $T$ dependence
corresponds to that of a $d+1$ dimensional gas, while the CFT lives in
$d$ dimensions!  By contrast conformal invariance in $d$ spacetime
dimensions implies that at very high temperatures $F$ must be
proportional to $T^d$ on dimensional grounds\foot{Here we are assuming
that in the high temperature limit, the leading term of the energy
density on the sphere behaves like the thermal energy density on the
plane.}. This suggests that the GFF Fock space is inconsistent with
the basic assumptions of a local $d$-dimensional conformal field
theory.

Notice that this problem does not arise for a standard free field of
$\Delta={d-2\over 2}$. Such a field obeys a $d$-dimensional equation
of motion $\nabla^2 {\cal O}=0$ on the boundary, so the derivatives
are not all independent i.e. they obey the relation
$\partial_i\partial^i = 0$, so the equivalent of \spp\ is $Z_1(q) =
{q^{d-2\over 2}\over (1-q)^d}- {q^{d+2\over 2}\over (1-q)^d}$, where
we subtracted a term to compensate for operators which are zero due to
the equations of motion. Effectively there is one less dimension for a
field of $\Delta={d-2\over 2}$ than one with $\Delta>{d-2\over 2}$ and
the free energy of the former grows like $T^d$, which is consistent
with the expectations for a $d$-dimensional CFT.

Going back to generalized free fields with $\Delta>{d-2 \over 2}$,
clearly one of our assumptions must be wrong.  Namely, as we will
argue shortly, the spectrum cannot have the structure of a freely
generated Fock space for arbitrarily large operators.  In other words,
a generalized free field cannot exist in an exact sense in a
consistent CFT since it has the wrong thermodynamic properties.

We will see that there are two important ways that the spectrum of the
theory has to be modified at large conformal dimension: first, we must
truncate the growth of multi-particle states and, second, a new sector
of states will have to be added to the theory, which are the analogue
of black hole microstates.

Before we proceed let us notice that the same argument about the growth of the
free energy with $T$ can be used to show that a non-gravitational theory in AdS
(for example a free scalar field in AdS) cannot have a holographic dual CFT
\SusskindDQ. The free energy of a free hot gas in AdS is the same as
\freeenergygff. In order for this theory to have a holographic dual CFT the
growth of the spectrum with $T$ has to be truncated by some mechanism in order
to be consistent with that expected from a CFT i.e. $F\sim T^d$ and not
\freeenergygff. In gravitational theories this is achieved by gravitational
collapse and the formation of black holes at high energies. See also \CastroCE\
for an interesting related discussion in 2-dimensional CFTs.

\subsec{Decoupling of the stress tensor and the need for large $c$}
\subseclab\swhygisc

We now present another reason why a generalized free CFT cannot be a
consistent theory by itself. A special operator present in any local
conformal field theory is the stress energy tensor $T_{\mu\nu}$.  It
is an operator of dimension $\Delta = d$ transforming in the traceless
symmetric 2-tensor representation of $SO(d)$.  Dropping Lorentz
indices for simplicity, the 2-point function of the stress energy
tensor has the following general form
$$
\langle T(x) T(y) \rangle  = {c\over |x-y|^{2d}}
$$
where the constant $c$ plays the role of a ``central charge'' of the
CFT\foot{ Here we are not careful about factors of order one in the
normalization of $c$ since our arguments are qualitative. The precise
relation between the constant in the 2-point function of the stress
tensor and the conformal anomaly $c$ can be found in \OsbornCR.}.

Recall that the OPE of $T_{\mu\nu}$ with a primary operator ${\cal O}$
is fixed by the Ward identities \OsbornCR.  In other words the
following 3-point function is exactly determined by the conformal
dimension of ${\cal O}$ to be
$$
C_{T{\cal O O}} \sim \langle T_{\mu\nu}{\cal O}{\cal O} \rangle \sim \Delta \langle
{\cal O} {\cal O}\rangle
$$
up to convention-dependent factors of order 1. This means that when we
consider the 4-point function of ${\cal O}$ and expand it in conformal
blocks there should be a contribution from the block of the stress
energy tensor. The coefficient with which this conformal block
contributes is proportional to $\Delta^2/c$. More precisely the
overall contribution from the stress tensor exchange to the 4-point
function is
\eqn\tdecouple{
{\Delta^2 \over c} {\bf G}_T(x_1,x_2,x_3,x_4) } where the function
${\bf G}_T$ is {\it completely} fixed by conformal invariance
(see \DolanUT\ for the explicit expression of ${\bf G}_T$ in
$d=4$). So as long as $\Delta>0$ this contribution cannot be zero or
cancelled by anything else.  However a simple analysis of the
factorized correlator \fourpointf\ shows that no such conformal block
appears! How is this possible?  The only explanation is if we assume
that the contribution \tdecouple\ vanishes because
$c=\infty$.

For a given and fixed CFT it does not make a lot of sense to assume
that $c = \infty$.  What is more reasonable is to understand this as a
limiting sequence of theories for which $c\rightarrow\infty$. So a
generalized free CFT can only be understood as a sector in a sequence
of theories of ever increasing $c$.

In general $c$ is related to the number of degrees of freedom of the
theory\foot{In a weakly coupled CFT this is clear, since $T_{\mu\nu}$
couples to all fields and $c$ is proportional to the 2-point function
of $T$ with itself. We give a few more arguments about the relation
between $c$ and the degrees of freedom in section \subcardy.}. Since
$c\rightarrow \infty$ we expect a $c$-dependent growth in the number
of degrees of freedom which is not apparent in the spectrum of a
generalized free CFT, i.e. the free energy computed in
section \spartition\ was finite and $c$-independent. This implies that
a generalized free CFT can only be a small part of a theory with a
very large number of degrees of freedom, with the peculiar property
that at low conformal dimension only a finite number of operators
remain.

From now on and for the rest of the paper we will use the term
generalized free CFT in the following sense: we assume that there is a
sequence of CFTs of ever increasing central charge with the property
that in the $c\rightarrow \infty$ limit a sector of a finite number of
generalized free operators develops at low conformal dimensions. It
may be more convenient to introduce a ``coupling constant'' $g$ such
that when $c\rightarrow \infty$ we have $g\rightarrow 0$\foot{If the
theory has more parameters than $c$, then we may have to tune them in
an appropriate way in order to end up with a CFT satisfying
factorization. For example in a gauge theory we may have to take the
't Hooft limit i.e. $N\rightarrow \infty\,,\, g_{YM}^2 N ={\rm
const}$.}. The coupling \tdecouple\ to the stress tensor suggests that
we define $g$ as
$$
g = {1 \over \sqrt{c}}
$$

Before we proceed let us see why the statement that factorization
for GFFs implies $c\rightarrow \infty$ is consistent with the fact that
correlators of {\it ordinary free fields} factorize even though the central
charge is finite. For a genuinely free field of $\Delta={d-2\over 2}$
there is a conformal block corresponding to the exchange of an
operator with the quantum numbers of a ``stress tensor''
i.e. $\Delta=d$ and symmetric traceless, which is the two-particle
operator $(:\partial_\mu {\cal O}\partial_\nu {\cal
O}: \,+\,\,\cdots$) where the dots denote the quadratic terms which
appear in the stress tensor of a conformally coupled scalar. Of course
this is not the stress tensor of the full CFT but only of the part of
the CFT consisting of the free field ${\cal O}$, which is decoupled
from the rest of the CFT since the field ${\cal O}$ is literally
free. Notice that for a GFF with $\Delta>{d-2\over 2}$ the same
two-particle operator has dimension greater than $d$ and thus it is
not conserved.

\subsec{Gravitational collapse and ``deconfinement''}

By now we have argued that while a generalized free CFT seems to
satisfy the conformal bootstrap in flat space, it suffers from two
pathologies: first if we assume that the Fock-space structure of the
generalized free CFT persists all the way to arbitrarily high
conformal dimension then the growth of states in the theory is not
consistent with the one expected for a $d$-dimensional local CFT.
Second, we found that for correlators to factorize we have to be able
to decouple the stress energy tensor for which we have to assume that
$c\rightarrow\infty$, in an appropriate limiting sense. The central
charge $c$ is related to the degrees of freedom of the theory, so if
$c\rightarrow \infty$ one would expect a $c$-dependent proliferation
of states which seems not to take place for a generalized free CFT.

The most natural way to avoid these problems (suggested by what
happens in large $N$ gauge theories) is to assume that a generalized
free CFT is a ``light'' (in the sense of low conformal dimension)
decoupled sector of a much bigger conformal field. If we probe the
generalized free CFT at sufficiently high conformal dimension we will
discover two things: that the multiparticle Fock space is truncated
and that a new big sector of states (with $c$-dependent entropy) will
kick in. The truncation of the multiparticle spectrum can be
understood as the analogue of gravitational collapse while the new
states are the ``black hole'' microstates.

That the free Fock space structure of a generalized free CFT has to be
truncated at large conformal dimension can be understood as a
dynamical effect in the following way: we saw that by taking
$c\rightarrow \infty$ we can decouple the stress energy tensor. What
happens though, if we look at correlators of operators with conformal
dimension of the order $\Delta \sim c$ at the same time that we take
the $c\rightarrow \infty$ limit?  Then it is clear that the ${1\over
c}$ suppression in equation \tdecouple\ is compensated by the $\Delta$
in the numerator and the conformal partial wave of the stress energy
tensor cannot be ignored even if $c\rightarrow \infty$. This means
that correlators of operators with conformal dimension of order $c$
cannot be factorizable.

More generally let us consider a many-particle state $:{\cal O} {\cal O}....
\partial{\cal O}....\partial\partial{\cal O}:$ of
total (naive) conformal dimension $\Delta_0$, as computed by the
Fock-space counting of conformal dimensions for multi-particle states
in a generalized free CFT. If we consider the contribution from the
exchange of the stress energy tensor between various constituents of
the multi-particle operator, we find that it will inevitably introduce
a correction $\delta \Delta$ to the naive conformal dimension
$\Delta_0$.  When the correction is of the same order as the naive
dimension i.e. when $\delta \Delta /\Delta_0 \sim 1$ then the freely
generated Fock-space structure will definitely not be reliable. This
happens when $\Delta_0 \sim c$.  Notice that this can happen even if
the operator ${\cal O}$ has low conformal dimension, provided that we
take sufficiently many insertions of ${\cal O}$ or $\partial$.  Thus
if we take sufficiently big multi-particle operators they will deviate
from the free-Fock space no matter how large $c$ is.

In the bulk the equivalent statement is that if we take a gas of
particles of total mass $M$ spread in a volume of the AdS radius, then
the gravitational correction to their energy is of the order $G_N
M^2/R_{AdS}^{d-2}$, where $G_N$ is the bulk Newton's constant. In the
limit $G_N R_{AdS}^{1-d}\rightarrow 0$ (which is equivalent to
$c\rightarrow \infty$) the backreaction is negligible as long as $M$
is kept fixed. If however we scale $M\sim G_N^{-1}R_{AdS}^{d-2}$ then
we find that the backreaction is of the same order as the bare mass
$M$ even in the $G_N\rightarrow 0$ limit. Such states cannot be made
arbitrarily big because at some point they will hit their
``Chandrasekhar'' bound and will undergo gravitational collapse
towards a black hole. See \deBoerWK, \ArsiwallaBT\ for related
discussions.

To summarize we do not expect the spectrum of such large operators to
arrange itself into a simple free Fock space as we naively assumed
above. The effects of gravity (exchange of $T_{\mu\nu}$ in the
boundary CFT\foot{And all kinds of other operators of course, we only
mention $T_{\mu\nu}$ because its coupling is universal.}) cannot be
ignored. Thus, we expect a modification of the simplistic
multiparticle Fock space at conformal dimensions of order $c$ and a
truncation due to ``gravitational collapse''.

What is more, in this regime (i.e. $\Delta > c$) we expect new sorts of states
to emerge corresponding to black hole states.  These states have a very
different structure from the original multiparticle operators as their conformal
dimensions and degeneracy is $c$-dependent (explaining why such states 
decouple in the limit $c \rightarrow \infty$). In particular, we expect the
entropy of these new states to diverge as $c\rightarrow \infty$ (unlike the
low-lying generalized free fields whose entropy is $c$ independent). This can
be motivated as follows: from general arguments we expect that at very high
temperature $T$ the expectation value of the stress tensor in a $d$-dimensional
CFT goes like
$$
\langle T_{00} \rangle \sim  \widetilde{c} \,T^d
$$
for some constant $\widetilde{c}$, from which we can derive that the entropy
density is $S \sim \widetilde{c}\, T^{d-1}$.  In section \subcardy\ we give some
indications that $c \rightarrow \infty$ implies $\tilde c \rightarrow \infty$.

A natural question\foot{We would like to thank J. de Boer for useful comments
along these lines.} is whether these operators can be thought, in some sense, as
multiparticle operators of GFF which, due to interactions, have received large
corrections to their dimensions. For this, the corrections would have to be
large enough so that even the qualitative scaling of their degeneracy with $c$
would be modified. In known examples, this would be analogous to assuming that
there is some way to think of the quark-gluon plasma of a large $N$ gauge theory
as a gas of glueballs with large corrections due to interactions, or a black
hole as a strongly-interacting gas of gravitons. Both of these possibilities
seem counter-intuitive. So perhaps it is more reasonable to assume that the
spectrum at high conformal dimension corresponds to a genuinely new phase of the
theory, in which new degrees of freedom emerge, and hence it is unreasonable to
describe this high temperature phase in terms of excitations of the low
temperature phase.

\subsec{Minimal generalized free CFT and gas of gravitons}

It is interesting to consider the smallest possible generalized free
CFT, which would be the one whose only single particle operator is the
stress energy tensor $T_{\mu\nu}$.  As we argued, if we assume that in
the $c\rightarrow \infty$ limit its correlators factorize then we
necessarily have to add to the spectrum the multi-particle states made
out of the stress energy tensor of the form $:T T:,\, :T\partial
T:,\,:TTT:,...$ to satisfy the conformal bootstrap. Such a CFT has the
same spectrum as a linearized free graviton gas around AdS. As we
explained before, such a CFT is inconsistent by itself and should be
thought of as the low-conformal dimension sector of a much larger
(sequence of) CFT with large $c$. In the bulk all these additional
states would represent the black hole microstates. Such a theory was
recently considered by E. Witten \WittenKT, for the case of
three-dimensional AdS bulk. In two-dimensional CFTs the module of the
stress energy tensor closes on itself and correlation functions of the
stress tensor are completely fixed by holomorphy, with the only input
being the value of the central charge $c$ i.e. they do not contain any
non-trivial dynamics. This is not the case for higher dimensional CFTs
where the conformal group is not large enough to constrain the
correlators of the stress energy tensor, which thus contain
non-trivial dynamical information.

A natural question is whether such a minimal generalized free CFT is
dual to (semi)-classical gravity in the large $c$ limit. From the CFT
point of view the question is whether the assumption that at
$c=\infty$ the spectrum consists solely of the stress-tensor and
multi-particle states is powerful enough to constrain the solutions of
the bootstrap equations at order $1/c$, to be those predicted by
tree-level gravity\foot{Obviously one can write down bulk theories
whose only light field is the graviton and which have higher
derivative corrections. The question is to what extent such bulk
theories can be related to a {\it consistent} boundary CFT.}. While
this is a very interesting question we have not been able to say
anything new about it, essentially because the CPW expansion of
stress-tensor four-point functions is still not known.

Notice that according to our previous discussion one can also see that
linearized gravity around AdS cannot be holographic unless the
spectrum of (approximately) free gravitons is truncated at some point
by gravitational interactions and the formation of black holes.

The stress tensor in dimension $d$ has $n= {d(d+1) \over 2} - 1$
independent components and single particle states associated to it are
$$
T_{ij},\, \partial_i T_{jk},\, \partial_i \partial_j T_{kl},\,...
$$
The degeneracy at each level is $n$ times the degeneracy coming from the
derivatives giving a (unconstrained) single-particle partition function
$$
\tilde{Z}_1(q) = {n \, q^d \over (1-q)^d}
$$
Where we used that the conformal dimension of $T_{\mu\nu}$ is $d$.  In
any theory the stress-tensor satisfies the conservation equation
$\partial^i T_{ij} = 0$ which implies $d$ relations between
derivatives.  It is straightforward to find a generating function for
these vanishing contributions (i.e. derivatives acting on $\partial^i
T_{ij}$)
$$
Z^s_1(q) = {d\, q^{d+1} \over (1-q)^d}
$$
The corrected single particle partition function is the difference of
these two
$$
Z_1(q) = \tilde{Z}_1(q) - Z^s_1(q) = {q^d \,(n  - d \, q)\over (1-q)^d} 
$$
This is quite similar to the free scalar case above because $n > d$
for $d > 2$ so in the limit $\beta \rightarrow 0$ we get 
$$
Z_1(q) = {n - d \over \beta^d}
$$
which will give a multi-particle partition function $\log Z(q) \approx T^d$ as in the
generalized scalar case.

\subsec{Summary}

Let us review what we have discussed so far.  We defined a ``large $c$
generalized free CFT'' by assuming it has a sector of
``single-particle operators'' $\{\CO_i\}$ defined by the fact that
their correlators factorize in an expansion in $1/c$.  Examining the
bootstrap equations for such a theory we find that such fields cannot
exist alone and in fact imply the existence of an entire tower of
``multiparticle states'' constructed from powers of $\CO_i$ and
derivatives.

At infinite $c$ the Hilbert space of such a ``generalized free CFT''
has the structure of a freely generated Fock space made out of the
basic building blocks ${\cal O}_i$.  Taking into account the
finiteness of $c$ we concluded that such theories can only exist as
small sectors of big conformal field theories.

The existence of the freely generated Fock space at infinite $c$
implies we are on the right track, as it is a necessary condition for
the existence of a weakly coupled bulk dual.  From the dual gravity
point of view, the factorized nature of the large $c$ CFT is due to
the fact that the correlators do satisfy a linear differential
equation in $d+1$ dimensions and, as such, constitute a genuinely free
theory implying a Fock space structure of the Hilbert space.

\newsec{Emergence of the holographic dimension}
\seclab\emergence

We will now review arguments which indicate that when a large $c$ conformal
field theory contains a sector of generalized free fields in its spectrum, then
that sector and its interactions\foot{When we move away from the strict
$c\rightarrow \infty$ limit.} can most naturally be described by an effective
higher dimensional theory which inevitably contains gravity\foot{By this we do
not necessarily mean ``classical gravity'' but also ``stringy gravity'' and
other possible exotic versions.}. In this section we basically review statements
made in \BanksDD, \BalasubramanianRI, \BenaJV, \wittentalk,
\HamiltonJU, \HamiltonAZ, \HamiltonFH. We start with what happens in the free
theory (i.e. at $c=\infty$).

A first hint towards the fact that a generalized free field should be
represented in terms of an ordinary free field living in higher
dimensions is the fact that the free energy of the former scales in
the right way, according to the analysis of section \spartition. The
free energy of a generalized free field is the same as that of a
thermal gas of free particles living in AdS$_{d+1}$.

A second important observation is that while the generalized free
fields are in a sense ``free'', i.e. they generate a Fock space of
excitations, they do not obey linear equations of motion on the
boundary, which seems to be counter-intuitive. A freely-generated Fock
space suggests that it is possible to superimpose excitations, which
is a property of linear systems.  The underlying linearity of a
generalized free field can be made more manifest as follows: let us
call ${\bf b}$ the $d$ Cartesian coordinates of the space on which the
CFT lives and introduce a new set of $d+1$ auxiliary coordinates
$({\bf x},z)$. We define the following family of operators
\eqn\bdhm{
\phi({\bf x},z) = \int d^d {\bf b}\, G({\bf x},z;{\bf b})\, {\cal O}({\bf b})}
where $G({\bf x},z;{\bf b})$ is a kernel, also called the ``transfer function'', which is defined by certain superposition of normalizable modes in AdS (see the references above for details). This kernel satisfies the equation
\eqn\bbp{
(\nabla^2_{d+1}-m^2)  G({\bf x},z;{\bf b}) = 0}
for a Laplacian defined on the manifold parametrized by the coordinates $({\bf
x},z)$ and equipped with the AdS$_{d+1}$ metric
\eqn\adsm{
ds^2 = {dz^2 + d{\bf x}^2\over z^2} } 
Here $d{\bf x}^2 = -dt^2 + dx_i^2$ is in Lorentzian signature. The mass $m$ and
conformal dimension $\Delta$ of the operator are related by the familiar formula
\eqn\deltam{
\Delta = {d\over 2} + \sqrt{{d^2 \over 4} + m^2}
} The operators $\phi({\bf x},z)$ are hermitian operators acting on
the Hilbert space of the conformal field theory i.e. they are still
operators ``in the CFT'', even though they are labeled by one
additional parameter $z$.  However they are not local operators. They
are constructed by smearing local operators with the kernel $G({\bf
x},z;{\bf b})$.

The important property of the operators $\phi({\bf x},z)$ is that they
satisfy a linear wave equation in the auxiliary $d+1$ dimensional
manifold
\eqn\bulkkg{
(\nabla^2_{d+1} -m^2 ) \phi = 0} Hence by lifting these operators in
one additional dimension we can make the underlying linearity of the
system manifest.

Correlation functions of the operators $\phi$ can be computed using
the correlators of ${\cal O}$. To the extent that the factorization of
correlators of ${\cal O}$ on the boundary is obeyed, we can see that
correlators of the field $\phi$ in the ``bulk'' will be the same as
those of a free massive field in AdS satisfying \bulkkg. For example
the 2-point function of the field in the bulk can be computed by
$$
\langle \phi({\bf x},z) \phi({\bf x}',z')\rangle = \int\int d^d {\bf b} \,d^d
{\bf b}'\, G({\bf x},z;{\bf b}) G({\bf x}',z';{\bf b}')\,
\langle {\cal O}({\bf b}){\cal O}({\bf b}')\rangle
$$
It is not hard to show using \basictwo\ that the 2-point function in
the bulk computed this way coincides with the (bulk-to-bulk) Green's
function of \bulkkg. In particular if this computation is performed in
Lorentzian signature one finds that\foot{For this equation to hold as
an operator equation (i.e. to hold inside correlators with additional
insertions) it is necessary to use the factorization of the
correlators of ${\cal O}$.}
\eqn\combulk{
[\phi(x,z),\phi(x',z')]=0 } if the two points are separated by a
spacelike distance, as measured by (the Lorentzian version of) the
metric
\adsm. Hence the field \bdhm\ has the behavior of a local free field
in AdS space.

What we have achieved so far is the following: we started with the
$d$-dimensional CFT and we have introduced an emergent AdS
space, which is used as a parameter space to label operators of the
form \bdhm. What we have gained by this is that we have constructed a
set of operators which obey linear equations of motion. In this sense
the emergent bulk geometry makes the linearity of the theory more
manifest.

The Fock space structure of the Hilbert space of the boundary
conformal field theory can now be understood more directly as the
standard Fock space of a free field in the bulk.  The importance of
this emergent AdS space will become more clear when we consider
turning on small interactions.

Applying this procedure to the stress energy tensor, which is an
operator present in any CFT, leads to a massless spin 2 field in the
AdS space which implies that the effective holographic theory will be
gravitational.

\subsec{Why this is not just group theory.}

One might think that the association of an AdS space with a conformal
field theory is somewhat trivial based on the identification of the
isometry group of AdS with the conformal group in one less
dimension. However this is not entirely correct. That the states in the
Hilbert space of a CFT fall into representations of $SO(2,d)$ is not
sufficient to guarantee that they should have an interpretation as
states living in a space with isometry group $SO(2,d)$.

When we think of ``emergent space'' what we usually imagine is that we
have an emergent manifold $M$ together with some quantum fields living
on it (at least at long wavelengths). This means that the states
should not only fall into representations of the isometry group of
$M$, but also that the types and degeneracies of representations that
appear in the Hilbert space must be consistent with those of
(approximately) local quantum fields living on $M$.

According to our previous discussions the introduction of the dual
spacetime is motivated by the fact that the low-lying Hilbert space of
the CFT can be represented in terms of free (or weakly interacting)
fields in the bulk, so it is important that the representations of
$SO(2,d)$ which appear in the spectrum of the CFT correspond to those
of a free field in AdS. In particular they should have a Fock space
structure and the correct extensive entropy. This requirement is
satisfied for generalized free CFTs but would not be true for CFTs at
finite central charge. For example it would not be true for the ${\cal
N}=4$ SYM with gauge group, say, $SU(2)$.

To summarize, the association of an AdS space\foot{As mentioned in
section 2, we restrict our attention to cases where there is at least some
semi-classical notion of AdS space, which might be ``stringy'' but not
``quantum'', since we have no (independent) formulation of what
quantum gravity means on an ``AdS space'' whose radius is comparable
to the Planck scale.} to a CFT is not only based on group theory but
also on the factorization of the low-lying operators, which is a
dynamical assumption going beyond symmetry considerations.

\subsec{Operators with spin}

So far we have focused our discussion on scalar operators, but the
same logic can be applied to operators of the CFT with
spin. Generalized free operators of higher spin can be associated to
higher spin bulk fields. Everything we discussed in the previous
subsection is readily generalizable.

Special care has to be taken if a generalized free operator of nonzero
spin saturates the unitarity bound of the conformal field theory. In
that case the boundary operator satisfies {\it first order}
equations\foot{Because the norm of the first conformal descendant
vanishes at the unitarity bound.} (for example for a conserved current
this would be the conservation equation). Notice that operators with
nonzero spin at the unitarity bound behave qualitatively differently
from scalar operators at the unitarity bound, which satisfy {\it
second order} equations \confeom.

The difference is that the first-order equations at the unitarity
bound of operators with spin do not imply that the operator is
``free'' in any sense. For example, consider a conserved current
$J^\mu$ in a 4-dimensional CFT. It is an operator of $\Delta=3$ and
$l=1$ which saturates the unitarity bound and obeys the first order
equation $\partial_\mu J^\mu=0$. Unlike the second order equation
$\nabla^2 {\cal O}=0$ at the unitarity bound for a scalar field of
$\Delta=1$, the first order conservation equation for a current does
not allow us to compute correlation functions of $J_\mu$ with
itself\foot{Of course the conservation equation constrains the form of
the correlation functions of $J_\mu$ but it is not powerful enough to
fully determine them, unlike the equation $\nabla^2 {\cal O}=0$ for a
scalar at the unitarity bound $\Delta=1$.}. In particular the
correlators of $J_\mu$ with itself do not factorize to products of
2-point functions. In other words the current $J_\mu$ is not a
generalized free field, even though it satisfies the first order
equation $\partial_\mu J^\mu=0$. It is a genuinely interacting field.

If we now impose the additional assumption that a conserved current is
``generalized free'' (in the sense that its correlators factorize)
then we have the following situation: as before, we try to construct
``bulk'' fields by applying a procedure similar to \bdhm. The goal is
to construct bulk fields which obey standard second order equations of
motion. In the process of doing this the first order conservation
equations on the boundary play important role. The resulting bulk
fields have gauge invariance. In the case of a generalized free
current on the boundary we get a free $U(1)$ gauge field in the bulk,
and for the case of the stress tensor we get a bulk spin two field
with the gauge invariance of a linearized graviton. This is the well
known relation between global symmetries in the CFT (which emerge when
currents hit the unitarity bound) and gauge symmetries in the bulk.

\subsec{Comments on bulk observables and background independence}

Before we close this section we would like to make some comments about
the construction of the bulk observables. While we can always define
the operator \bdhm\ as a (non-local) operator in the CFT, the fact
that it obeys \bulkkg\ and \combulk\ is not an exact statement but
only true at infinite $c$ keeping certain other ``parameters'' of the
problem fixed. To be more precise, if we want \bulkkg\ and \combulk\
to be satisfied as operator equations, then they must hold even 
when they are inserted in correlation functions evaluated on states of
the CFT. If these states have a large enough mass to backreact on the
geometry (i.e. if they have energy of order $c$) then we do not
expect \bulkkg\ and \combulk\ to be true.

To see an example where this issue arises, let us consider a CFT with
classical gravity dual, which is placed at finite temperature. Then
we expect that correlators of gauge invariant operators can be
approximated by evaluating gravity correlators on the background of an
AdS-Schwarzschild black hole. This means that if we want to ``uplift''
the generalized free fields $\CO$ off the boundary and into regular
free fields in the bulk, for the purpose of computing finite
temperature correlators, we should be using a different kernel
in \bdhm, namely the kernel corresponding to propagation on a black
hole background.

From the boundary point of view the deviation from \bulkkg\
and \combulk\ in these situations can be intuitively understood as
follows: these relations were derived from the assumption of
factorization of correlators of the boundary fields. As we explained,
factorization is not an exact statement but only true in a $1/c$
expansion. If we evaluate correlators on states with $c$-enhancing
factors (for example if their conformal dimension is of order $c$)
then the $1/c$ suppression of interactions will no longer hold.

Moreover \combulk\ may receive corrections if the points $x,x'$ are
brought very close to each other, that is, close enough so that the
$1/c$ suppression of interactions can be compensated. This
means that the observables \bdhm\ are not infinitely localizable in
the ``bulk'' unless we take $c\rightarrow\infty$ first.

The conclusion from this is the following: the effective bulk theory
which we introduced in order to make manifest the linearity of the
generalized free fields is not static and fixed but rather dynamical,
in the sense that it depends on the saddle point/state of the CFT on
which one is evaluating correlation functions. This is consistent with
our expectation about a gravitational theory not having well defined
(background independent) local observables.

It seems quite nontrivial that (at least in theories with classical
gravitational duals) there are many classes of states/ensembles of the
CFT on which correlators of operators can be easily computed by
evaluating correlators of free fields propagating in some dual
classical geometry which has to obey certain equations of
motion\foot{i.e. the Einstein equations and generalizations.}.

\newsec{Including  interactions}
\seclab\interactions

In section \emergence\ we constructed operators in the generalized
free CFT which behave like free fields in an emergent AdS
space. The factorized correlators of this generalized free CFT have to
be understood as the limiting form of correlators in a sequence of
CFTs where $c\rightarrow \infty$. If we move a little bit away from
the strict $c\rightarrow\infty$ limit we expect that the boundary CFT
is no longer ``generalized free'', i.e. correlators of single particle
operators no longer factorize. This indicates that there are certain
effective\foot{Not to be confused with the interactions of the
underlying fundamental fields of the theory which may be strong even
in the strict $c\rightarrow \infty$ limit.}  interactions between
them. We want to understand to what extent we can represent these
interactions as local interactions in the emergent bulk space.

More specifically, our goal is to argue that the $1/c$ expansion of a
generalized free CFT can be naturally organized in terms of structures
which coincide with ``Witten diagrams'' in anti de Sitter space.

\subsec{Scaling of Correlators with $c$}

We now wish to consider the leading finite $c$ corrections.  Thus we
expand the CFT data $({\Delta_i, C^k_{ij}})$ in the small ``coupling
constant'' $g \equiv {1\over \sqrt{c}}$ around the point $g=0$. The
choice of ${1\over \sqrt{c}}$ as the expansion parameter is suggested
by the coupling of the stress-tensor to other operators. If we rescale
the stress tensor so that its 2-point function is order 1, i.e if we
define $\widetilde{T} = {1\over \sqrt{c}} T$, then the Ward identities
fix the coupling to any other conformal primary to be of the form
$\langle \widetilde{T} {\cal O} {\cal O}\rangle \sim
{1\over \sqrt{c}} \Delta$, where $\Delta$ is the conformal dimension
of ${\cal O}$. This suggests that the natural expansion parameter is
${1\over \sqrt{c}}$, though one can imagine more complicated
situations where various sectors of the CFT have different effective
coupling. Since we want to focus on qualitative aspects we will ignore
such complications and, in fact, much of what we say below is insensitive to 
the origin of the $g > 0$ perturbation.

The corrections to the CFT data will thus take the form
\eqn\perturba{
\Delta_i \,=\, \Delta_i^{(0)} +\, g\Delta_i^{(1)}\,\, +\,\, g^2 \Delta_i^{(2)}+...
}
\eqn\perturbb{C_{ij}^k\, = \,C_{ij}^{k (0)} + \,g\, C_{ij}^{k(1)}\,\,+ \,\,g^2 C_{ij}^{k(2)}+...}
These corrections to the conformal dimensions and to the OPE
coefficients must be such that the bootstrap conditions are satisfied
order by order in $g$. 

To proceed let us first define the {\it connected} correlators
$\langle {\cal O}_{i_1}(x_1)...{\cal O}_{i_n}(x_n)\rangle^c$ of single
trace operators by the following formal relation
\eqn\defconcor{\eqalign{
&\sum_{n=1}^\infty\int dx_1..dx_n
 J_{i_1}(x_1)..J_{i_n}(x_n)\,\,\, \langle {\cal O}_{i_1}(x_1)...{\cal
 O}_{i_n}(x_n)\rangle\cr &\qquad=\exp\left(\sum_{n=1}^\infty\int
 dx_1..dx_n J_{i_1}(x_1)..J_{i_n}(x_n)\,\,\, \langle {\cal
 O}_{i_1}(x_1)...{\cal O}_{i_n}(x_n)\rangle^c \right)}} 
Since 1-point
functions vanish due to conformal invariance, we have for the 2- and
3-point functions
$$
\langle {\cal O}_1(x_1) {\cal O}_2(x_2)\rangle^c\equiv \langle {\cal O}_1(x_1) {\cal O}_2(x_2)\rangle$$
$$
 \langle {\cal O}_1(x_1) {\cal O}_2(x_2)
{\cal O}_3(x_3)\rangle^c\equiv \langle {\cal O}_1(x_1) {\cal O}_2(x_2)
{\cal O}_3(x_3)\rangle$$
The connected 4-point function is
\eqn\examplconb{\eqalign{
& \langle {\cal O}_1(x_1) {\cal O}_2(x_2) {\cal O}_3(x_3){\cal
O}_4(x_4)\rangle^c\equiv  \langle {\cal O}_1(x_1) {\cal O}_2(x_2) {\cal
O}_3(x_3){\cal O}_4(x_4)\rangle \cr &- \langle {\cal O}_1(x_1) {\cal
O}_2(x_2)
\rangle\,\,\langle{\cal O}_3(x_3){\cal O}_4(x_4)\rangle 
 -  {\rm permutations}}}
and so on. 

In order to proceed we will make the simplifying assumption that the
connected correlators of single-particle operators scale as
\eqn\cscaling{
\langle {\cal O}_{i_1}(x_1) \,...\, {\cal O}_{i_n}(x_n)\rangle^c \sim g^{n-2}
}plus subleading corrections, which scale with higher powers of $g\sim
{1\over \sqrt{c}}$.  From the CFT point of view there is no a priori
reason to assume this specific scaling, but we choose it as it is the
one corresponding to theories that we are familiar with and leads to
simple results\foot{One can construct examples with more complicated
scaling, for example by taking products of CFTs with different ranks
of the gauge groups and turning on interactions between them. Our goal
is not to consider the most general case, but rather to find examples
which capture the essential physics without adding unnecessary
complications.}.  It would be interesting to study which of our
following statements will be modified if a different scaling is
assumed. The scaling \cscaling\ holds in the usual large $N$ expansion
of gauge theories \tHooftJZ\ (where $c\sim N^2$ and thus $g\sim
{1\over N}$) and in large $N$ symmetric orbifold CFTs in two
dimensions \LuninYV, \LuninPW, \PakmanZZ. From a string theory point
of view this scaling means that each time we add a vertex operator to
a genus zero diagram we get a power of the string coupling constant
which in this case would be identified with $g$.

In the rest of this paper we will only discuss the leading terms of
the connected correlators \cscaling\ which should be thought of as
{\it tree level} correlators of the dual AdS theory. We will make the
crucial assumption that at this order the conformal bootstrap can be
solved purely within the low-lying generalized free field sector. In
other words we will assume that to leading order in the $g$ expansion,
there are no operators dual to black hole microstates running in the
decomposition of the correlators of light fields. It would be
interesting to analyze in more detail what constraints this assumption
implies about the degeneracies and couplings of heavy operators and
also how this condition is modified at higher orders in the
$g$-expansion.

The scaling assumption \cscaling\ basically fixes the $g$ scaling of
various OPE coefficients. For example from \cscaling\ we find that the
OPE coefficient between three single trace operators starts at order
$g={1\over \sqrt{c}}$. Similarly the OPE coefficient between two
single particle states ${\cal O}_1$, ${\cal O}_2$ and the composite
two-particle operator $:{\cal O}_1 {\cal O}_2:$ starts at order 1
(from the disconnected correlator) and the next correction can start
at order $g^2={1\over c}$, as indicated by the scaling of the
connected correlator $\langle {\cal O}_1 {\cal O}_1 {\cal O}_2 {\cal
O}_2\rangle$.  The leading terms of the OPE of two single particle
operators $\CO_1(x)$ and $\CO_2(0)$ will have the following schematic form
\eqn\opeexp{
\CO_1 \CO_2 \sim {1 \over x^{\Delta_1 + \Delta_2}} \left[ g \,
\CO_k \, x^{\Delta_k} +
(\delta_{i\{1}\delta_{2\}j}(1+g^2\log x) + g^2 )  \, \CO_{n,l}^{(ij)} \, 
x^{\Delta_i + \Delta_j + 2n + l} \right]
}
So a consequence of \cscaling\ is the vanishing of the order $g$ terms
in the multiparticle contribution to the single-particle OPE.  Note we
only indicate the $g$-dependence above; in principle each term has a
free coefficient.  The logarithmic terms arise from corrections to the
conformal dimensions of the two-particle operators.

\subsec{Conformal bootstrap in perturbation theory and holographic interactions}

We now wish to study solutions of the bootstrap equations
perturbatively in $g$ around the generalized free field solution. Our
goal is to argue that such a perturbation is most naturally described
in terms of Witten diagrams in the AdS space introduced in the
previous section.

First of all we consider the converse problem: i.e. we start with the
free bulk theory constructed in section 5 and consider a perturbation
by adding local interactions in the bulk. If we assume that
correlators on the boundary are still related to correlators in the
bulk by the usual AdS/CFT prescription then the perturbed CFT
correlators will be related to ``Witten diagrams'' in the
bulk \BanksDD. For example turning on an interaction of the form $\int
d^{d+1}x\, \sqrt{g} \,\phi_1 \phi_2\phi_3$ will correspond to a
nonzero boundary 3-point function $\langle{\cal O}_1 {\cal O}_2 {\cal
O}_3 \rangle$, where the three operators ${\cal O}_1,{\cal O}_2,{\cal
O}_3$ on the boundary dual to fields $\phi_1,\phi_2,\phi_3$ in the
bulk.

From the symmetries of the problem it is clear that if the interacting
correlators are computed from Witten diagrams they will automatically
satisfy some of the consistency requirements of the CFT, such as
having an expansion consistent with an OPE and crossing symmetry. This
does not imply full consistency from the CFT point of view, as one
still has to check the unitarity of the theory. In other words a
Witten diagram expansion does not guarantee that the OPE coefficients
that will be computed from Witten diagrams will necessarily be real
(or their square positive)\foot{For example consider a 4-point
function of single trace operators $\langle {\cal O}_1(x_1){\cal
O}_2(x_2){\cal O}_1(x_3){\cal O}_2(x_4)\rangle$ which we compute by
using effective Witten diagrams in the bulk. Let us say that in the
channel $(12)\rightarrow(34)$ a single trace operator ${\cal O}_3$
appears. The corresponding CPW comes multiplied by the coefficient
$p=(C_{12}^3)^2$, which is of order $1/c$.  Unitarity implies that $p$
must be positive, which (presumably) imposes certain constraints on
the effective Witten diagrams used to compute the 4-point
function. In \HeemskerkPN\ constraints from unitarity were not
relevant because they only considered the exchange of two-particle
operators in the intermediate channel, which means that the relevant
constraints would only appear at higher order in $1/\sqrt{c}$.}. Of
course it would be extremely interesting to use this approach to find
constraints on possible effective actions in AdS spaces, but we will
not do it in this paper (but see e.g. \FitzpatrickZM).

The logic of our approach is the opposite and we want to understand
the emergence of the interactions in the bulk starting from the
CFT. In other words we want to explain why a physicist studying the
perturbations of a ``generalized free CFT'' would naturally introduce
the concept of a ``Witten diagram''. From the CFT point of view
interactions between operators are naturally described in a conformal
partial wave expansion. To understand why in generalized free CFTs
this expansion can be reorganized in terms of Witten diagrams we have
to clarify the precise relation between conformal partial waves and
Witten diagrams.

\fig{Witten diagram and conformal partial wave.}
{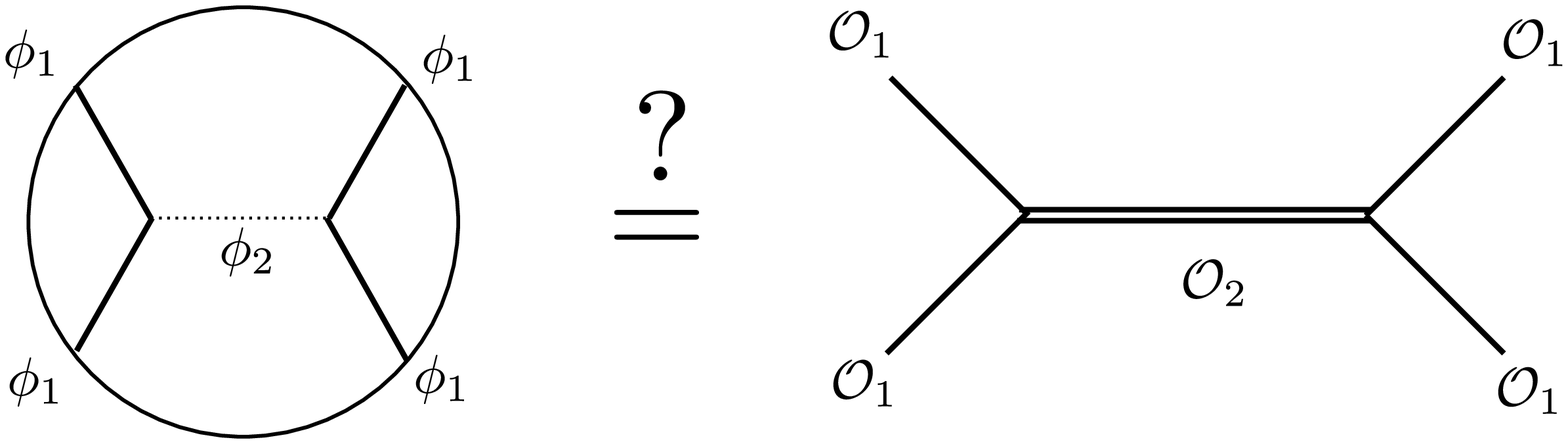}{4.truein}
\figlabel{\wittenvscpw}

Let us consider a basic exchange Witten diagram with 4 external
$\phi_1$ lines and the exchange of a particle of type $\phi_2$.
Intuitively one would expect that it is dual to the conformal block
corresponding to the exchange of the operator ${\cal O}_2$ (and all of
its descendants) between four external ${\cal O}_1$ operators, as
depicted in figure 3.  However a careful comparison of the two shows
that they are not exactly the same \LiuTH. The exchange Witten diagram
can be explicitly evaluated and expanded in a basis of conformal
blocks. One then finds that it is equal to the conformal block of the
exchange of ${\cal O}_2$ plus additional conformal blocks
corresponding to the exchange of 2-particle states of the form $:{\cal
O}_1 \partial...\partial {\cal O}_1:$. The exchange of these operators
is not represented by additional diagrams in the bulk, instead they
are already encoded in the basic exchange Witten diagram.

This raises the questions: what is the precise relation between Witten
diagrams and conformal blocks? What is the role of these additional
2-particle states?  Why are they automatically included in the Witten
exchange diagram? We will address these questions in the next
subsection. Notice that the relation between Witten diagrams and
conformal blocks has been discussed in several works in the
past \LiuTY, \FreedmanBJ, \LiuTH, \DHokerPJ, \DHokerNI, \DHokerJP, \HoffmannTR,
\HoffmannMX, \DolanUT, \DiazNM.

\subsec{CFT interpretation of Witten diagrams: ``dressed'' conformal blocks}

These questions were partly addressed by the authors
of \HeemskerkPN, \HeemskerkTY. They considered the 4-point function of
an operator ${\cal O}$ in an ${1\over N}$ expansion and showed that
the solutions of crossing symmetry, under certain technical
assumptions about the spins of the intermediate fields, are in
one-to-one correspondence with the possible contact Witten diagrams
that can be written in the bulk for the dual field $\phi$. This
provides evidence that the tree level interactions of the generalized
free field ${\cal O}$ can be described in terms of an effective action
in AdS for the dual field $\phi$.

Here we develop a related perspective by focusing on the CFT
interpretation of the scalar exchange Witten diagrams.  Our goal is to
explain why, in perturbed generalized free CFTs, it is more natural to
reorganize the ${1\over c}$ expansion into an expansion in certain
{\it linear combinations} of conformal partial waves. These linear
combinations of conformal partial waves coincide with the Witten
diagrams. Hence we will try to argue that even if we did not know
about AdS/CFT, we would still have a motivation to introduce the
Witten diagrams in order to describe the perturbative expansion of
generalized free CFTs.

The main point is that a single conformal partial wave, once expanded
in the crossed channel, has undesirable behavior: it cannot be written
as a superposition of conformal blocks in the crossed channel,
consistently with the scaling assumption \cscaling\ for the perturbed
correlators. We will explain this point below.

Of course this is not a problem in principle, since we are supposed to
solve the bootstrap conditions by summing over all conformal partial
waves in the direct channel and writing this sum in terms of another
sum of conformal partial waves in the crossed channel. That is, the
bootstrap conditions apply to the sums and not to individual
terms. However, relating two infinite sums over conformal partial
waves makes solving the bootstrap equations difficult and it would be
useful if we could break down the problem into smaller pieces. In
particular it would be helpful if we could redefine the conformal
blocks in such a way that each individual ``redefined block'' had nice
factorization properties in all channels, consistent with the large
$c$ scaling. If that was possible we could then combine these
redefined blocks to get the most general solution of the bootstrap
equations. As we will argue, this is possible for generalized free
CFTs, and these ``redefined blocks'' are precisely the Witten
diagrams.

We will now explain how this works in the case of a 4-point function
and we will discuss higher order correlators in a later subsection.
We consider the correlation function of four single-particle scalar
operators
$$
\langle{\cal O}_1(x_1) {\cal O}_2(x_2) {\cal O}_3(x_3) {\cal O}_4(x_4)\rangle
$$
To avoid certain technical complications we assume that the conformal
dimensions are different generic real numbers\foot{i.e. no linear
combination of the dimensions with integer coefficients is equal to an
integer number. Otherwise logarithmic terms may appear due to
corrections to the dimensions of double-trace operators.}. This
implies that at leading order (order $g^0$) the correlator vanishes,
so we do not have to deal with disconnected contributions to the
perturbed correlator.  According to our scaling assumption the 4-point
function will become nonzero at order $g^2$.  Moreover we will assume
that the conformal dimensions of the single-trace operators do not
receive any corrections as we turn on $g$. This is a harmless
simplifying assumption in order to avoid technical complications. It
is true, for example, for chiral primaries in supersymmetric
CFTs\foot{Though in this case it is more likely that the conformal
dimensions will be integer numbers and thus logarithmic corrections
will have to be considered.}.

Let us consider the conformal partial wave ${\bf G}_i^{12,34}$, which
was defined in \cblockdef, corresponding to the exchange of the
operator ${\cal A}_i$ and its descendants in the channel
$(12)\rightarrow (34)$ and the conformal partial wave ${\bf
G}_j^{14,23}$ for the exchange of an operator ${\cal A}_j$ in the
channel $(14)\rightarrow (23)$.  The bootstrap condition is
\eqn\bootstrapc{
\sum_i C_{12}^i C_{34}^i {\bf G}_i^{12,34} = \sum_j C_{14}^j C_{23}^j  {\bf G}_j^{14,23}}
where $C_{ij}^k$ are the OPE coefficients. 

This is a complicated equation for the OPE coefficients $C^k_{ij}$. How can we
solve it? In a general CFT the problem is too difficult and, apart from free
(and generalized free) CFTs, has only been solved for certain simple
two-dimensional CFTs. However, in a perturbed generalized free CFT we can use
the assumption \cscaling\ to derive some additional information about the kind
of operators which appear in the summation over $i$ and $j$.  The OPE between
$O_1$ and $O_2$ is constrained by the scaling \cscaling\ and has a low order
expansion (in $g$) given in \opeexp.  At order $g^0$ we can only have 2-particle
operators of the form $:O_1 \partial...\partial O_2:$ and $:O_2 \partial
...\partial O_1:$ while at order $g$ we can only have new single particle operators
$O_i$. At order $g^2$ we can have corrections to the coefficients of operators
which appeared at lower order (including the order 1 coefficient) and new
2-particle operators of the form $:O_i \partial...\partial O_j:$ with
$(i,j)\neq(1,2)$.  In particular we can have 2-particle operators of the form
$:O_3\partial...\partial O_4:$. Similar results are true for the OPEs in the
crossed channel\foot{Notice that because we chose to use four different
operators we do not have any contribution from the correction to conformal
dimensions of double trace operators. While these corrections are in general
nonzero, they enter at higher order in perturbation theory. This would not be
the case if we had considered four-point functions of the same operators on the
external legs.}.

Because of this scaling, when we consider the double OPE in the
$(12)\rightarrow (34)$ channel we see that there is no possible
contraction at order $g^0$ or $g^1$. At order $g^2$ we have two kinds
of contractions: the order $g^2$ 2-particle operator on one side can
contract with a leading order operator on the other side, or the order
$g$ single-particle operators on both sides can contract together.  So
the solution of the bootstrap equations has to be constructed by using
only the operators mentioned above.

Using these operators we would like to find basic building blocks,
i.e. basic solutions of the bootstrap equations, which we will then be
able to combine to get more complicated solutions. Let us start with
the exchange of just one single-particle operator ${\cal O}_m$ in the
$(12)\rightarrow (34)$ channel, i.e. we take
$C_{12}^i,C_{34}^i \sim \delta^{im}$.  Then the LHS of \bootstrapc\ is
equal to a single conformal block ${\bf G}_m^{12,34}$ in this channel.
Now we ask whether we can choose the coefficients $C_{14}^j,C_{23}^j$
in such a way that the equation is solved.

For this let us try to see whether a single conformal block in the
direct channel can be expanded in the basis of conformal blocks in the
crossed channel as
\eqn\false{
{\bf G}_m^{12,34} =? \sum_{n,\,{\rm allowed}} K_{mn} {\bf
G}_{n}^{14,23}} where $K_{mn}$ is a matrix to be determined. While
trying to solve this equation we additionally have to make sure that
the operators labeled by $n$, running in the crossed channel, are
consistent with the assumptions about the OPE between $(13)$ and
$(24)$ i.e. that they are only single-particle operators at order $g$
and double-particle operators at orders $g^0$ and $g^2$. With these
additional restrictions it is easy to check that it is not possible to
write a single conformal block in the direct channel in terms of the
allowed blocks in the crossed channel i.e. it is not possible to find
$K_{mn}$ which satisfies \false\ \foot{For example when the conformal
block ${\bf G}_m^{12,34}$ is expanded in the crossed channel, it has
logarithmic terms in the conformal cross-ratios. This logarithmic
behavior is inconsistent with an OPE expansion in the crossed channel,
since we argued that due to the large $c$ scaling
assumption \cscaling, and for 4-different operators on the external
legs, there are no contributions of anomalous dimensions in the
correlator at this order, which might introduce logarithmic
dependence.}.

This means that if we start with a single conformal block in the
direct channel, we have to ``dress it up'' with other allowed
2-particle operators to make it have a nice (i.e. consistent with the
OPE and the scaling \cscaling) expansion in the crossed channel.

This can be done in more than one ways, but we would like to find the
minimal modification of a basic conformal block so that it can solve
the bootstrap conditions by itself. So we start with the conformal
block ${\bf G}_m^{12,34}$ corresponding to the exchange of $m$ in the
$(12)\rightarrow (34)$ channel and we add to it a superposition of
other conformal blocks ${\bf G}_n^{12,34}$ corresponding to the
exchange of other operators $n$ in the same channel, and we try to
adjust this superposition in such a way that when expanded in the
$(14)\rightarrow (23)$ channel it can be expressed as a superposition
of the allowed (that is, allowed by the scaling \cscaling)
operators. So we define the object
$$
{\bf W}^{12,34}_m(x_1,x_2,x_3,x_4) = {\bf
G}^{12,34}_m(x_1,x_2,x_3,x_3) + \sum_{n\,\in \,{\rm\,direct\,channel}}
c_m^n\,\, {\bf G}_n^{12,34}(x_1,x_2,x_3,x_4)
$$
where the sum over $n$ runs only over the allowed
operators\foot{i.e. allowed by the assumption \cscaling\ for the
scaling with $g$ of the connected correlators.} in the direct
channel. We want to choose the coefficients $c_m^n$ so that we can
write
$$
{\bf W}^{12,34}_m(x_1,x_2,x_3,x_4) = \sum_{n\in\,\rm \,crossed\,channel} d_m^n\,\, {\bf G}_n^{14,23}(x_1,x_4,x_2,x_3)
$$
for appropriate coefficients $d_m^n$. Here the index $n$ runs over the
allowed operators in the crossed channel. This is a problem for the
unknown numbers $c_m^n,d_m^n$.

To find the simplest possible solution we first make the assumption
that there is no other single particle operator involved in the sums,
apart from ${\cal O}_m$ in the direct channel, but we can have
2-particle operators running in both channels, allowed
by \cscaling. Even after imposing these restrictions, there is still
more than one solution to our problem. Motivated by \HeemskerkPN, we
try to look for the simplest possible solution by restricting the
maximum spin of the 2-particle operators which are exchanged. If we
demand that all 2-particle (primary) operators involved in the
summations above are scalars then we drastically reduce the number of
solutions of the bootstrap equations.

However even with these constraints the solution is still not
unique. It turns out that with the constraints that we imposed in the
last paragraph, there is still a one-parameter family of solutions.
To understand more intuitively why there is an ambiguity let us
consider the problem we are trying to solve: we assume the exchange of
the operator ${\cal O}_m$ (and of its descendants) and ask how we can
dress it up by adding contributions of 2-particle scalar operators in
such a way that the combination has nice factorization in the crossed
channel.  The ambiguity comes from the fact that there is actually a
combination of the 2-particle states which by itself (without adding
the contribution of ${\cal O}_m$) solves the bootstrap.  This solution
is the one corresponding to the basic contact Witten diagram in the
bulk.  The explicit expansion of the contact Witten diagram in terms
of CPWs can be found in appendix B. This solution can be added with
arbitrary overall coefficient to any other solution of the original
problem, containing the exchange of ${\cal O}_m$. In a sense this is
like trying to solve an inhomogeneous linear problem, where the
solution is defined up to an ambiguity of adding any solution of the
corresponding homogeneous problem.

As usual this ambiguity is fixed by imposing appropriate boundary
conditions.  The ``homogeneous solution'' (the solution involving only
2-particle scalar operators) is unique \HeemskerkPN\ and is exactly
equal to the contact Witten diagram in the bulk. With appropriate
boundary conditions, the``dressed conformal block'' should become
unambiguously defined and as we will see it will precisely coincide
with the exchange Witten diagram.

What is the relevant``boundary condition'' is not obvious a priori. We
can guess what it should be by using the intuition from the Witten
diagram picture: we have to identify a qualitative difference between
an exchange and a contact diagram. One difference is that the contact
diagram is ``harder'' at very high energies. In the case of the
exchange diagram the propagator of the exchanged particle suppresses
the diagram if the momentum of the intermediate particle is very
large. Such a suppression does not take place in the contact
diagram. In CFT language this difference can be translated into a
statement about the singularity structure of the 4-point functions.
This singularity becomes obvious in the Lorentzian continuation of the
4-point function. Since these points are somewhat technical they will
be not be explained here but we refer the reader to the relevant
papers \PolchinskiRY, \GiddingsJQ, \GaryAE, \HeemskerkPN, \GaryMI. The
important point is that the 4-point function, as computed by tree level
Witten diagrams, has a singularity as a function of the conformal cross
ratios when there is a common point in the light-cones of the 4
operators. The strength of this singularity is qualitatively different
between a contact and an exchange Witten diagram.

The conclusion is that if we demand that the solution has as soft a
singularity as possible then it is unique: there is a unique way to
dress up a conformal partial wave by 2-particle operators of spin zero
so that it solves the bootstrap equation keeping the short distance
singularity as soft as possible. This unique solution is the exchange
Witten diagram.

\subsec{Summary: CFT interpretation of the basic exchange Witten diagram}

Let us review the previous section: the basic scalar exchange Witten
diagram is equal to a single conformal block dressed up with
2-particle scalar operators in such a way that its expansion in the
crossed channel is consistent with an OPE, large $c$ scaling, is as
simple as possible (i.e. it only involves scalar 2-particle
operators), and the singularity in the Lorentzian regime is as weak as
possible. The explicit form of the 2-particle coefficients of this
solution are expressed in appendix B.

\fig{Conformal partial wave expansion of a scalar exchange Witten diagram in the direct channel. The coefficients of the various CPWs are computed in appendix B. }
{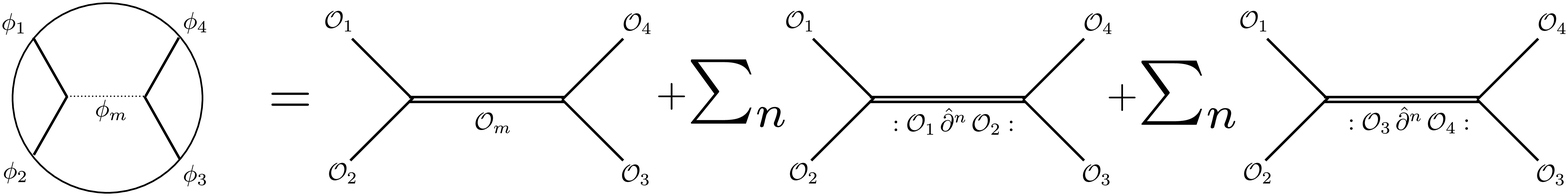}{6.truein}
\figlabel{\wittencpwb}

\fig{Conformal partial wave expansion of a scalar exchange Witten diagram in the crossed channel.}
{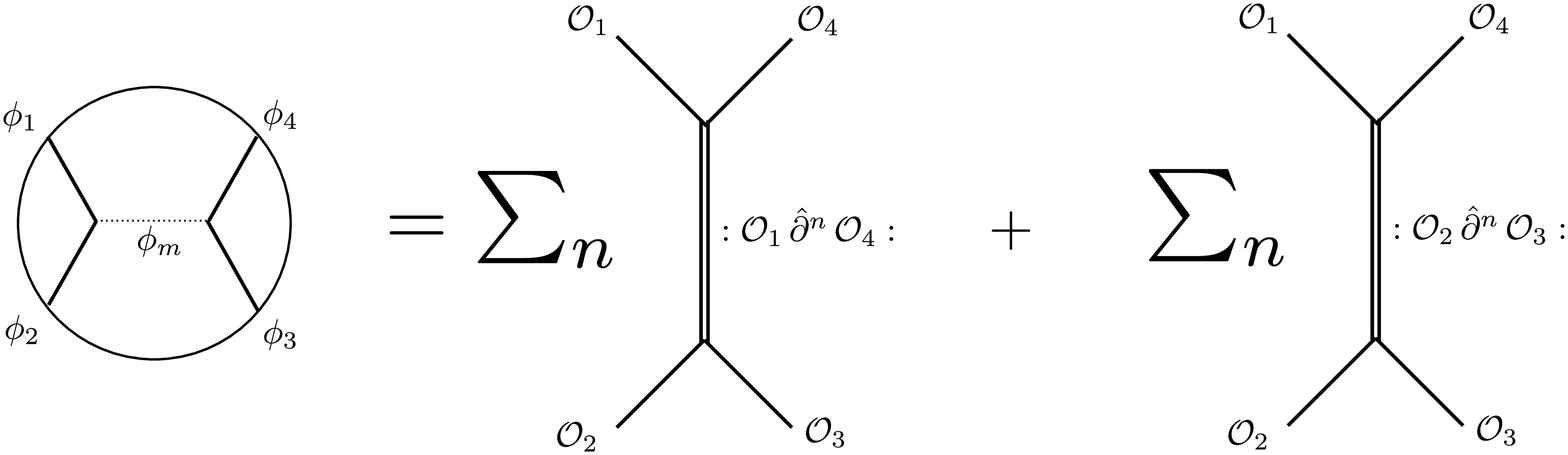} {4.truein}
\figlabel{\wittencpwc}

Solutions of the same problem where we include higher spin
intermediate 2-particle operators correspond to Witten exchange
diagrams with derivative interactions (or can be reinterpreted as
having mixing with contact Witten diagrams with derivative
interactions \HeemskerkPN, since there is always an ambiguity related
to field redefinitions \LiuTH).  It would be interesting to
demonstrate this more explicitly. Finally solutions of the problem
where no single particle operator is involved at all (i.e.  the
``homogeneous'' solutions mentioned above) correspond to contact
Witten diagrams as demonstrated in \HeemskerkPN.

Similar statements should hold for exchange diagrams of intermediate
fields with spin, such as the graviton which is dual to the stress
energy tensor. Though we have not checked this explicitly, we expect
that the graviton exchange diagram\foot{In Einstein-Hilbert gravity.}
is equal to the conformal partial wave of the stress-tensor exchange,
dressed up with 2-particle states in such a way that the total
contribution factorizes into allowed 2-particle states in the crossed
channel.

\subsec{Dressing the OPE}

In the previous section we explored the utility of the bulk in
providing a perturbative solution of crossing symmetry.  The CFT data,
however, is specified not by the conformal partial waves but rather by
the OPE coefficients and conformal dimensions.  Ideally then we would
like to argue that bulk interactions provide a natural way to
``dress'' the OPE of a large $c$ CFT so that it manifestly satisfies
crossing symmetry.  This extends, to the interacting theory, the
intuition that the bulk provides a manifest realization of the
``weak-coupling'' of the GFF.

Imagine we try to find a solution to crossing symmetry by perturbing
some OPE coefficients of the factorized theory.  For instance we can
try turning on a non-trivial single-particle OPE
$C_{12}^k \sim \CO(g)$ for some $k$ and ask if this is a consistent
deformation of the CFT.  This deformation is, of course, constrained
by crossing symmetry of $\langle \CO_1 \CO_2 \CO_1 \CO_2 \rangle$ but
this constraint is difficult to analyze because corrections to the
conformal dimension of $\CO_1 \CO_2$ two particle states enter at
leading order (ie.  $g^2$) and generate logarithmic terms in the
expansion.

In order to avoid this subtlety we consider simultaneously deforming
both $C_{12}^k$ and $C_{34}^k$ (subject to the constraints on the
conformal dimension mentioned in the previous section).  From the
previous section and Appendix B we see that this, by itself, is {\it
not} a consistent deformation of the CFT.  Rather any such deformation
must be accompanied by a deformation of $C_{12}^{:34\,n:}$ and
$C_{34}^{:12\,n:}$ where $:12\,n:$ and $:34\,n:$ indicate
multiparticle states of the form $\CO_1\nabla^n \CO_2$ and
$\CO_3 \nabla^n \CO_4$, respectively.  These deformations must occur
in a precise ratio given by comparing \coefexcone\ and
\coefexctwo.  Moreover, we must also deform $C_{14}^{:23\,n:}$ and
$C_{23}^{:14\,n:}$ by an amount given by comparing \horrcoef\ and \coefexctwo.  

From \coefexcone-\horrcoef\ it is thus possible to extract a
prescription for a one parameter family of consistent deformations of
the GFF OPE coefficients.  Of course from the bulk this is nothing
more than the two parameter deformation given by
$g_{12k} \phi_1 \phi_2 \phi_k$ and $g_{34k} \phi_3 \phi_4 \phi_k$ but
the CFT is only constrained by the product $g_{12k}\, g_{34k}$ (we
could extract constraints associated only to $g_{12k}$ but we would
have to consider $\langle
\CO_1 \CO_2 \CO_1 \CO_2\rangle$ or some more complicated combination of
correlators).

This suggests that the OPE coefficients are somehow not the most
natural basis to perturb a large $c$ CFT with but rather we would like
to work with ``dressed'' coefficients $\tilde{C}_{12}^k$ which are
particular linear combinations of the original OPE which correspond,
in some sense, to a bulk three-point vertex.  It may be that
correlators in a large $c$ CFTs enjoy an alternate OPE expansion that
manifestly solves crossing symmetry (perhaps not unrelated to the
proposal of \PolyakovGS\foot{We would like to thank S. Rychkov for
bringing this paper to our attention.} or that of \PenedonesUE). In
such an expansion the holographic nature of perturbed GFF correlators
may well be more manifest.

As mentioned in the previous section, however, the above prescription
or deformation, is not unique.  We also have the freedom to switch on
single-multiparticle interactions without any single-single particle
interactions.  We could e.g. switch on $C_{12}^{:34:}$ so long as we
also switch on $C_{34}^{:12:}$, $C_{14}^{:23:}$ and $C_{23}^{:14:}$ in
a way constrained by the function $a_n$ of \coefcont\ (i.e. the ratio
of these coefficients is fixed by the ratio of the
$a_n(\Delta_i, \Delta_j, \Delta_k, \Delta_l)$ with its arguments
appropriately permuted).  As a consequence of this freedom it is far
from obvious that there is a natural way to modify the OPE expansion
so that it corresponds to a bulk vertex expansion (as the former is
always cubic while the latter can involve higher point vertices).

It would be interesting to explore this construction further but for
now we wish merely to point out that the constraints coming from the
CPW expansion enjoy a natural reformulation in terms of deformations
of OPE coefficients.  Note further that if one performs the minimal
deformation (consistent with having only a three-point vertex in the
bulk) this defines a consistent CFT including multi-particle
correlators as well.  More general solutions, however, are possible
corresponding to introducing four and higher point vertices in the
bulk.

\subsec{Conformal bootstrap for higher $n$-point functions}

We now sketch what happens when we consider higher $n$-point functions
of single-particle operators, but still at tree level i.e. focusing on
the leading terms in the scaling \cscaling. Before doing so let us
mention some basic facts. In a general conformal field theory the
bootstrap conditions correspond to checking equation \bootstrapb\ for
all primary operators on the external legs. For a CFT which is a
perturbation of a ``generalized free CFT'' there is an alternative
formulation of the bootstrap conditions: instead of checking that the
4-point function of {\it all} conformal primaries factorizes correctly
in all channels, we can check that the $n$-point function of {\it
single particle} operators factorizes correctly in all possible OPE
decompositions\foot{Of course this condition only guarantees the
consistency of the low-lying sector of generalized free fields and not
of the entire CFT}. In this way we have to check the factorization of
correlators of fewer operators (only the single-particle ones) but the
trade-off is that we have to consider all $n$-point functions instead
of just the 4-point functions. Let us refer to this equivalent formulation
as the ``single-particle bootstrap''.

If we check that the ``single-particle bootstrap'' is true for all
single particle operators on the external legs, then we can show that
the standard bootstrap is also satisfied for all possible
multi-particle operators on the external legs of 4-point functions:
this can be done by constructing the latter as a limit of the former.
For example the factorization properties of a correlator of 4
two-particle operators can be derived by the factorization properties
of a correlator of 8 single-particle operators by taking the limit of
pairs of insertions coming together and subtracting singular terms. As
we will see later the ``single-particle bootstrap'' condition is more
natural to impose for a generalized free CFT. This is also suggested
by the fact that Witten diagrams in supergravity are dual to
correlators of single particle operators\foot{The correlators of
multi-particle operators can be computed in AdS/CFT by taking limits
of higher-point Witten diagrams of single-particle operators.}.  This
is again a special property we expect for CFTs with a gravity dual at
large $c$; a generic CFT does not have a sub-basis of
``single-particle'' operators from which the full set of operators can
be generated.

Let us explain in more detail what the ``single-particle bootstrap'' condition
means. An $n$-point function of single-particle operators can be evaluated by
performing successive OPEs between the operators. For this we need to choose an
order in which to perform the OPE. The different orderings can be encoded in
trivalent tree diagrams with (single-particle) operators on the external lines.
For example let us consider a 6-point function $\langle{\cal O}_1...{\cal
O}_6\rangle$. Let us do the OPE in the following order $(1\cdot 2)\rightarrow
a\,,\,(a\cdot 3) \rightarrow b\,,\, (b\cdot 4)\rightarrow c\,, (c\cdot
5)\rightarrow 6$. This is depicted in the first graph in figure 3. In the same
figure we also show two other possible orderings of doing the OPE.  \fig{Three
of the possible different ways to compute a 6-point function by successive OPEs.
Summation over the intermediate operators $a,b,c$ is implied.} 
{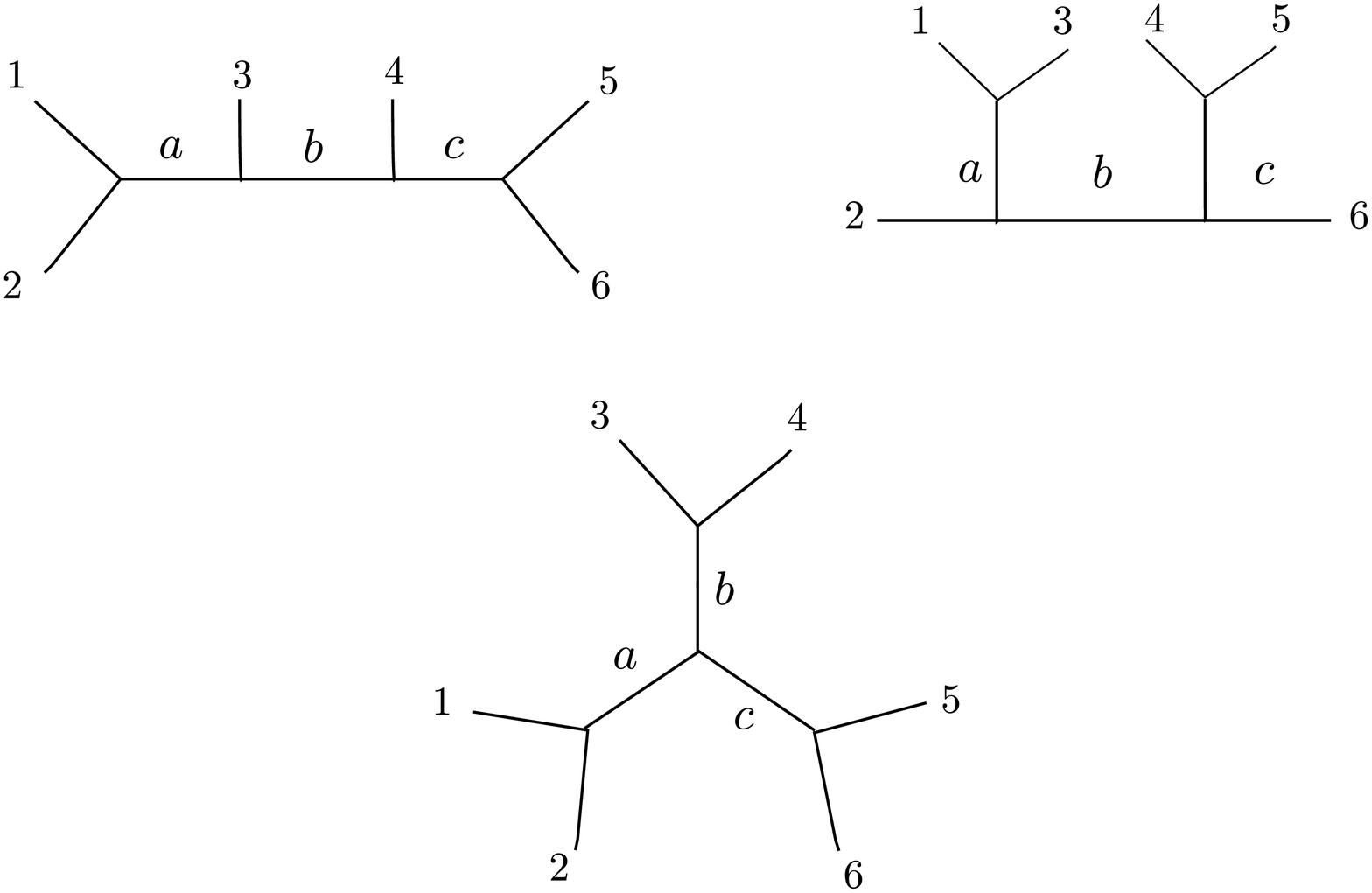}{4.truein}
\figlabel{\spartb}
 
Here it is convenient to introduce a generalization of the conformal
partial waves for higher $n$-point functions $\langle{\cal
O}_1(x_1)...{\cal O}_n(x_n)\rangle$ i.e. to introduce special
functions which encode parts of CFT correlators which depend on
kinematics and not on the dynamics of any specific CFT. First we
choose a given order of doing the successive OPEs to compute the
$n$-point function. This ordering can be represented by a trivalent
graph ${\bf T}$. The external legs of the graph are attached to the
operators $\{ {\cal O}_i(x_i)\}$ whose correlator we want to
compute. Each of the internal legs is labeled by a conformal primary
operator of given dimension and spin. At each vertex of the tree we
perform the OPE and we keep only the operator indicated by the
corresponding internal line of the graph together with all of its
conformal descendants. This procedure gives a contribution to the
$n$-point function which is equal to a certain product of OPE
coefficients $C_{ij}^k$ times a special function of the external
positions $x_1,...,x_n$ which only depends on the choice of the tree
${\bf T}$ and the conformal dimensions and spins of the operators
appearing on the legs of the tree. This special function ${\bf G}^{\bf
T}_{\{{\cal O}_i\},\{a,b,c...\}}(x_1,...,x_n)$ is the $n$-point
analogue of the conformal partial wave. Its functional dependence on
the coordinates of the external legs is completely fixed by
kinematics. While we do not have explicit expressions for these
generalized CPWs we hope it is clear from their definition that in
principle these functions do exist and they are indeed determined by
kinematics of the conformal group.

If we sum over all possible intermediate operators (not necessarily
single-particle ones) we reproduce the full $n$-point function. Hence
the $n$-point function can be written as a sum of the form
\eqn\gencpw{\langle {\cal O}_1(x_1)...{\cal O}_n(x_n)\rangle = \sum_{a,b,c...} 
[C C...]^{\bf T}{\bf G}^{\bf T}_{\{{\cal O}_i\},\{a,b,c...\}}(x_1,...,x_n) }
where the generalized conformal partial waves ${\bf G}$ depend only on
kinematics and the conformal dimensions and spins of the involved
operators, and all the dynamics is contained in the 3-point function
coefficients. The notation $[C C ...]^{\bf T}$ refers to a product of
OPE coefficients for the operators ${\cal O}_i$ and $a,b,c...$ where
the way that the indices are contracted is dictated by the tree graph
${\bf T}$.

The conformal bootstrap conditions are based on the idea that we can
do the OPE in any order we want and the resulting $n$-point function
will be the same. In equations this means that we must
have\eqn\genboot{\sum_{a,b,c...} [C C ...]^{\bf T_i}{\bf G}^{\bf
T_i}_{\{{\cal O}_i\},\{a,b,c...\}}(x_1,...,x_n) = \sum_{a,b,c...} [C C
...]^{\bf T_j}{\bf G}^{\bf T_j}_{\{{\cal
O}_i\},\{a,b,c...\}}(x_1,...,x_n) } for all possible trees ${\bf
T}_i,{\bf T}_j$.

These equations are generalizations of the basic conformal bootstrap
equations \bootstrapb\ for higher $n$-point functions. As we said, if
we impose \bootstrapb\ for all possible operators on the external legs
then the general condition \genboot\ will be satisfied.

Conversely, in a generalized free CFT if we impose \genboot\ for all
$n$, but only for single-particle operators on the external legs (and
arbitrary operators on the internal legs), then the conformal
bootstrap \bootstrapb\ will be automatically satisfied for all
many-particle operators.

In the previous section we argued that the solution to the bootstrap
equations for the 4-point function can be naturally reorganized in
terms of tree-level Witten diagrams. This reorganization was natural
because each Witten diagram solved the bootstrap conditions by
itself. Now we want to sketch how the same reorganization might be
performed for higher $n$-point functions of single-particle
operators. In other words we want to consider the leading order
$n$-point function and show how it can be naturally divided into
certain parts, each of which solves the bootstrap conditions by
itself. We want to argue that these parts coincide with tree-level
Witten diagrams.

Let us consider the correlator $\langle {\cal O}_1(x_1)...{\cal
O}_n(x_n)\rangle$ and isolate from the sum \gencpw\ a term ${\bf T}$ which
contains only single-particle operators in all the intermediate lines. We want
to understand how the generalized $n$-point CPW ${\bf G}^{\bf T}_{\{{\cal
O}_i\},\{a,b,c...\}}(x_1,...,x_n)$ is related to the corresponding Witten
diagram with trivalent vertices. It is not hard to see that the $n$-point CPW
has nice factorization properties (i.e. consistent with an OPE and the scaling
\cscaling) in the ``direct channel'' but not in the crossed channels.

From the discussion of the 4-point function in the previous subsection
it is clear what we have to do to fix this problem. We add to the
$n$-point CPW a series of $n$-point CPWs where in the intermediate
channel we have multi-particle operators running, in such a way that
the resulting ``dressed'' CPW has nice factorization properties in all
channels. Presumably this can be done in a unique way if we restrict
the spin of the intermediate operators and demand that the resulting
amplitude has the correct singularities in the Lorentzian regime.  We
expect that the dressed $n$-point CPW thus constructed should coincide
with the corresponding trivalent tree level Witten diagram.

In addition to these trivalent diagrams, we also have to include the
contributions from quartic and higher interactions in order to
reproduce the full $n$-point function, in analogy to what happened for
the 4-point function.

\subsec{Generalized free CFTs and expansion in Witten diagrams}

To summarize: in a perturbed generalized free CFT, the solution to the
bootstrap equations at ``tree level'' i.e. leading order in the $1/c$
expansion, can be expressed as follows: to each single particle
operator we assign a dual field $\phi_i$. For each 3-point function
between single particle operators include a coupling $ C_{ijk}\int
d^{d+1}x \,\,\phi_i \phi_j \phi_k$. With this bulk Lagrangian we
compute all exchange Witten diagrams in all possible channels. These
should reproduce the part of the $n$-point functions with single
particle operators running in any possible factorization channel. To
reproduce the remaining part of the $n$-point function we add quartic
and higher interactions terms (possibly with higher derivatives).

Formally any perturbed generalized free CFT can be reorganized in this
way.  In particular we have not assumed that the bulk theory
corresponds to a large macroscopic AdS space. To see what this means
let us consider for example the ${\cal N}=4$ SYM at large $N$ and
small 't Hooft coupling where we know that the dual theory is IIB
string theory in a highly curved AdS space. According to what we have
discussed it is possible to reorganize the $1/N$ expansion of this
theory in terms of ``Witten diagrams'', i.e. in terms of dressed
conformal blocks. On a basic level the 1/N expansion of the ${\cal
N}=4$ SYM can be written in terms of Feynman diagrams. Since the
theory remains conformal in the $1/N$ expansion, the Feynman diagram
expansion can be regrouped and organized into a conformal partial wave
expansion. Finally, as we argued above, this conformal partial wave
expansion can be further reorganized in terms of dressed conformal
blocks (or ``Witten diagrams''), since this basis is more natural for
the $1/N$ expansion of large $N$ gauge theories. The main difference
between the ``Witten diagram expansion'' of the ${\cal N}=4$ at weak
and strong 't Hooft coupling is that in the first case we have a very
large number (exponentially growing with conformal dimension\foot{but
the number is $N$-independent and in that sense it is still a small
number of exchanged operators in comparison to the $N^2$ fundamental
fields of the Lagrangian.}) of intermediate single particle states,
while in the second case we only have a small number corresponding to
the chiral primaries of the CFT which survive at strong coupling.

However, our perspective is not to distinguish between CFTs which have
classical gravity duals from those with ``stringy'' duals. For us the
main distinction is to be made between theories whose low-lying
spectrum has an effective description in terms of a finite number
(with respect to the $c$-scaling) of degrees of freedom in AdS from
those where such a dual description is not useful at all. The latter
case would correspond to large $c$ CFTs with a low-lying spectrum
whose degeneracy grows with $c$, as explained in section \cexamples.

Before closing this section let us mention that in this section we
have only discussed the conformal bootstrap of the low-lying sector of
the CFT which consists of the (perturbed) generalized free fields. In
principle the bootstrap conditions have to be checked for {\it all}
local operators of the CFT, even those whose conformal dimension is of
order $c$ and which are not generalized free fields in any sense. In
other words, even if we show that the correlators of low-lying
operators satisfy the bootstrap conditions it is not a sufficient
condition to guarantee the existence of such a CFT but only a
necessary one.

\newsec{The CFT at finite temperature}
\seclab\finitetemp

In this section we will discuss some issues about the finite temperature
behavior of the theories under consideration.

\subsec{Factorization requires an infinite number of degrees of freedom}
\subseclab\subcardy

As we argued in section \gff\ we can have factorization of correlators
only if $c\rightarrow \infty$, where $c$ is defined by the 2-point
function of the stress energy tensor. In a certain sense $c$ is
sensitive to the number of fields in the theory i.e. the number of
degrees of freedom\foot{See appendix C for a discussion of the notion
of central charge and the number of degrees of freedom in higher
dimensional CFTs.}. In this section we want to give some additional
explanations of this statement and thus argue that a sector with
factorized correlators can only emerge in a theory with an infinite
number of degrees of freedom\foot{Of course we are talking only about
the nontrivial version of factorization, that is for operators which
are not ordinary free fields i.e. for operators with
$\Delta>{d-2 \over 2}$. Obviously for operators with $\Delta
={d-2 \over 2}$ we have factorization even for finite number of
degrees of freedom, but this case is of no interest to us as these
fields do not have a holographic description.}. The equivalent
statement in the bulk is that gravity can be decoupled only if $G_N
R_{AdS}^{1-d}\rightarrow 0$. In this limit the entropy of a black hole
of finite size in AdS units diverges, as can be seen from the formula
$S={A \over 4 G_N}$, indicating a very large number of degrees of
freedom.

A more intuitive way to count the number of degrees of freedom of a
CFT is the following: consider the CFT on ${\bf R}^{d-1}$ and at
temperature $T$. By scale invariance the entropy density must be of
the form
\eqn\defentr{
s \sim \widetilde{c}\, T^{d-1} } with some convention-dependent
constant of proportionality of order one.  The constant
$\widetilde{c}$ can be thought of as counting the degrees of
freedom. In two dimensions Cardy's formula implies that $\widetilde{c}
= c$ (in the appropriate conventions). In higher dimensional theories
there is no such direct relation between $c$ and
$\widetilde{c}$. Notice that if the CFT has a classical gravitational
dual\foot{i.e. a gravitational dual theory governed by the
Einstein-Hilbert action.} then again it is true that $\widetilde{c} =
c$, but more generally this relation does not have to be
satisfied. For example in the ${\cal N}=4$ SYM the ratio
$\widetilde{c}/c$ varies continuously between the value $4/3$ at weak
't Hooft coupling to the value 1 at strong coupling. This directly
shows that there cannot be a simple general relation between the two
constants in higher dimensional CFTs.

However, we would like to claim that if for a sequence of theories
$c\rightarrow \infty$, then we must also have
$\widetilde{c} \rightarrow \infty$. Ideally it would be useful if we
could find some model-independent bounds for the ratio ${\widetilde{c}
\over c}$. Let us say that we could argue that independent of the CFT, there are two
positive real numbers $\lambda_1,\lambda_2$ such that
$$
\lambda_1 \leq {\widetilde{c}
\over c}\leq \lambda_2
$$
Then our claim would follow immediately. We have not been able to show such a
statement. Notice however that if one considers the most general weakly coupled
CFT with $n_s$ scalars, $n_f$ fermions and $n_v$ vectors, and one scans all
possible values of $n_s,n_f,n_v$ one then finds that the ratio ${\widetilde{c}
\over c}$ is indeed bounded by two numbers $\lambda_1,\lambda_2$ \KovtunKW,
while for strongly coupled CFTs with gravity duals the ratio is equal to one.
These facts are an indication that such a bound may actually exist.

In any case, without such a model-independent bound, our only hope is
to find an indirect argument about the relation between $c$ and
$\widetilde{c}$.

In two dimensional conformal field theories we have $c=\widetilde{c}$
due to the Cardy formula, which is usually proved by using modular
invariance. As we explained in section \cardyone\ there is an
alternative derivation of the two dimensional Cardy formula which may
be more relevant in generalizing to higher dimensional CFTs: for any
CFT in any dimension the following equation must hold
\eqn\cardyb{
{\partial \langle T_{00}\rangle_{\beta} \over \partial \beta} =-
{1\over \beta}\int d^d x \, \, \langle
T_{00}(x) \,T_{00}(0)\rangle_{\beta}^c} where the integral is over
${\bf R}^{d-1}\times {\bf S}^1$ and the superscript $c$ in the 2-point
function stands for ``connected''\foot{i.e. $ \langle
T_{00}(x) \,T_{00}(0)\rangle_{\beta}^c \equiv \langle
T_{00}(x) \,T_{00}(0)\rangle_{\beta} - \langle
T_{00}(0)\rangle_\beta^2$.}. The left hand side of equation \cardyb\
is related to the constant $\widetilde{c}$ by the following form of
the thermal expectation value
$$
\langle T_{00}\rangle_\beta \sim {\widetilde{c} \over \beta^d}
$$
which can be derived by basic thermodynamics from \defentr.

In two dimensions the thermal 2-point function $\langle T(z)
T(0)\rangle_\beta$ on ${\bf R}\times {\bf S}^1$ can be computed
exactly using the exponential map, or from holomorphy. Then using
formula \cardy\ we can derive $\widetilde{c}=c$. In higher dimensions
we cannot compute the thermal 2-point function $ \langle
T_{00}(x) \,T_{00}(0)\rangle_{\beta}$ exactly. We can, however, still
try to see how it scales with $c$. We consider the OPE of the stress
tensor with itself. It has the following general form, suppressing
Lorentz indices
\eqn\stressope{
T(x) T(0) = {c \over |x|^{2d}} +...+ {T(0) \over |x|^d}+...
}
Evaluating this at finite temperature we find schematically
$$
\langle T_{00}(x) T_{00}(0)\rangle_\beta = 
{c \over |x|^{2d}} +...+ {
\widetilde{c} \over \beta^d}{1 \over |x|^d}+...
$$
Inserting into \cardy\ we find an equation relating $c$ and
$\widetilde{c}$, along with other 1-point functions. Unlike in 2-d we
cannot use this relation to fix the ratio $\widetilde{c}/c$ because we
need to know the 1-point functions of other operators which appear on
the RHS of \stressope. However unless certain nontrivial cancellations
take place, we can see that the statement that $c\rightarrow \infty$
suggests that $\widetilde{c}\rightarrow \infty$. Even though it is
definitely not a proof, this argument suggests that the central
charge $c$ in the 2-point function of the stress tensor can go to
infinity only if we have an infinite number of degrees of freedom
(i.e.  only if $\widetilde{c}\rightarrow \infty$).

\subsec{Hawking-Page transition ?}

What happens to a generalized free CFT when we turn on a small
temperature?  As we explained, a generalized free CFT should be
thought as a low-lying sector in a sequence of CFTs of increasing
central charge $c$. In the limit $c\rightarrow \infty$ and at low
conformal dimension (i.e. of order $c^0$) we have a finite number
(i.e. their number does not scale with $c$) of generalized free
fields, together with their multiparticle states.  On the other hand
at large enough conformal dimension the spectrum has to be modified
and the entropy at high enough temperature should become $c$
dependent.

Intuitively, and judging from what happens in large $N$ gauge theories, one
would expect that a generalized free CFT has two phases, one at low
temperatures, which can be effectively described as a finite temperature
Fock-space of generalized free field excitations and one at high temperatures
where the generalized free fields are no longer a good description and where the
underlying fundamental degrees of freedom are deconfined. One would expect the
low temperature phase to have an entropy density which scales like $c^0$ while
the high temperature phase has an entropy density of order $c$.  More precisely,
we have shown that so long as the GFF's free Fock space structure approximately
holds, i.e. $\Delta \ll c$, the theory is ``confined''  as its entropy $S \sim
T^d$ is $c$-independent.  At sufficiently high temperatures, however, conformal
invariance implies that this must give way to a ``deconfined'' holographic phase
with extensive entropy $S \sim c \, T^{d-1}$, dominated by states with $\Delta
\gg c$.

Moreover at $\Delta \sim c$ we expect the freely generated Fock space
structure to be modified by interactions.  Recasting these
estimates\foot{Trading the temperature for the mean conformal
dimension $\langle
\Delta\rangle$ in both the confined and the deconfined ensemble.  This is of
course only valid in the regimes $\Delta \ll c$ and $\Delta \gg c$,
respectively.} using thermodynamic relations the confined phase has $S \sim
\Delta^{d \over d+1}$ while the deconfined phase is characterized by $S \sim
c^{1 \over d} \Delta^{d-1 \over d}$.  The cross-over between these entropies
occurs at $\Delta \sim c^{d+1}$, far beyond the regime of validity of the free
Fock space approximation.  The high temperature entropy is valid if $T =
\left({\Delta \over c}\right)^{1/d} \gg 1$ so it does not necessarily hold in
the regime where the Fock space structure starts breaking down, $\Delta \sim c$.
Nonetheless, even slightly beyond this regime (e.g. $\Delta \sim
c^{1+\epsilon}$) the temperature is parametrically large (at large $c$) and the
black hole entropy is significantly larger than the (uncorrected) Fock space
entropy.  In this regime, 
$$
{S_{deconfined} \over S_{confined} } \sim c^{1 \over d + 1}
$$
suggesting that the transition from the confined to the deconfined phase is not
likely smooth and thus has the character of a phase transition.  In particular
it is hard to imagine that corrections to the GFF spectrum are sufficient to
account for this parametrically large entropy difference.

Thus quite generally we expect that the two phases are separated by a phase
transition.  Bulk intuition would suggest that this occurs at some finite
temperature of order one as it does in  the Hawking-Page phase transition
\WittenZW, \SundborgUE, \AharonySX. This is indeed what happens in weakly
coupled large $N$ gauge theories with fields in the adjoint of the gauge group
and in large $N$ symmetric orbifold CFTs in two dimensions as we reviewed in
section \examples.

However there are indications that the story may be more complicated and the
phase structure of generalized free CFTs may involve additional possibilities
than those which are realized in large $N$ gauge theories. In particular it
seems that there are examples of generalized free CFTs where the analogue of the
Hawking-Page phase transition does not take place at temperatures of order 1 but
at a temperature which goes to infinity as we take $c\rightarrow \infty$. This
suggests the peculiar result that there are no standard black holes in the bulk
(i.e. whose temperature is of the order of the AdS scale).

\subsec{Extending the ``Cardy regime''?}

From the discussion in the last subsection it is clear that
in the ``generalized Cardy regime'', ${\Delta \over {\tilde c}} \gg
1$, the dual CFT exhibits black hole-like entropy suggesting the bulk
is dominated, in this regime, by a large AdS black hole (which can be
stringy in nature \AharonySX, \FestucciaSA).  On the other hand, arguments from
general relativity \HawkingDH\ imply that the transition to a black
hole dominated phase occurs at a temperature of order one (in AdS
units).  From the relation between $c$ and $\tilde c$, however, it
follows that the generalized Cardy regime corresponds to a much higher
temperature $T \sim
\left({\Delta \over c}\right)^{1 \over d} \gg 1$.
Thus CFTs which admit Einstein-Hilbert gravity duals\foot{Recall our general
approach incorporates bulk duals that may include many other degrees
of freedom beyond pure GR -- e.g.  bulks that are highly curved in
``string units''.} require that the generalized Cardy formula
discussed in the previous section to hold already at $T \sim
\CO(1)$.  

In the black hole phase $S \sim c\, T^{d-1}$ and $\Delta \sim c\, T^d$ so if
this is to hold already at $T \sim {\cal O}(1)$ then the CFT must have $e^c$
states with $\Delta \sim c$.  This extended Cardy regime is, in fact, evident in
the examples discussed in Section \examples.  For instance, in ${\cal N}=4$, the
entropy $S \sim N^{1/2} E^{3/4}$ can be derived from general CFT
considerations at high-temperature, $T \sim (E/N^2)^{1/d} \gg 1$.  In the free
theory, however, it is not hard to see that already at $E \sim N^2$ we have $S
\sim N^2$ so the entropy formula seems to apply already at $T \sim {\cal O}(1)$.
From the growth $S \sim N^2$ at $E \sim N^2$ we see that, in the canonical
ensemble, the phase transition alluded to above occurs at $T \sim {\cal O}(1)$
because the free energy goes as $F \sim \alpha N^2 - \gamma N^2 T$ with $\alpha,
\gamma$ order one numbers.  In symmetric orbifold theories
the existence of twist sectors with twist $k$ and a mass gap $1/k$ results in an
``effective temperature'' in these sectors that grows with $k$ even while the
real temperature (measured in e.g. units of the CFT circle) is order one.

While ${\cal N}=4$ SYM and symmetric orbifolds seem to exhibit this ``extended
Cardy regime'' there is no reason for it to hold more generally.  On the other
hand $T \sim 1$ is well-outside the Cardy regime in 2-d CFTs yet, in many known
examples, Cardy's formula seems to hold in this regime nonetheless.  In
particular this seems to hold for free bosons in two-dimensional theories which
provide a rather simple example.

\newsec{Discussions}
\seclab\discussions

We have tried to highlight the basic features of conformal field theories
with holographic duals. We emphasized the importance of operators
whose correlators factorize as the basic ingredient for the emergence
of the bulk. We also explained that such operators with factorized
correlators can only be realized in a limiting sense as a small
low-lying sector of very big conformal field theories.

In our paper we have not addressed an important point. We argued that the
interactions between generalized free fields can be related to effective bulk
interactions between fields in AdS but we have not discussed whether the bulk
fields and their interactions admit an independent (dual) formulation. For
instance, in the case of large $N$ gauge theories we generally expect that there
is a dual two-dimensional worldsheet CFT which can reproduce the interactions
between spacetime fields by the usual perturbative string-theory rules. Is there
such a dual theory for more general generalized free CFTs? Or should generic
dual theories that emerge from this construction be thought of as entirely
``effective'', lacking an underlying organizing principle.  The problem with the
latter viewpoint is that in general we expect that the number of fields in the
AdS space will be quite large (though $c$-independent), similar to what happens
in the ${\cal N}=4$ SYM at weak 't Hooft coupling or in the $O(N)$ vector
models\foot{In other words, typically we do not expect a generalized free CFT to
have a gap between single trace operators of spin two and higher as in
\HeemskerkPN, which would lead to a dual with a large macroscopic AdS
spacetime.}.  In these cases the bulk theory is not a conventional effective
field theory with a small number of fields so without an organizing principle
(such as that provided by perturbative string theory) it is not so clear how to
handle the bulk theory. On the other hand the answer cannot be that there is
always a dual string theory, as can be seen from the example of M-theory
backgrounds (for example, AdS$_4\times$S$^7$ or AdS$_7\times$S$^4$) which
presumably have no description in terms of a perturbative string theory, or from
examples of higher-spin gravity.

Our paper focused on the holographic representation of conformal field
theories. However, conformal invariance does not play a fundamental
role in our discussions; it was merely a simplifying assumption. More
generally let us consider a non-conformal QFT with a local operator
${\cal O}$ whose correlators factorize to products of 2-points
functions (in an appropriate limiting sense). As we argued in the text
it would be natural to represent this field in a higher dimensional
space, whose geometry has to be chosen so that the field will satisfy
a linear wave equation on that background and that the bulk propagator
will coincide with the boundary 2-point function.

An interesting problem would be to examine a generalized free CFT
whose low-lying spectrum consists only of the stress energy tensor and
multiparticle states. If such a CFT existed it would be dual to pure
gravity in AdS. It would be interesting to study to what extent the
conformal bootstrap can constrain the tree-level (i.e. leading order
in an $1/ c$ expansion) interactions between the stress energy tensor
to be the same as those predicted by tree level Einstein-Hilbert
gravity in the bulk\foot{In other words, to what extent AdS theories
with higher-derivative corrections are inconsistent if the only light
particle in the spectrum is the graviton. For such an analysis it
might be important to also analyze the consistency of the theory at
finite temperature, as was partially done
in \HofmanUG, \deBoerPN, \deBoerGX, \KulaxiziJT. We thank J. de Boer
for discussions on these issues.}. Unfortunately this problem is
difficult in practice since the conformal partial waves for operators
with nonzero spin on the external legs are not explicitly known. Along
these lines, it might be interesting to explore the use of analyticity
methods, in the spirit of the BCFW relations and their generalizations
for AdS \RajuBY, to simplify the bootstrap problem for correlators of
the stress energy tensor.

A related, but perhaps simpler, problem would be to consider the simplest perturbed GFF
we can construct.  This would include a single GFF scalar $\CO$ which factorizes
to order $c^0$ and which at order ${1\over \sqrt{c}}$ receives only corrections
mediated by the stress tensor.  Thus one could attempt to construct a consistent
CPW expansion by relating a graviton exchange diagram to a CPW expansion.  This
would presumably include stress-tensor and perhaps some multi-particle
contributions.  Despite requiring higher-spin intermediate operators this
approach depends only on the CPW expansion of scalar four-point functions and is
hence accessible with current technology.  In the bulk this should correspond to a free
scalar minimally coupled to gravity.

In this paper we focused on the leading terms in the large $c$ expansion
and their representation in terms of tree-level interactions in the bulk. It
would be interesting to further study the subleading terms corresponding to loop
diagrams. This was also recently explored in \PenedonesUE.

It would be satisfying to find examples of CFTs with generalized free
sectors which are not large $N$ gauge theories (or symmetric orbifolds
in two dimensions). This would clarify the fact that the holographic
correspondence is not intimately related to gauge invariance but
rather to the fact that generalized free fields can be realized
linearly in a higher dimensional AdS space.

Finally a natural question is the following: we saw that free (or
weakly interacting) fields in AdS can be thought of as a
representation of generalized free fields of a CFT which makes the
linearity more manifest. What is the analogue for flat space/de Sitter
fields?  What is the class of quantum systems for which free fields in
flat space/de Sitter provide an effective (approximately) linearized
description via a holographic map?

\bigskip  
 \centerline{\bf Acknowledgments}

\noindent We would like to thank R. Brustein, F. Dolan, M. Gaberdiel, 
R. Gopakumar, N. Iizuka, E. Kiritsis, M. Kulaxizi, S. Minwalla,
A. Petkou, N. Prezas, S. Raju, S. Rychkov, M.  Shigemori for useful
discussions. We would like to especially thank J. de Boer and
E. Verlinde for useful comments and stimulating discussions. We would
also like to thank J. de Boer, R. Gopakumar, E. Kiritsis for very
helpful comments on the draft. K.P. would like to thank the University
of Crete, the University of Amsterdam, CERN and CEA Saclay for
hospitality during the completion of this work. S.E. would like to
thank the University of Amsterdam and the CEA Saclay for hospitality
during the completion of this work.  The research of S.E. is partially
supported by the Netherlands Organisation for Scientific Research
(NWO) under a Rubicon grant.

\appendix{A}{Conformal Partial Waves}

In this appendix we review some of the results of \DolanUT\ which are
necessary for our discussions. We focus on four dimensional conformal
field theories. We consider a 4-point function of scalar operators
${\cal O}_i, i=1,2,3,4$ and an operator ${\cal O}_k$ of spin $l$ which
is exchanged in the $(12)\rightarrow (34)$ channel. Then the conformal
partial wave corresponding to ${\cal O}_k$ and all of its descendants
is
\eqn\cpwdefi{
{\bf G}_k^{12,34} (x_1,x_2,x_3,x_4) = {1\over |x_{12}|^{\Delta_1+\Delta_2} |x_{34}|^{\Delta_3+\Delta_4}} \left({x_{24} \over x_{14}}\right)^{\Delta_{12}} \left({x_{14} \over x_{13}}\right)^{\Delta_{34}} 
\overline{G}_k^{12,34}(u,v)
}
where we introduced the conformal cross-ratios
$$
u = {x_{12}^2\, x_{34}^2 \over x_{13}^2\, x_{24}^2}\qquad,\qquad v =  {x_{14}^2 \,x_{23}^2 \over x_{13}^2\, x_{24}^2}
$$
which can also be parametrized as
$$
u = z \overline{z}\,\qquad ,\qquad v = (1-z)\,(1-\overline{z})
$$
If we work in Euclidean signature then we have $\overline{z}=z^*$. The
functions $\overline{G}_k^{12,34}(u,v)$ written in terms of the
variables $z,\overline{z}$ have the form
\eqn\cpwa{\eqalign{
& \overline{G}_k^{12,34}(u,v) =\left(-{1\over 2} \right)^l
 {(z \overline{z})^{{1\over 2}(\Delta_k - l)} \over
 z-\overline{z}}{\bf \Bigg[}z^{l+1} \,_2F_1\left({\Delta_k
 -\Delta_{12}+l \over 2}, {\Delta_k+\Delta_{34}+l \over
 2},\Delta_k+l,z\right) \times \cr & \times \,_2F_1\left({\Delta_k -\Delta_{12}-l
 -2 \over 2}, {\Delta_k+\Delta_{34}-l -2 \over
 2},\Delta_k+l-2,\overline{z}\right)
 - \left(z\leftrightarrow \overline{z} \right)
{\bf \Bigg]}
}}

A 4-point function of scalar operators can be expanded in CPWs in the
$(12)\rightarrow (34)$ channel
$$
\langle {\cal O}_1(x_1) {\cal O}_2(x_2) {\cal O}_3 (x_3) {\cal O}_4 (x_4)
\rangle = \sum_k C_{12}^k C_{34}^k {\bf G}_k^{12,34}(x_1,x_2,x_3,x_4)
$$
This expansion converges when $x_1$ and $x_2$ can be enclosed by a
sphere which does not contain $x_3$ or $x_4$, in other words when the
conformal cross ratios are within a certain distance around the point
$u=0, \,v=1$.  In terms of the variables $z,\overline{z}$ the region
of convergence of this expansion corresponds to the disk $|z|<1$.

In the other limit, where $x_1$ and $x_4$ can be enclosed by a sphere
not containing the other insertions, we can expand in the
$(14)\rightarrow (23)$ channel as
$$
\langle {\cal O}_1(x_1) {\cal O}_2(x_2) {\cal O}_3 (x_3) {\cal O}_4 (x_4)
\rangle = \sum_k C_{14}^k C_{23}^k {\bf G}_k^{14,23}(x_1,x_4,x_2,x_3)
$$
which converges around the point $u=1,\, v=0$, or the disk $|1-z|<1$
on the $z$-plane.

The statement of the conformal bootstrap is that the two expansions
are equivalent, in the sense that we have
$$
\sum_k C_{12}^k C_{34}^k {\bf G}_k^{12,34}(x_1,x_2,x_3,x_4)= \sum_k C_{14}^k C_{23}^k {\bf G}_k^{14,23}(x_1,x_4,x_2,x_3)
$$
in the overlapping region of convergence i.e. the region satisfying
$|z|<1$ and $|1-z|<1$ simultaneously. This condition can also be
written as
\eqn\bootap{
\sum_k C_{12}^k C_{34}^k \overline{G}_k^{12,34}(u,v)= \sum_k C_{14}^k C_{23}^k\, {u^{(\Delta_3+\Delta_4)/2}\over v^{(\Delta_2+\Delta_3)/2}}\, \overline{G}_k^{14,23}(v,u)
}

\appendix{B}{CPW expansion of Witten diagrams}

In this appendix we present the CPW expansion of some basic Witten
diagrams.  These are not new results, but to our knowledge explicit
expressions for the CPW coefficients in the case of operators with
different conformal dimensions
(i.e. equations \coefcont, \coefexctwo, \horrcoef) have not been
presented in the literature.

\subsec{Contact Witten diagram}

We only discuss the case of AdS$_5$ i.e. we take $d=4$. The conformal
dimension $\Delta$ is related to the mass of a field in the bulk by
the formula\foot{We assume that $m^2$ is large enough so that there is
no ambiguity in the choice of the branch of the square root.} $\Delta
= 2 +\sqrt{4 + m^2}$. The AdS metric is
$$
ds^2 = {dz_0^2 + d\vec{z}^2 \over z_0^2}
$$
The bulk-to-boundary propagator has the form
$$
K_\Delta(z,x) =
{\Gamma(\Delta) \over \pi^2\Gamma(\Delta-2)} \left({z_0 \over z_0^2 +
(\vec{z}-\vec{x})^2}\right)^\Delta
$$
and the bulk-to-bulk propagator
$$
G_\Delta(z,w) =  {\Gamma(\Delta)\over 2^{\Delta+1} \pi^2 \Gamma(\Delta-1)}\,\,s^\Delta \,\,\,_2F_1
\left({\Delta\over 2},{\Delta+1\over 2};\Delta -1,s^2\right)
$$
where we introduced the geodesic distance in AdS
$$
s = {2 z_0 w_0 \over z_0^2 +w_0^2 + (\vec{z}-\vec{w})^2}
$$

Let us call ${\bf W}^{1234}(x_1,x_2,x_3,x_4)$ the basic contact Witten
diagram for 4 scalar operators ${\cal O}_i$ of conformal dimension
$\Delta_i$.
\eqn\contwitdef{{\bf W}^{1234}(x_1,x_2,x_3,x_4) =
\int {d^5 z \over z_0^5} K_{\Delta_1}(z,x_1)K_{\Delta_2}(z,x_2) K_{\Delta_3}(z,x_3) K_{\Delta_4}(z,x_4)}
For simplicity we take the conformal dimensions to be generic real
numbers\foot{In order to avoid logarithmic terms due to anomalous
dimensions of two-particle operators.}. By massaging the expression for
this Witten diagram we find that it can be decomposed in conformal
blocks in the $(12)\rightarrow (34)$ channel as
\eqn\starcpw{\eqalign{{\bf W}^{1234}(x_1,x_2,x_3,x_4) = &
\sum_{n=0}^\infty a_n(\Delta_1,\Delta_2,\Delta_3,\Delta_4) {\bf G}^{12,34}_{
\Delta_1+\Delta_2 + 2n} (x_1,x_2,x_3,x_4)\cr
 +& \sum_{n=0}^\infty a_n(\Delta_3,\Delta_4,\Delta_1,\Delta_2) {\bf G}^{12,34}_{
\Delta_3+\Delta_4 + 2n} (x_1,x_2,x_3,x_4)}}
where the coefficients $a_n$ are given by
\eqn\coefcont{\eqalign{&a_n(\Delta_1,\Delta_2,\Delta_3,\Delta_4) =   {(-1)^n\over \pi^6}{\Gamma(\Delta_1+n)\Gamma(\Delta_2+n)
\Gamma(\Delta_1 +\Delta_2 +n-2)  \over 2 \Gamma(\Delta_1-2)\Gamma(\Delta_2-2)\Gamma(\Delta_3-2)\Gamma(\Delta_4-2) }\times \cr &\times
{ \Gamma\left(\Sigma-\Delta_3+n\right) \Gamma(\Sigma-\Delta_4+n) \Gamma(\Sigma+n-2)\Gamma(\Sigma-\Delta_1-\Delta_2-n)
\over\Gamma(n+1)\Gamma(\Delta_1+\Delta_2+2n)\Gamma(\Delta_1+\Delta_2+2n-2)}}}
where $\Sigma = {1\over 2}(\Delta_1+\Delta_2+\Delta_3+\Delta_4)$. Our
assumption that the conformal dimensions are generic means that none
of the Gamma functions will blow up for any $n$. 

The physical interpretation of this expansion is that in the
$(12)\rightarrow (34)$ channel, the contact Witten diagram corresponds
to the exchange of 2-particle conformal primaries $:{\cal O}_1
(\nabla^2)^n {\cal O}_2:$ and $:{\cal O}_3 (\nabla^2)^n {\cal O}_4:$
for $n=0,1,...$, together with their descendants as predicted by
conformal invariance.

The same diagram can be expanded in the crossed channel as
\eqn\starcpwcrossed{\eqalign{{\bf W}^{1234}(x_1,x_2,x_3,x_4) = &
\sum_{n=0}^\infty a_n(\Delta_1,\Delta_4,\Delta_2,\Delta_2) {\bf G}^{14,23}_{
\Delta_1+\Delta_4 + 2n} (x_1,x_4,x_2,x_3)\cr
 +& \sum_{n=0}^\infty a_n(\Delta_2,\Delta_3,\Delta_1,\Delta_4) {\bf G}^{14,23}_{
\Delta_2+\Delta_3 + 2n} (x_1,x_4,x_2,x_2)}}
where now we have the exchange of operators $:{\cal O}_1 (\nabla^2)^n
{\cal O}_4:$ and $:{\cal O}_2 (\nabla^2)^n {\cal O}_3:$.  So we can
explicitly see (though perhaps not surprisingly) that the basic
contact diagram is consistent with an OPE expansion in all
channels\foot{Notice that in older literature the statement has
appeared that a single Witten diagram is not fully consistent with an
OPE expansion. However as we can see from the explicit expansion
presented above this is not true. A single Witten diagram is
consistent with an OPE expansion by itself. We believe that the reason
for this confusion is that the 2-particle {\it conformal primary
states} of the form $:{\cal O}_1 (\nabla^2)^n {\cal O}_2:$ etc. were not
properly taken into account in earlier works.}  and is a solution of
the bootstrap equations.

As a test that we got the expansion \coefcont\ right, let us check the
bootstrap condition \bootap\ numerically by summing up the conformal
blocks with coefficients given by \starcpw, \coefcont
, \starcpwcrossed\ in the direct and crossed channels. To do this in
practice, we regularize the infinite sums by taking $n$ up to an
integer $N_{max}$. In the following graph we plot the ratio ${{\rm
direct} \over {\rm crossed}}(N_{max})$ evaluated along the real axis
$z=\overline{z}$ (see Appendix A for the meaning of the variables
$z,\overline{z}$), for increasing values of $N_{max}$. In the range of
the $z$-axis which is plotted, both expansions are convergent so the
ratio should approach the value $1$ for large $N_{max}$. We see that
this is indeed the case.

\fig{Plot of the ratio ${{\rm direct} \over {\rm crossed}}(N_{max})$ as a
function of $z=\overline{z}$ for $N_{max}=5$ (red), $N_{max}=10$
(green) and $N_{max}=25$ (blue). We see that the ratio converges to 1
as we increase $N_{max}$. } {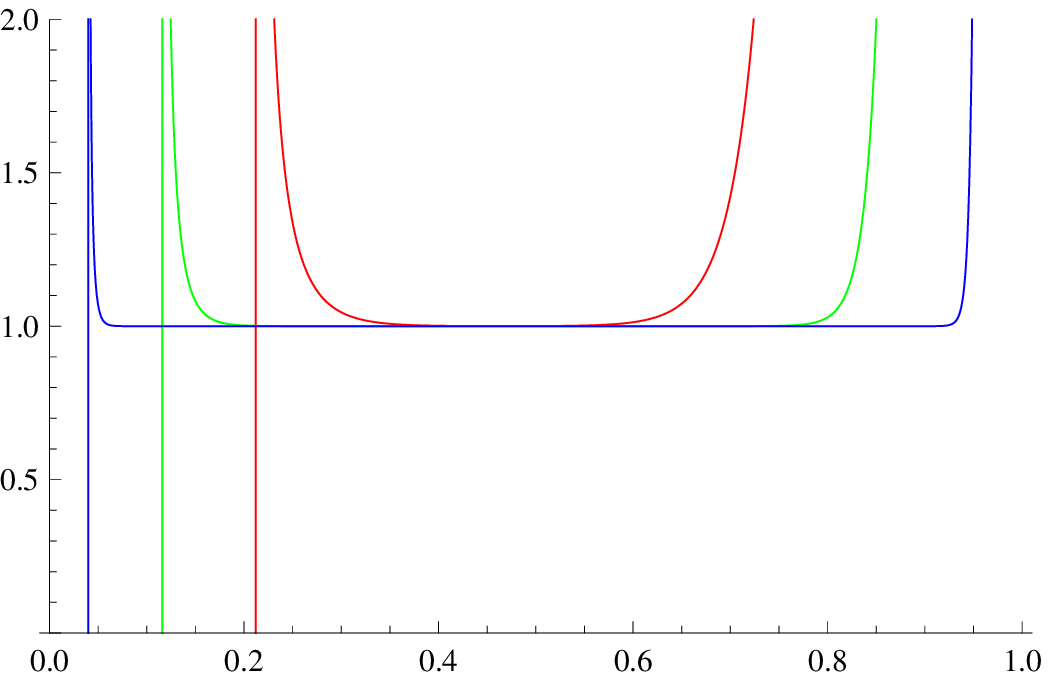}{4.truein}
\figlabel{\curvature}

Let us see what this CPW expansion bodes for the OPE.  By construction
\eqn\coefope{
\lambda \, a_n(\Delta_1, \Delta_2, \Delta_3, \Delta_4)  = 
C_{12}^{:12\,n:} C_{34}^{:12\,n:} \left\langle \CO_{n}^{(12)}
\CO_{n}^{(12)} \right\rangle}
in a hopefully obvious notation. Here $\lambda$ is the coefficient of
the bulk 4-pt vertex (which we had previously set to 1).  As a
consequence of the freedom allowed by $\lambda$ the real physical
content of the expression $a_n$ is not in its value but in the ratio
of $a_n$ with different arguments (i.e.
$a_n(\Delta_1, \Delta_2, \Delta_3, \Delta_4)/a_n(\Delta_3, \Delta_4, \Delta_1,
\Delta_2)$).

Because we know $C_{12}^{:12\,n:}$ and $\langle \CO_{n}^{(12)} \CO_{n}^{(12)}
\rangle$ to first order (from factorization) we can read off the leading $g^2$
component of $C_{34}^{:12\,n:}$ from $a_n$.  More precisely we can
read off the ratio of this coefficient and e.g. the coefficient
$C_{12}^{:34\,n:}$ fixed by $a_n$ with another order of the arguments.

\subsec{Exchange Witten diagram}

Let us now consider the basic exchange Witten diagram ${\bf
W}_k^{12,34}(x_1,x_2,x_3,x_4)$ with operators ${\cal O}_i,\,
i=1,...,4$ of dimensions $\Delta_i$ on the external legs and the
operator ${\cal O}_k$ of dimension $\Delta_k$ dual to the field
exchanged between $(12)$ and $(34)$. We have
$$
{\bf W}_k^{12,34}(x_1,x_2,x_3,x_4) = \int {d^5 z \over z_0^5}
\int{d^5 w \over w_0^5} K_{\Delta_1}(z,x_1)K_{\Delta_2}(z,x_2) G_{\Delta_k}(z,w) K_{\Delta_3}(w,x_3) K_{\Delta_4}(w,x_4)
$$

Let us first present its expansion in conformal partial waves in the
direct (i.e. $(12)\rightarrow (34)$ channel). We have
\eqn\wittenexchangedirect{\eqalign{ {\bf
W}_k^{12,34}(x_1,x_2,x_3,x_4) =&\,
 b(\Delta_1,\Delta_2,\Delta_3,\Delta_4,\Delta_k)\, {\bf
 G}_k^{12,34}(x_1,x_2,x_3,x_4) \cr +& \sum_{n=0}^\infty
 c_n(\Delta_1,\Delta_2,\Delta_3,\Delta_4,\Delta_k)\, {\bf
 G}_{\Delta_1+\Delta_2+2n}^{12,34}(x_1,x_2,x_3,x_4) \cr
 +& \sum_{n=0}^\infty
 c_n(\Delta_3,\Delta_4,\Delta_1,\Delta_2,\Delta_k)\, {\bf
 G}_{\Delta_3+\Delta_4+2n}^{12,34}(x_1,x_2,x_3,x_4) } } with the
 following coefficients
\eqn\coefexcone{\eqalign{
& b(\Delta_1,\Delta_2,\Delta_3,\Delta_4,\Delta_k) =
{\Gamma\left({\Delta_1+\Delta_2 -\Delta_k \over
2}\right)\Gamma\left({\Delta_3+\Delta_4 -\Delta_k \over 2}\right)
\Gamma\left({\Delta_1+\Delta_2 +\Delta_k \over 2}-2\right)\Gamma\left({\Delta_3+\Delta_4 +\Delta_k \over 2}-2\right)\over 8\pi^6 \Gamma(\Delta_1-2) \Gamma(\Delta_2-2) \Gamma(\Delta_3-2) \Gamma(\Delta_4-2)}\times \cr &
\times {
\Gamma\left({\Delta_k+\Delta_1-\Delta_2 \over 2} \right)\Gamma\left({\Delta_k-\Delta_1+\Delta_2 \over 2} \right)
\Gamma\left({\Delta_k+\Delta_3-\Delta_4 \over 2} \right)\Gamma\left({\Delta_k-\Delta_3+\Delta_4 \over 2} \right) 
\over\Gamma(\Delta_k) \Gamma(\Delta_k-1) }}}
and
\eqn\coefexctwo{\eqalign{
& c_n(\Delta_1,\Delta_2,\Delta_3,\Delta_4,\Delta_k)
={(-1)^{n+1}\over \pi^6}
{\Gamma(\Delta_1+n)\Gamma(\Delta_2+n)\Gamma\left(\Delta_1+\Delta_2+n-2\right)
\over 2 \Gamma(\Delta_1-2)\Gamma(\Delta_2-2)\Gamma(\Delta_3-2)\Gamma(\Delta_4-2)} \times \cr & \times
{\Gamma(\Sigma-\Delta_3+n) \Gamma(\Sigma-\Delta_4
+n)\Gamma(\Sigma+n-2)\Gamma(\Sigma-\Delta_1
-\Delta_2-n)\over \Gamma(n+1)\Gamma(\Delta_1+\Delta_2+2n) \Gamma(\Delta_1+\Delta_2+2n-2)
(\Delta_1+\Delta_2-\Delta +2n) (\Delta_1+\Delta_2+\Delta-4 +2n)}}}

In the direct channel (i.e. $(12) \rightarrow (34)$) the exchange
Witten diagram corresponds to the exchange of the conformal block of
the operator ${\cal O}_k$, together with the exchange of the conformal
blocks of 2-particle operators of the form $:{\cal O}_1 (\nabla^2)^n
{\cal O}_2:$ and $:{\cal O}_3 (\nabla^2)^n {\cal O}_4:$.

Note that once more we can relate $b_n$ and $c_n$ to a product of OPE
coefficients.  $b_n$ is proportional to the product of two unknown
(order $g$) coefficients $C_{12}^k C_{34}^k$ but from its structure
there appears to be a natural factorization into two components
depending only on $(\Delta_1,
\Delta_2)$ and $(\Delta_3, \Delta_4)$.  Of course, as in the case with $a_n$,
the expansion is only fixed up to a coefficient (in this case the
product of two three-point vertices $g_{12k}\, g_{34k}$ in the bulk)
so the physical content is actually in the ratio of the $b_n$ to the
$c_n$.  This translates into a statement about the ratios of
e.g. $C_{12}^k$ and $C_{12}^{:34\,n:}$ (the latter being fixed by
$c_n$ via a relation analogous to \coefope).

Expanding the Witten diagram in the crossed channel is more
complicated. Starting with the expressions of \LiuTH\ and following
the method of \HoffmannTR, \HoffmannMX\ we find that the same exchange
Witten can be expanded in the crossed channel (i.e. $(14)\rightarrow
(23)$) as
\eqn\wittenexchangecrossed{\eqalign{ {\bf
W}_k^{12,34}(x_1,x_2,x_3,x_4) =& \sum_{n=0}^\infty
 d_n(\Delta_1,\Delta_4,\Delta_2,\Delta_3,\Delta_k)\, {\bf
 G}_{\Delta_1+\Delta_4+2n}^{14,23}(x_1,x_4,x_2,x_2) \cr
 +& \sum_{n=0}^\infty
 d_n(\Delta_2,\Delta_3,\Delta_1,\Delta_4,\Delta_k)\, {\bf
 G}_{\Delta_2+\Delta_3+2n}^{14,23}(x_1,x_4,x_2,x_3) } } which
 corresponds to the exchange of two-particle operators $:{\cal O}_1
 (\nabla^2)^n{\cal O}_4:$ and $:{\cal O}_2 (\nabla^2)^n{\cal O}_3:$. We have not
been able to find an explicit closed form for the coefficients $d_n$ but they
can be determined by solving the following equations

\eqn\horrcoef{
\sum_{n=0}^\infty
 d_n(\Delta_1,\Delta_4,\Delta_2,\Delta_3,\Delta_k)\, \overline{G}_{\Delta_1+\Delta_4+2n}(u,v) = h\,  u^{\Delta_1+\Delta_4 \over 2}  \sum_{n,m=0}^\infty I_{mn} {u^m (1-v)^n\over m! n!}
}

\noindent where $\overline{G}(u,v)$ are related to the CPWs by equation \cpwdefi\
and

$$
h =  {\Gamma({\Delta_1+\Delta_2+\Delta_3+\Delta_4-4\over 2})
\Gamma({\Delta_1+\Delta_2+\Delta_k-4 \over 2})\Gamma({\Delta_3+\Delta_4+\Delta_k-4 \over 2}) \Gamma({\Delta_3-\Delta_4-\Delta_1+\Delta_2\over 2})\over 8 \pi^6 \Gamma(\Delta_1-2)\Gamma(\Delta_2-2)\Gamma(\Delta_3-2)\Gamma(\Delta_4-2) \Gamma(\Delta_k-1)}
$$
\bigskip
\noindent The coefficients $I_{mn}$ are given by the following integral

\eqn\bbbb{\eqalign{
& I_{mn} = {1\over 2\pi i} \int_{-i \infty}^{+i\infty} ds  {\Gamma({\Delta_1+\Delta_2 \over 2}-s) \Gamma({\Delta_3+\Delta_4
\over 2} - s) \Gamma({\Delta_k \over 2} -s) \Gamma({\Delta_4-\Delta_3 \over 2}+s)
\Gamma({\Delta_1-\Delta_2 \over 2}+
s)
\over  \Gamma({\Delta_1+\Delta_2 +\Delta_3+\Delta_4+\Delta_k-4 \over 2}-s)}\times\cr
& \times {(-s)_m(s+{\Delta_4-\Delta_3 \over 2})_n
(s+{\Delta_1-\Delta_2 \over 2})_n \over ({\Delta_1-\Delta_2-\Delta_3
+\Delta_4 \over 2}+1)_n}\,
_3F_2\left(\matrix{{\Delta_1+\Delta_2+\Delta_k-4 \over 2},
{\Delta_3+\Delta_4+\Delta_k-4 \over 2}, {\Delta_k \over 2}-s\cr
{\Delta_1+\Delta_2+\Delta_3+\Delta_4+\Delta_k -4 \over
2}-s, \Delta_k-1} ;1
\right) }}
where $(a)_m\equiv {\Gamma(a+m) \over \Gamma(a)}$ is the Pochhammer
symbol. Notice that the integral is convergent but we have not been
able to compute it in closed form\foot{One can close the contour to
the right and pick up the residues from sequences of poles, as was
done in \LiuTH, however we have not been able to evaluate the
resulting infinite sums in closed form.}. Hence we have not been able
to find the coefficients $d_n$ explicitly. However they can be
computed, in principle, by solving equation \horrcoef. This can be
done most efficiently by setting $v=1$ and considering \horrcoef\ as
an equation for two functions of $u$. The coefficients $d_n$ can then
be determined order by order starting from matching the lowest order
terms in $u$ and going up. Having determined the coefficients $d_n$
this way, we can then reconsider the dependence on $v$. The fact that
the equation \horrcoef\ remains true for all $u,v$, with the determined coefficients
$d_n$, is a nontrivial result which depends on the fact
that the Witten diagram has a consistent CPW expansion.

\appendix{C}{Central charges in higher dimensional CFTs}

Two dimensional CFTs are characterized by a constant $c$ appearing in
the most singular term of the $T \cdot T$ OPE, in the trace anomaly
and also in the 1-point function of $T$ in a thermal background.  We
have
\eqn\twodimcentral{\eqalign{
 T_{zz}(z) \cdot T_{zz}(0) &= {c \over 2 z^4} + {2 T(z) \over z^2} +
 {\partial T_{zz}(0) \over z}+... \cr T_{z\bar z} &=- {c\over 12}
 R \cr \langle T_{00} \rangle_\beta &= {\pi \over 6}{c \over \beta^2}
 } } with $R$ the Ricci scalar of the manifold.  The 1-point function
 in the last line is evaluated on ${\bf R}\times S^1$ at an inverse
 temperature $\beta$.  Moreover, in a two-dimension QFT with coupling
 constants $g_i$ there exists a function $c(g_i)$ defined via the two
 point function of the stress tensor even away from a conformal point;
 the latter can be shown to flow monotonically along RG flows and to
 coincide with $c$ at fixed points in the flows \ZamolodchikovTI,
\ZamolodchikovGT.  This gives a very direct interpretation of $c$ as counting
the available (massless, in the conformal case) degrees of freedom of the
theory.

In higher dimension, unfortunately, the story is not so straight
forward.  The analogues of the $c$'s appearing in the three lines in
eqn \twodimcentral\ are not generally equal.  In four dimensions, for
instance, several distinct quantities appear.  The stress tensor has a
canonical normalization fixed by the Ward identity
$$
\partial_\mu \langle T_{\mu\nu} (x) \phi (x_1)...\phi (x_n) \rangle = \sum_i
\delta(x - x_i) \langle \phi(x_1) ... \partial_\nu \phi(x_i) ...\phi(x_n)\rangle
$$
With this normalization the two point function takes the following
form
$$\langle T_{\mu\nu}(x) T_{\rho\sigma}(0) \rangle = C_T
f_{\mu\nu\rho\sigma}(x)$$ where $f_{\mu\nu\rho\sigma}(x)$ is a
four-tensor fixed by conformal invariance and independent of the
details of the CFT (for the exact form see e.g.
\OsbornCR).  

If the conformal field theory is placed on a curved manifold scale
invariance is broken by the curvature of the manifold and this is
reflected in the non-vanishing VEV of $T^\mu_{\,\,\mu}$ (which becomes
$T_{z\bar z}$ in 2d)
$$
\langle T^\mu_{\,\,\mu} \rangle = {c\over 16 \pi^2}({\rm Weyl})^2 
-{a \over 16 \pi^2}({\rm Euler})
$$ 
where the two terms correspond to the square of the Weyl tensor and
the Euler density (see \OsbornCR\ for more details).  Note that unlike
the two dimensional case the trace anomaly is characterized by two
constants, $c$ and $a$, rather than one. Of these two, $a$ is
considered a candidate for a higher-dimensional analog to
Zamolodchikov's two-dimensional $c$-theorem.  In supersymmetric
theories this result has even been partially demonstrated
\IntriligatorJJ\ using the fact that these coefficients are related via anomaly
computations to the R-symmetry charges. It can be shown that the coefficients $c$
and $C_T$ are proportional \OsbornCR.

Other interesting constants appear in the 3-point function of stress
energy tensors. In two dimensions the 3-point function is proportional
to $c$. In four dimensions the form of the 3-point function is fixed
by conformal invariance up to 3 constants $A,B,C$ \OsbornCR. A linear
combination of these constants is proportional to $C_T$ (and thus to
the anomaly $c$). Another linear combination is proportional to the
anomaly $a$. The third linear combination is an independent quantity.

Finally if we formulate the CFT at finite temperature then, as
discussed in the paper, operators can admit a temperature-dependent
VEV and conformal invariance constrains the energy density to be of
the form
$$\langle T_{00} \rangle \sim {{\tilde c} \over \beta^d}$$ where
${\tilde c}$ is related by standard thermodynamic arguments to the
entropy density.  Note that, in contrast with the two-dimensional
case\foot{Where $\tilde{c}=c$ characterizes the CFT.}, the constant
$\tilde{c}$ is not protected and, in known examples, varies as a
function of the CFT moduli.  In ${\cal N}=4$ SYM at large $N$, for
instance, it is known to vary by a factor of $3/4$ as the theory goes
from weak to strong coupling. 

Thus in a general 4d CFT we have the following a priori independent
constants: $c,a,\widetilde{c}$, as well as the undetermined linear
combination of the $A,B,C$ constants in the 3-point function.
However, as we argue in the paper, there is reason to believe that as
one is sent to infinity the others must scale in essentially the same
way\foot{The relative scaling of $a, c,A,B,C$, and $C_T$ in the limit
$C_T \rightarrow \infty$ follows, to some extent, from the
inequalities derived in \HofmanAR.}  (particularly $C_T$ and ${\tilde
c}$ with which we are mostly concerned).

\listrefs

\end